%% file: main.tex
\documentclass[review]{elsarticle}

\usepackage[colorlinks,citecolor=blue,linktoc=all,linkcolor=cyan]{hyperref}
\usepackage{graphicx}

\usepackage[T1]{fontenc}
\usepackage{dsfont}               
\usepackage{mathrsfs}             
\usepackage{slashed}              
\usepackage{amsmath}
\usepackage{amssymb}
\usepackage{amsbsy}
\usepackage{amsfonts}
\usepackage{physics} 
\usepackage{diffcoeff} 
\usepackage{subcaption} 
\usepackage{orcidlink}
\numberwithin{equation}{section}
\numberwithin{table}{section}
\numberwithin{figure}{section}

\newcommand{\MSbar}{\overline{\text{MS}}}
\newcommand{\N}{NLO}

\newcommand{\NLR}{(NLO+LRR)$\times$RGR}
\newcommand{\NN}{NNLO}

\newcommand{\NNLR}{(NNLO+LRR)$\times$RGR}
\newcommand{\LambdaQCD}{\Lambda_\text{QCD}}

\journal{Progress in Particle and Nuclear Physics}

\usepackage{titlesec}
\usepackage{sectsty}
\titleformat{\section}{\normalfont\Large\bfseries}{\thesection}{1em}{}
\titleformat{\subsection}{\normalfont\large\bfseries}{\thesubsection}{1em}{}
\titleformat{\subsubsection}{\normalfont\normalsize\bfseries}{\thesubsubsection}{1em}{}

\begin{document}

\begin{frontmatter}

\title{Mapping Parton Distributions of Hadrons with Lattice QCD}

\author[mysecondaryaddress]{Huey-Wen Lin \corref{mycorrespondingauthor}}
  \cortext[mycorrespondingauthor]{Corresponding author}
  \ead{hwlin@pa.msu.edu}

  \address[mymainaddress]{Department of Physics and Astronomy, Michigan State University, East Lansing, MI 48824}

\begin{abstract}
The strong force which binds hadrons is described by the theory of quantum chromodynamics (QCD).
Determining the character and manifestations of QCD is one of the most important and challenging outstanding issues necessary for a comprehensive understanding of the structure of hadrons.
Within the context of the QCD parton picture, the parton distribution functions (PDFs) have been remarkably successful in describing a wide variety of processes.
However, these PDFs have generally been confined to the description of collinear partons within the hadron.
New experiments and facilities provide the opportunity to additionally explore the three-dimensional structure of hadrons, which can be described by generalized parton distributions (GPDs), for example.

In recent years, a breakthrough was made in calculating the Bjorken‐$x$
dependence of PDFs in lattice QCD by using large‐momentum effective theory (LaMET) and other similar frameworks.
The breakthrough has led to the emergence and rapid development of direct calculations of Bjorken-$x$--dependent structure.
In this review article, we show some of the recent progress made in lattice QCD in PDFs and GPDs and discuss future challenges.
\end{abstract}


\end{frontmatter}

\newpage

\thispagestyle{empty}
\tableofcontents


\newpage
\input{sec1-intro}

\newpage
\input{sec2-methodology}

\newpage
\input{sec3-PDFs}

\newpage
\input{sec4-GPDs}

\newpage
\input{sec5-summary-future}

\newpage
\section*{Acknowledgements}
HL thanks the MILC Collaboration for sharing the lattices used to perform many of the parton-distribution studies quoted in this review. 
The work of HL are partially supported by the US National Science Foundation under grant PHY 1653405 ``CAREER: Constraining Parton Distribution Functions for New-Physics Searches'', grant PHY 2209424,
by the U.S.~Department of Energy under contract DE-SC0024582, and by the Research Corporation for Science Advancement through the Cottrell Scholar Award.

\bibliography{refs}
\end{document}

%% file: sec1-intro.tex
\section{Introduction}\label{intro}

Gluons and quarks are the underlying degrees of freedom in quantum chromodynamics (QCD) that explain the properties of hadrons, such as the nucleon and pion.
Fully understanding how they contribute to the properties of hadrons (such as their mass or spin structure) helps to decode the Standard Model that rules our physical world. 
Many mysteries remain after decades of experimental effort;
for example, what is the origin of the proton mass?
How are sea quarks and gluons, and their spins, distributed in space and momentum inside the nucleon?
A great deal has been learned about hadron PDFs from analysis of hard-scattering experiments and others since the 1960s (see example reviews and latest results in Refs.~\cite{Achenbach:2023pba,Amoroso:2022eow,Ethier:2020way,Lin:2017snn,Ji:2020ect,Constantinou:2020hdm,Lin:2023kxn}), and these measurements have provided a standard against which theoretical calculations can be judged. 
However, greater experimental precision is required to answer these remaining QCD mysteries;
hence, the exploration of hadron structure continues: in the US, physics targets are being pursued by multiple DOE laboratories, Brookhaven National Lab (BNL) and Jefferson Lab (JLab), as well as the future Electron-Ion Collider (EIC)~\cite{AbdulKhalek:2021gbh,Accardi:2012qut,Burkert:2022hjz}.
Worldwide, facilities such as GSI in Germany and J-PARC in Japan will join the effort, and future facilities are being considered, such as an EIC in China and the LHeC at CERN. 
The pursuit of PDFs has led to collaborations of theorists and experimentalists working side-by-side to take advantage of all available data, evaluating different combinations of input theories, parameter choices and assumptions, resulting in multiple global-PDF determinations.
Comparison of these different global-fit determinations of the PDFs is important to reveal hidden uncertainties in PDF data sets.
Often, in kinematic regions where experimental data are plentiful or overconstrained, such as the mid-$x$ region of the PDFs, there is consistency among different PDF data sets.
However, in the regions where data are sparse or suffer from complicated nuclear effects, such as at small- or large-$x$ or for heavy-flavor PDFs, disagreements are seen.
For more details, we refer readers to the non-technical review in Ref.~\cite{Lin:2017snn}. 
A nonperturbative approach from first principles can provide the necessary inputs to fill gaps in the experimental data or add information to constrain global fits.

Lattice quantum chromodynamics (LQCD) is an ideal theoretical tool to study the parton structure of hadrons, starting from quark and gluon degrees of freedom.
LQCD is a regularization of continuum QCD using a discretized four-dimensional spacetime;
it contains a small number of natural parameters, such as the strong coupling constant and quark masses.
LQCD discretizes four-dimensional continuum QCD to allow the study of the strong-coupling regime of QCD, where perturbative approaches converge poorly.
As in continuum QCD, we calculate an observable of interest through a path integral:
\begin{equation}
 \langle 0| O (\overline{\psi},\psi,A)|0 \rangle =\frac{1}{Z}\int [\dl A][\dl{\overline{\psi}}][\dl \psi]
O(\overline{\psi},\psi,A) e^{i\int\!\dl x^4\mathcal{L}_\text{QCD}(\overline{\psi},\psi,A)},
\end{equation}
where $\mathcal{L}_\text{QCD}$ is the sum of the pure-gauge and fermion Lagrangian, $O$ is the operator that gives the correct quantum numbers for our observable, and $Z$ is the partition function of the space-time integral of the QCD Lagrangian.
It is straightforward to carry out this path integral numerically within a finite spacetime volume and under an ultraviolet cutoff (the lattice spacing $a$).
Unlike continuum QCD, LQCD works in Euclidean spacetime (rather than Minkowski), and the coupling and quark masses can be set differently than those in our universe.
The theory contains two scales that are absent in continuum QCD, one ultraviolet (the lattice spacing $a$) and one infrared (the spatial extent of the box $L$);
this setup keeps the number of degrees of freedom finite so that LQCD can be solved on a computer.
For observables that have a well-defined operator in the Euclidean path integral for numerical integration, we can find their values in continuum QCD by taking the limits $a \to 0$, $L \to \infty$ and quark mass $m_q \to m_q^\text{phys}$. 

In order to make predictions using QCD on the lattice, we calculate observables corresponding to vacuum expectation values of operators $O$, taking the form
\begin{equation}
\langle O \rangle  =
  \frac{1}{Z} \int [\dl U][\dl \psi][\dl{\overline{\psi}}]
              e^{-S_F(U, \psi, \overline{\psi})-S_G(U)}O(U, \psi,\overline\psi) \nonumber 
\end{equation}
where
\begin{equation}\label{eq:Z}
 Z  =  \int [\dl U][\dl \psi][\dl{\overline{\psi}}] e^{-S_F-S_G},
\end{equation}
$S_G$ is the gauge action and $S_F=\overline{\psi} M \psi$ is the fermion action with Dirac operator $M$.
The bilinear structure of the fermion action allows the integration over the fermion fields to be done explicitly, bringing down a factor of $\det M$.
This means that the anticommuting fermion fields (impossible to simulate on a computer) are integrated out, leaving an integrand that depends only the values of the gauge fields: 

\begin{equation}
\langle O \rangle  = \frac{1}{Z} \int [\dl U] \det M\, e^{-S_G(U)} O(U). \nonumber 
\end{equation}
In the early days of lattice QCD, computational resources were insufficient to compute the fermionic determinant.
Instead, the determinant was approximated by a constant. 
This is equivalent to removing quark loops from the Feynman diagrams of a perturbative expansion, and this technique became known as the ``quenched approximation''.
Although quenching retains many of the important properties of QCD, such as asymptotic freedom and confinement, it introduces an uncontrollable systematic error, and modern calculations keep at least the up, down, strange and charm quarks in the sea.

The discrete integral we have derived can be evaluated numerically using Monte-Carlo methods.
Monte-Carlo integration uses random points within the domain of gauge configurations to approximately evaluate the integral.
The ``importance sampling'' technique is introduced to perform this task more efficiently: instead of choosing points from a uniform distribution, they are chosen from a distribution proportional to the Boltzmann factor $e^{-S_\text{eff}(U)}$ which is $\det M\, e^{-S_G(U)}$, which concentrates the points where the function being integrated is large.

\begin{figure}[tb]
\includegraphics[width=0.55\linewidth]{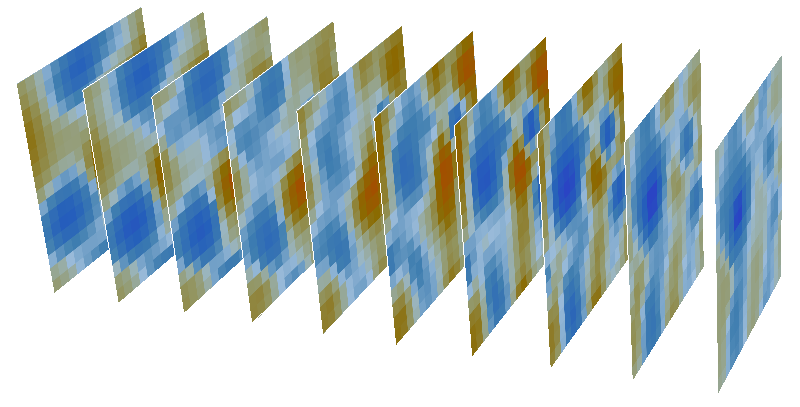}
\includegraphics[width=0.4\linewidth]{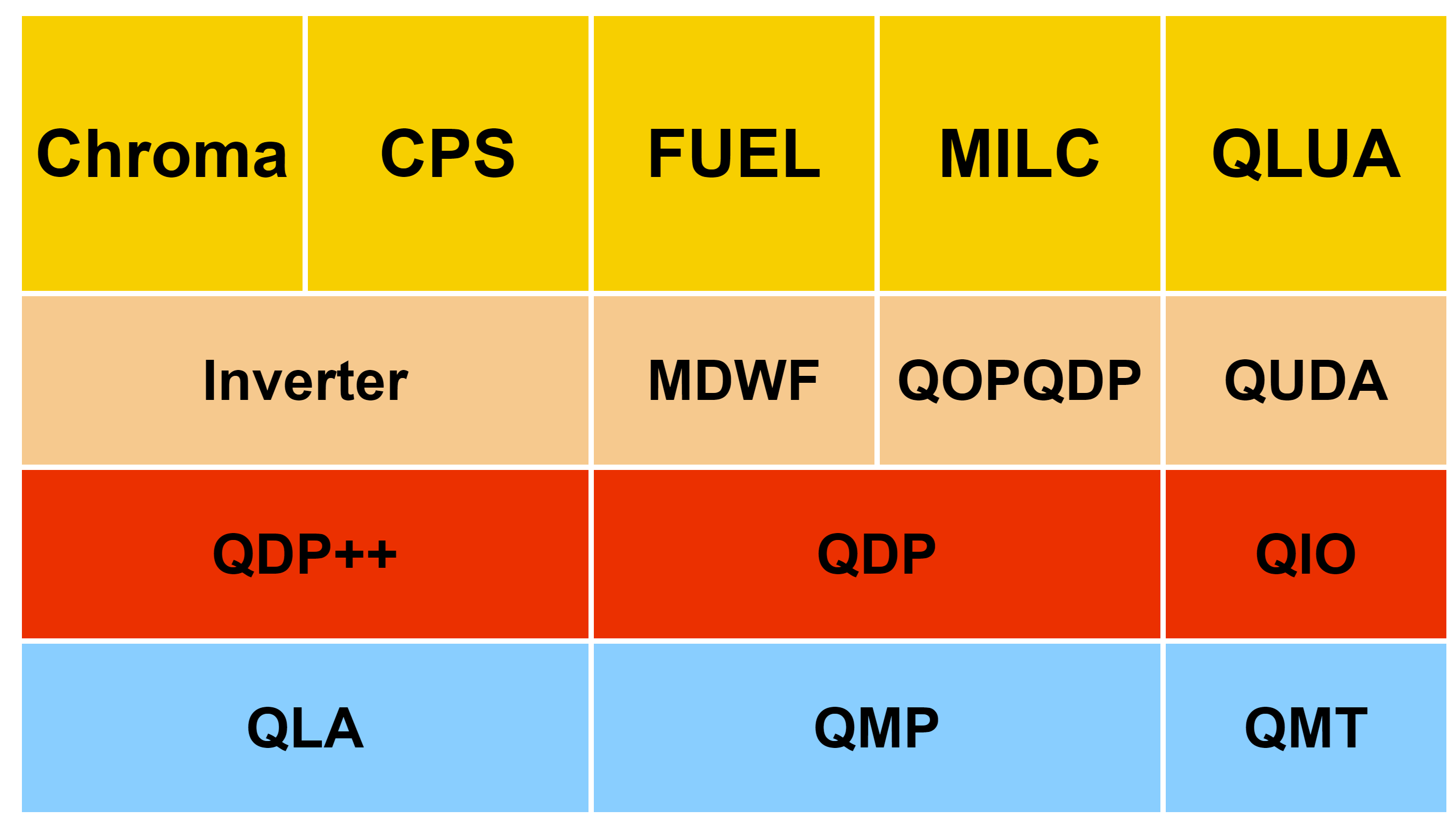}
\caption{
(left)
Illustration of an example four-dimensional topological QCD vacuum of the $N_f=2+1+1$ gauge field configurations (generated by MILC collaboration) represented in two-dimensional frame. 
(right) The SciDAC Layers showing the modular software architecture; taken from USQCD software website, \url{https://usqcd-software.github.io}.
\label{fig:lattice101}}
\end{figure}

Using this method, we accumulate an ensemble of gauge-field configurations generated using a Markov-chain technique. 
Based on the current state of the gauge configuration, a new configuration is selected.
A transition probability $P([U^\prime] \leftarrow [U])$ is determined based solely on this new configuration and the current one.
The new configuration is added to the ensemble or rejected, according to the resulting probability.
If the ``detailed balance'' condition
\begin{equation}
P([U^\prime] \leftarrow [U]) e^{-S_\text{eff}(U)} =
    P([U] \leftarrow [U^\prime]) e^{-S_\text{eff}(U^\prime)}
\end{equation}
is satisfied by this update procedure, then the canonical ensemble is a fixed point of the transition probability matrix.
Under this condition, repeated updating steps will bring the gauge-field distribution to the canonical ensemble.
One can then save each of the gauge-field configurations;
that is, effectively taking snapshots of QCD vacuum (see the right-hand side of Fig.~\ref{fig:lattice101} for an illustration).
These calculations require high-performance supercomputer centers with software developed by and shared among the community.
The right-hand side of Fig.~\ref{fig:lattice101} shows an example of commonly used software from the US lattice community (i.e. USQCD collaboration) and their dependencies.
Once the proper QCD vacuum is obtained, one can move on to calculate QCD observables, such as spectroscopy and structure of hadrons.
One will then repeat the calculations as many times as the available QCD ensembles; these forms the statistical errors in the lattice-QCD hadronic observables.

\begin{figure}[htbp]
\centering
\centering
\includegraphics[width=0.8\textwidth]{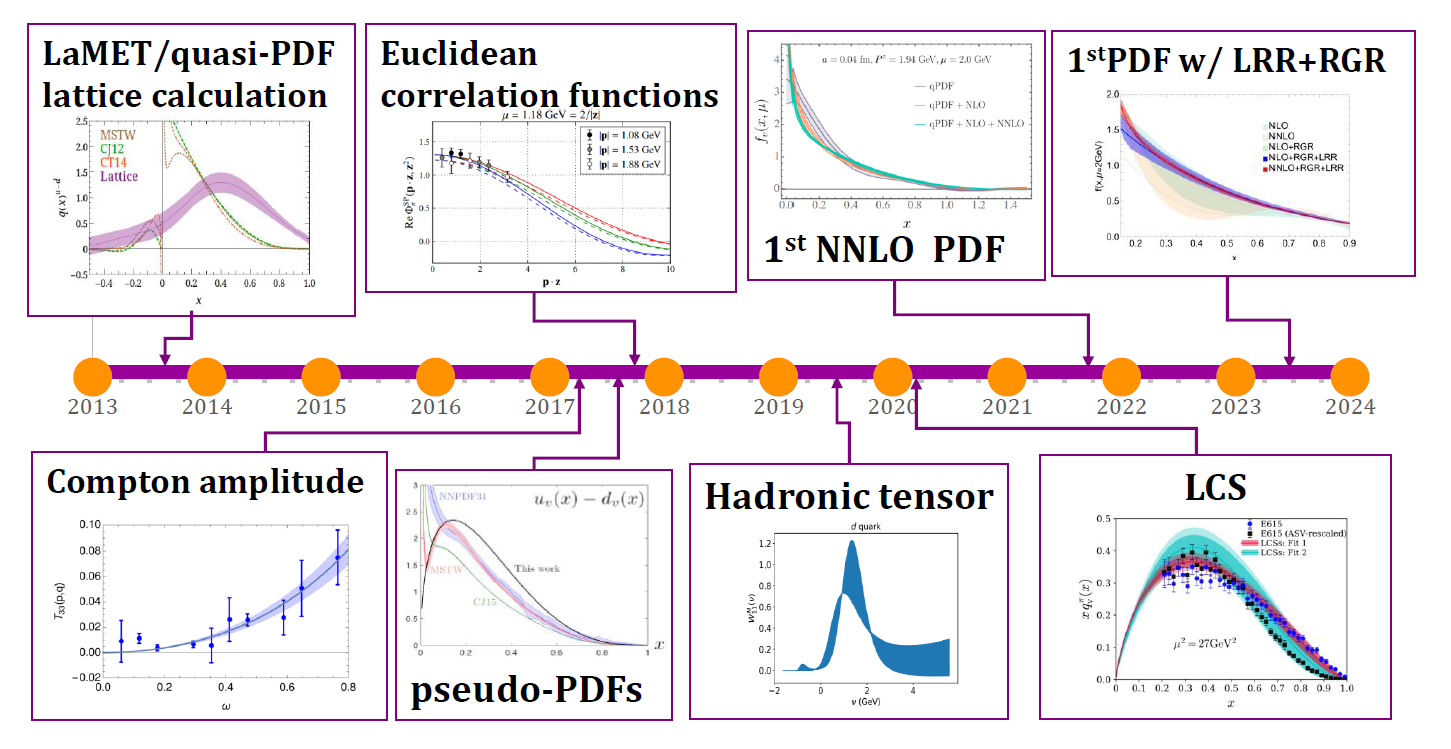}
\caption{ A timeline showing the rapid development of the lattice methods for $x$-dependent hadron structure calculations, starting from the quasi-PDF (or LaMET method) in unpolarized and polarized isovector nucleon calculation, toward first of the many methods used in different approaches and hadron structure~\cite{Lin:2014zya,Chambers:2017dov,Radyushkin:2017cyf,Bali:2017gfr,Liu:2020okp,Sufian:2020vzb,Gao:2021hxl,Zhang:2023bxs}
}\label{Fig:timeline-method}
\end{figure}

\begin{figure}[htbp]
\centering
\centering
\includegraphics[width=0.8\textwidth]{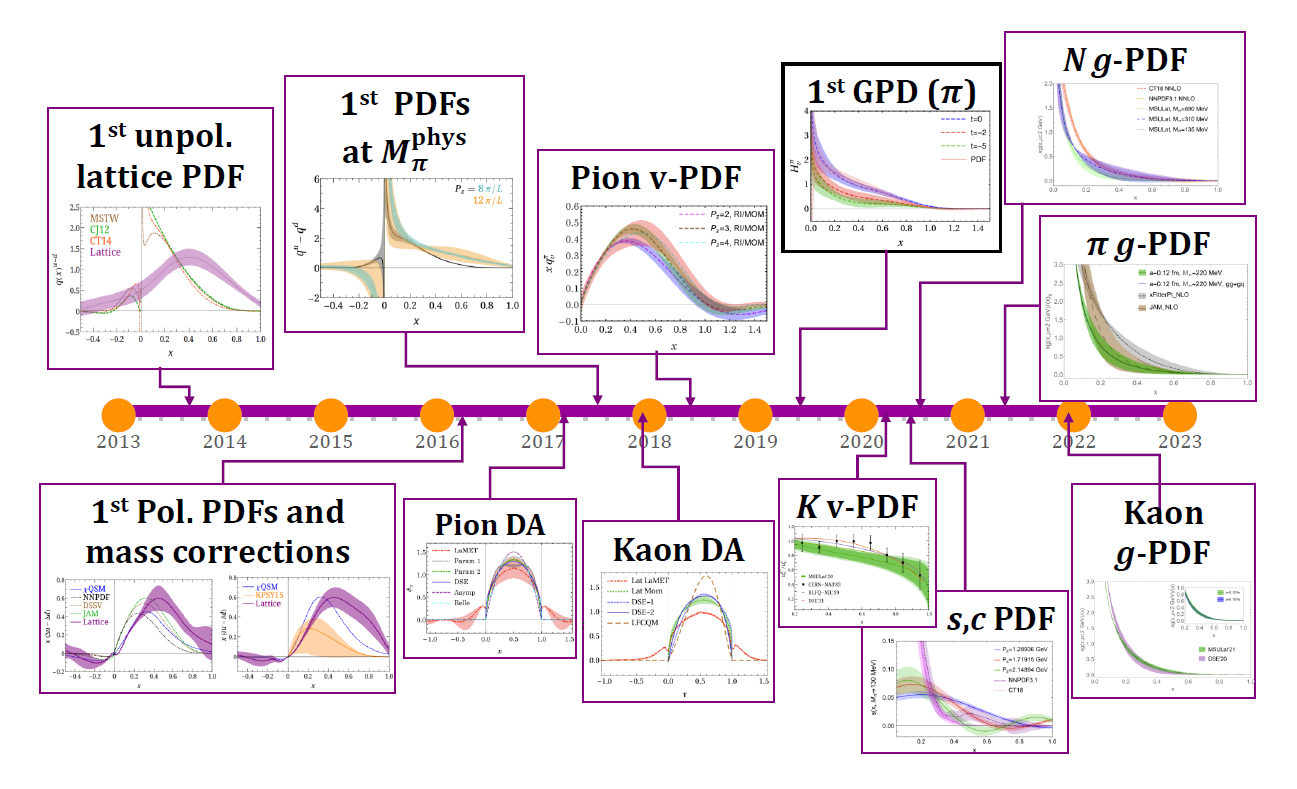}
\caption{
Timeline from 2013 to 2023 of the first $x$-dependent observables calculated on the lattice~\cite{Lin:2014zya,Chen:2016utp,Zhang:2017bzy,Lin:2017ani,Zhang:2017zfe,Zhang:2018nsy,Chen:2019lcm,Lin:2020ssv,Zhang:2020dkn}.
\label{Fig:timeline-observables}
}
\end{figure}

For a long while, lattice PDF calculations were limited to lattice observables, ``moments'' only, that is, where the $x$ dependence of the parton distributions is integrated out.
A breakthrough came in 2013 when
Large-momentum effective theory (LaMET)~\cite{Ji:2013dva,Ji:2014gla} enables computation of the Bjorken-$x$ dependence of hadron PDFs on a Euclidean lattice.
LaMET relates equal-time spatial correlators, whose Fourier transforms are called quasi-PDFs, to PDFs in the limit of infinite hadron momentum.
For large but finite momenta accessible on a realistic lattice, LaMET relates quasi-PDFs to physical ones through a factorization theorem, the proof of which was developed in Refs.~\cite{Ma:2017pxb,Izubuchi:2018srq,Liu:2019urm}.
Since LaMET was proposed, a lot of progress has been made in the theoretical understanding of the formalism~\cite{Xiong:2013bka,Ji:2015jwa,Ji:2015qla,Xiong:2015nua,Ji:2017rah,Monahan:2017hpu,Stewart:2017tvs,Izubuchi:2018srq,Xiong:2017jtn,Wang:2017qyg,Wang:2017eel,Xu:2018mpf,Ishikawa:2016znu,Chen:2016fxx,Ji:2017oey,Ishikawa:2017faj,Rossi:2017muf,Ji:2017rah,Briceno:2018lfj,Hobbs:2017xtq,Jia:2017uul,Xu:2018eii,Jia:2018qee,Spanoudes:2018zya,Rossi:2018zkn,Liu:2018uuj,Ji:2018waw,Radyushkin:2018nbf}.
The method has been applied in lattice calculations of PDFs for the up and down quark content of the nucleon~\cite{Lin:2014zya,Chen:2016utp,Lin:2017ani,Alexandrou:2015rja,Alexandrou:2016jqi,Alexandrou:2017huk,Chen:2017mzz,Lin:2018pvv,Alexandrou:2018pbm,Chen:2018xof,Alexandrou:2018eet,Lin:2018qky,Liu:2018hxv,Wang:2019tgg,Lin:2019ocg,Liu:2020okp,Lin:2019ocg,Zhang:2019qiq,Alexandrou:2020qtt},
$\pi$~\cite{Chen:2018fwa,Izubuchi:2019lyk,Lin:2020ssv,Gao:2020ito} and $K$~\cite{Lin:2020ssv} mesons,
and the $\Delta^+$~\cite{Chai:2020nxw} baryon.
Despite limited volumes and relatively coarse lattice spacings, previous state-of-the-art nucleon isovector quark PDFs, determined from lattice data at the physical pion mass, have shown reasonable agreement~\cite{Lin:2018pvv,Alexandrou:2018pbm} with phenomenological results extracted from the experimental data.
Encouraged by this success, LaMET has also been extended to twist-three PDFs~\cite{Bhattacharya:2020cen,Bhattacharya:2020xlt,Bhattacharya:2020jfj},
as well as gluon \cite{Fan:2018dxu,Fan:2020cpa},
strange and charm distributions~\cite{Zhang:2020dkn}.
It was also applied to meson distribution amplitudes~\cite{Zhang:2017bzy,Chen:2017gck,Zhang:2020gaj,Hua:2020gnw,LatticeParton:2022zqc,Gao:2022vyh,Baker:2024zcd,Blossier:2024wyx,Cloet:2024vbv}
and generalized parton distributions (GPDs)~\cite{Chen:2019lcm,Alexandrou:2020zbe,Lin:2020rxa,Alexandrou:2019lfo}.
Attempts have also been made to generalize LaMET to transverse momentum dependent (TMD) PDFs~\cite{Ji:2014hxa,Ji:2018hvs,Ebert:2018gzl,Ebert:2019okf,Ebert:2019tvc,Ji:2019sxk,Ji:2019ewn,Ebert:2020gxr},
to calculate the nonperturbative Collins-Soper evolution kernel~\cite{Ebert:2018gzl,Shanahan:2019zcq,Shanahan:2020zxr}
and soft functions~\cite{Zhang:2020dbb} on the lattice. 
Alternative approaches to access $x$-dependent structure in lattice QCD are also proliferating; for example,
the Compton-amplitude approach (or ``OPE without OPE'')~\cite{Aglietti:1998ur,Martinelli:1998hz,Dawson:1997ic,Capitani:1998fe,Capitani:1999fm,Ji:2001wha,Detmold:2005gg,Braun:2007wv,Chambers:2017dov,Detmold:2018kwu,QCDSF-UKQCD-CSSM:2020tbz,Horsley:2020ltc,Detmold:2021uru},
the ``hadronic-tensor approach''~\cite{Liu:1993cv,Liu:1998um,Liu:1999ak,Liu:2016djw,Liu:2017lpe,Liu:2020okp},
the ``current-current correlator''~\cite{Braun:2007wv,Ma:2017pxb,Bali:2017gfr,Bali:2018spj,Joo:2020spy,Gao:2020ito,Sufian:2019bol,Sufian:2020vzb}
and the pseudo-PDF approach~\cite{Radyushkin:2017cyf,Balitsky:2019krf,Orginos:2017kos,Karpie:2017bzm,Karpie:2018zaz,Karpie:2019eiq,Joo:2019jct,Joo:2019bzr,Radyushkin:2018cvn,Zhang:2018ggy,Izubuchi:2018srq,Joo:2020spy,Bhat:2020ktg,Fan:2020cpa,Sufian:2020wcv,Karthik:2021qwz,HadStruc:2021wmh,Fan:2021bcr,HadStruc:2022yaw,Salas-Chavira:2021wui,Fan:2022kcb}.
A few works have started to include lattice-QCD systematics, such as finite-volume effects~\cite{Lin:2019ocg,Sufian:2020vzb} and lattice-spacing dependence for quark~\cite{Lin:2020fsj,Alexandrou:2020qtt,Karpie:2021pap,Zhang:2020gaj,Lin:2020ssv,Gao:2022iex}
and gluon~\cite{Fan:2022kcb,Salas-Chavira:2021wui,Fan:2021bcr} distributions, in their $x$-dependent structure calculations.
Most lattice calculations of PDFs use next-to-leading-order (NLO) matching~\cite{Xiong:2013bka,Ma:2014jla,Ji:2017rah,Ji:2020ect}, but recently some lattice calculations of the valence pion PDF~\cite{Gao:2021dbh} have incorporated NNLO matching~\cite{Chen:2020ody,Li:2020xml}.
Figures~\ref{Fig:timeline-method} and \ref{Fig:timeline-observables} shows the important milestones that has been made in lattice QCD in $x$-dependent hadron structure in the past decade. 
We refer interested readers to these recent reviews for more details on lattice moments~\cite{Lin:2022nnj,Lin:2017snn,Constantinou:2020hdm,Constantinou:2022yye}.

In this review, we will focus on the $x$-dependent parton-distribution results from two popular methods: the LaMET (or sometimes also called ''quasi-PDF'') and pseudo-PDF method.
We will briefly describe the methods currently used on the lattice in Sec.~2, focusing on selected PDFs and GPDs results that demonstrate the lattice progress in Sec.~3 and 4, and summary of the state-of-the-art and remaining challenges in the future in Sec.~5.

%% file: sec2-methodology.tex
\section{Methodology}

\subsection{$x$-dependent Parton Distribution Methodologies in Lattice QCD \label{sec:Methodology}}

In this section, we briefly review recent approaches to determining the $x$-dependence of parton distributions from lattice QCD.
For this review, we will only focus on those new methods whose lattice-calculated matrix elements can be treated similarly to the experimental cross-sections to be combined in the global-QCD analysis.
These will one day be used to extract the best knowledge from both available lattice-QCD and experimental data.
Modern moment methods, such as Refs.~\cite{Davoudi:2012ya} and \cite{Shindler:2023xpd}, that extend  the traditional moment method beyond the leading few moments are not included in this review.

\subsubsection{Hadronic Tensor}

Parton distributions can be determined from hadronic tensors provided that the higher-twist contributions, which have different $Q^2$ dependence than the leading-twist, can be subtracted.
Using the hadronic-tensor method in the Euclidean path-integral approach has the advantage that no complicated lattice renormalization is required if conserved vector currents are used in the current-current correlation, and the renormalization for the local currents is well understood on the lattice. 
Furthermore, hadronic tensors can be calculated in any momentum frame of the nucleon, since the structure functions are frame-independent.
One can choose the nucleon momenta and momentum transfers judiciously to yield the desired coverage of $x$ for a given $Q^2$.
However, the inverse Laplace transform that is needed to convert the hadronic tensor from Euclidean space to Minkowski space can be a challenge~\cite{Liu:1993cv,Liu:1999ak}.
Three numerical approaches, the Backus-Gilbert method~\cite{Hansen:2017mnd}, improved maximum entropy, and fitting with model spectral functions, are suggested to tackle this inverse Laplace-transform problem~\cite{Liu:2016djw}.
The hadronic-tensor approach also provides a promising avenue to directly evaluate of the lepton-nucleon scattering cross sections across quasi-elastic, resonance, shallow-inelastic, and deep inelastic regions, which gives its special applications among the available $x$-dependent methods. 
Examples of recent results using hadronic tensor can be found in Ref.~\cite{Liang:2023uai,Liang:2019frk}.

\subsubsection{Compton Amplitude Method}
\label{Sec:ComptonAmplitude}

Both unpolarized and polarized PDFs are accessible, theoretically and
experimentally, through the forward Compton scattering amplitude
\begin{equation}
\label{eq:Compton}
T_{\mu\nu}(p,q,s) =
\int \text{d}^4z\, e^{iqz} \langle p,s |T J_\mu(z) J_\nu(0)|p,s\rangle
\end{equation}
at large virtual photon momenta $q^2=-Q^2$.
Here $T$ is the time-ordering operator, $J_\mu(z)$ and $J_\nu(0)$ are vector currents at spacetime points $z$ and $0$ respectively, and the external states are hadronic states with momentum $p$ and spin $s$.
The Compton-amplitude method proposes to directly calculate $T_{\mu\nu}(p,q)$ in lattice QCD.
The disconnected contributions can be calculated by application of the Feynman-Hellmann technique to lattice QCD~\cite{Horsley:2012pz,Chambers:2014qaa,Chambers:2015bka,Chambers:2017dov}.
The Compton amplitude will be dominated by the leading twist-two contributions, provided one works at sufficiently large $Q^2$.
Varying $Q^2$ allows one to test the twist expansion and even isolate twist-four contributions~\cite{QCDSF-UKQCD-CSSM:2020tbz}.
Moreover, one can distinguish between contributions from up, down and strange quarks, connected and disconnected, by appropriate insertions of the electromagnetic current.
Recent example results of the Compton-amplitude method can be found in Ref.\cite{CSSMQCDSFUKQCD:2021lkf,Hannaford-Gunn:2024aix}.

\subsubsection{Lattice Cross Sections}

A lattice cross section is defined as a single-hadron matrix element of a time-ordered, renormalized, nonlocal operator ${\cal O}_n(z)$: ${\sigma}_{n}(\nu,z^2,p^2)=\langle p| {T}\{{\cal O}_n({z})\}|p\rangle$ with four-vectors $p$ and $z$ effectively defining the collision kinematics; 
the choice of ${\cal O}_n$ determines the dynamical features of the lattice
cross section (LCS)~\cite{Ma:2017pxb}. 
An example class of good LCS can be constructed in terms of a
correlation of two renormalizable currents,
${\cal O}_{j_1j_2}(z)\equiv z^{d_{j_1}+d_{j_2}-2} Z_{j_1} Z_{j_2}\, j_1(z) j_2(0)$,
where $Z_j$ is the renormalization constant of the current $j$ with dimension $d_j$.
There are many choices for the current, such as a vector quark current $j_q^V(z) = \overline{\psi}_q(z)\gamma\cdot{z}\, {\psi}_{q}(z)$, or a tensor gluonic current, $j_g^{\mu\nu}(z)\propto F^{\mu\rho}(z){F_{\rho}}^\nu(z)$.
Different combinations of the two currents could help enhance the lattice cross sections' flavor dependence.
If $z^2$ is sufficiently small, the lattice cross section constructed from two
renormalizable currents can be factorized into PDFs~\cite{Ma:2017pxb},
\begin{equation}
\label{eq:fac}
{\sigma}_{n}(\nu,z^2,p^2) = \sum_{a}\int_{-1}^1 \frac{\dl x}{x}\, f_{a}(x,\mu^2)
K_{{n}}^{a}(x\nu,z^2,x^2p^2,\mu^2) +O(z^2\Lambda_\text{QCD}^2)\, ,
\end{equation}
where $\mu$ is the factorization scale,
$K_n^{a}$ are perturbatively calculable hard coefficients,
and $f_{a}$ is the parton distribution. 
Note that a subset of LCS, including the short-distance factorization of spatial current-current correlators, was explored by Braun and Mueller in 2007~\cite{Braun:2007wv}. 
Parton distributions can be extracted from fits of LQCD data for various LCS $\sigma_{n}(\nu,z^2,p^2)$ with corresponding perturbatively calculated coefficients $K_n^{a}$ in Eq.~\ref{eq:fac}.
Recent example results using the lattice-cross-sections method can be found in Ref.~\cite{Sufian:2020vzb}.

\subsubsection{Pseudo-PDF}

\begin{figure}[tb]
\includegraphics[width=0.32\textwidth]{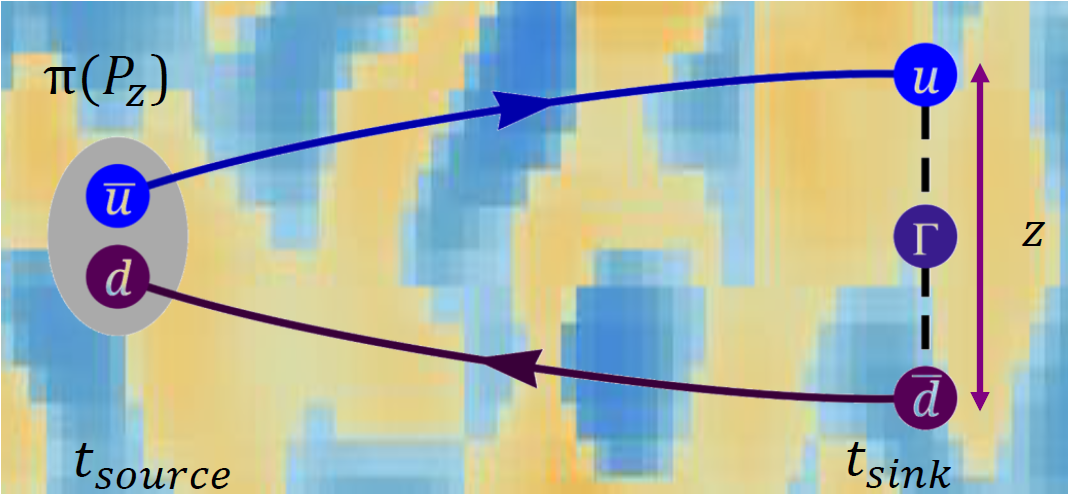}
\includegraphics[width=0.31\textwidth]{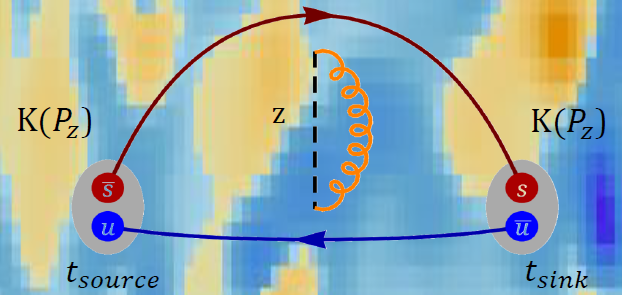}
\includegraphics[width=0.36\textwidth]{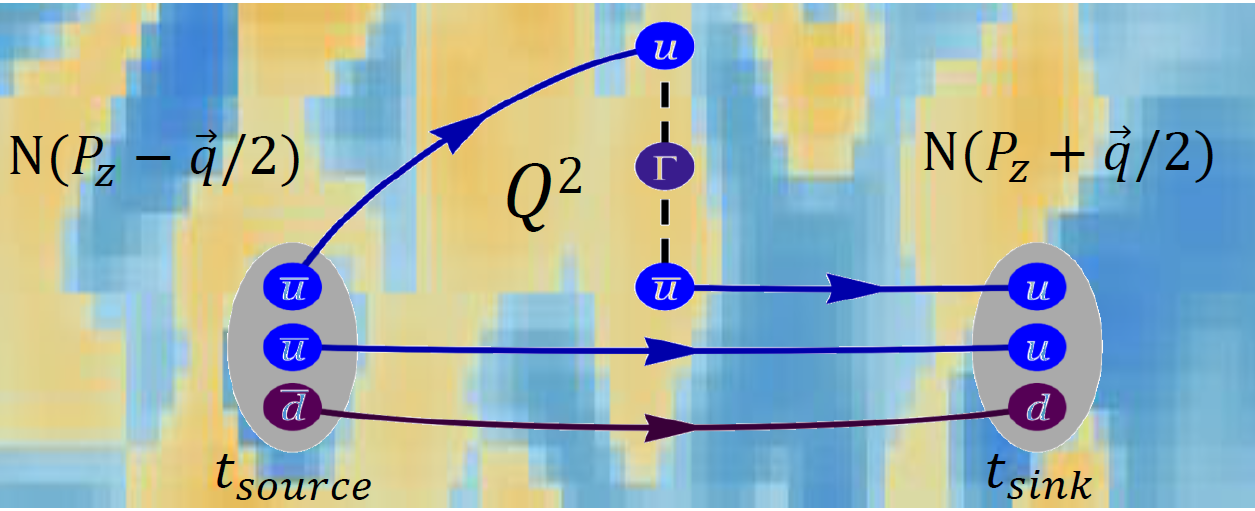}
\caption{
Illustration of examples pion distribution amplitude (left), kaon gluon PDF (middle) and nucleon GPDs in Breit frame (right) matrix-element calculation depicted on top of the LQCD vacuum.
\label{fig:feyman-diag}}
\end{figure}

For the pseudo-PDF method, one will first calculate three-point correlators, as shown in Fig.~\ref{fig:feyman-diag} and then extract the ground-state matrix elements of the desired hadron $h(z,P_z)=\langle H(P_z)|{\cal O}(z)|H (P_z)\rangle$, where $P$ is the hadron momentum and $z$ is the Wilson displacement of the quark and antiquark (or gluon) fields in ${\cal O}(z)$.
One can then introduce $\overline{h}(\nu,z^2) \equiv h(z,p_z)$ as a function of the Lorentz invariants $\nu=z\cdot p$ (Ioffe time~\cite{Ioffe:1969kf,Braun:1994jq}) and $z^2$, where $z$ and $p$ are general 4-vectors and compute the reduced Ioffe-time pseudo-distribution (RpITD)~\cite{Radyushkin:2017cyf,Orginos:2017kos,Zhang:2018diq,Li:2018tpe}
\begin{equation}
\mathscr{M}(\nu,z^2)=\frac{h(zP_z,z^2)/h(0\cdot P_z,0)}{h(z\cdot 0,z^2)/h(0\cdot 0,0)}.
\label{eq:RITD}
\end{equation}
By construction, the renormalization of ${\cal O}(z)$ and kinematic factors are canceled in the RpITDs, and the ultraviolet divergences are removed.
The RpITD double ratios employed here are normalized to one at $z=0$, and the lattice systematics are reduced due to the double ratio.
These RpITDs will be input into the pseudo-PDF framework detailed in Ref.~\cite{Radyushkin:2017cyf,Balitsky:2019krf} to obtain the desired parton distribution:
\begin{equation}
\label{eq:matching-gg}
\mathscr{M}(\nu,z^2) = \int_0^1 \dl x
 \frac{f(x,\mu^2)}{\langle x^n \rangle_f}
 R_{f{_1}f{_2}}(x\nu,z^2\mu^2),
\end{equation}
where $\mu$ is the renormalization scale in the $\overline{\text{MS}}$ scheme,
${\langle x^n \rangle_f} $ is the moment of $f=\{q,xg\}$ (with $n=0$ and 1, respectively),
$R_{f_1f_2}$ are parton-in-parton (could be quark-in-quark, gluon-in-gluon or quark-gluon) matching kernels, and
$f(x,\mu^2)$ is the desired parton distribution. 
One can obtain the desired parton distribution $g(x,\mu^2)$ by fitting the RpITD through the matching condition in Eq.~\ref{eq:matching-gg} with phenomenologically motivated form commonly used in global analyses
\begin{equation}
f(x,\mu) = \frac{x^A(1-x)^C}{B} k(x),
\label{functional}
\end{equation}
for $x\in[0,1]$ and zero elsewhere with $k(x)$ additional polynomial functions, $B$ to normalize the area to 1.
Recent example results using the pseudo-PDF method can be found in Refs.~\cite{HadStruc:2021wmh,Fan:2021bcr,HadStruc:2022yaw}.

\subsubsection{LaMET/Quasi-PDF}

Large Momentum Effective Theory (LaMET)~\cite{Ji:2013dva,Ji:2014gla} also takes the matrix elements calculated in Fig.~\ref{fig:feyman-diag} and Fourier transforms the renormalized matrix elements $h^R$,
\begin{equation}\label{eq:qPDF}
\widetilde{F}(x,P_z) = \int \frac{\dl z}{2\pi} e^{-i x z P_z} P_z h^R(z,P_z).
\end{equation}
In order to prevent unphysical oscillations in the quasi-PDF, we first extrapolate the renormalized matrix element $h^\text{R}(z,P_z)$ to infinite distance before Fourier transforming.
There is no unique way of performing this step; Refs.~\cite{Ji:2020brr,Gao:2021dbh,Gao:2022uhg} propose a model used for extrapolation:
\begin{equation}\label{eq.Extrapolation}
 h^\text{R}(z,P_z)\to \frac{Ae^{-m^\prime z}}{|zP_z|^d}\quad\text{as $z\to\infty$},
\end{equation}
where $A$, $m^\prime$ and $d$ are fitting parameters. 
This extrapolation model is inspired by the PDF having the functional form $f(x,\mu)\sim x^{d-1}$ at small $x$, which corresponds to the anticipated large-distance behavior in the renormalized matrix element in the above equation.
Note that the extrapolation mainly affects the small-$x$ region and has very little impact on moderate to to large $x$, where LaMET makes predictions. 
With the current $P_z^\text{max}$ ranging 2--3~GeV, the systematics due to different $z\to\infty$ extrapolations are smaller than the LaMET matching systematic in the small-$x$ region. 
With the renormalized coordinate-space matrix element computed and extrapolated to infinite length of Wilson-line displacement, the quasi-PDF is computed via Eq.~\ref{eq:qPDF}.

The lightcone parton distribution in the $\overline{\text{MS}}$ scheme at scale $\mu$ is then convolved with a perturbative hard matching kernel, up to power corrections that are suppressed by the hadron momentum:
\begin{equation}
\label{eq:matching}
\tilde{f}(x,\xi,t,P_z) = \int \frac{\dl y}{|y|}
  C\left(\frac{x}{y}, \frac{\xi}{y}, \frac{yP_z}{\mu}\right)f(y,\xi,t,\mu)
 + \mathcal{O}\left(\frac{\Lambda_\text{QCD}^2}{x^2 P_z^2},         
                    \frac{\Lambda_\text{QCD}^2}{(1-x)^2P_z^2}\right),
\end{equation}
where $\Lambda_\text{QCD}$ is the QCD scale (0.3--0.5~GeV, depending on the lattice), and the matching kernel $C$ connects finite quasi-distribution calculated at $P_z$ to lightcone parton distribution and has been computed up to NNLO~\cite{Li:2020xml,Izubuchi:2018srq}.
In the zero-skewness limit $\xi=0$ for GPDs (more details in Sec.~4), the matching kernel $C$ is the same as the matching kernel for the PDF~\cite{Chen:2018xof,Liu:2018uuj}.
There have been many developments in multiple renormalization schemes to treat the short- and long-distance physics properly and additional improvements to deal with the renormalon and divergences; more details are given in the later subsections.
There are many more lattice parton-distribution calculations done using LaMET method; rather than listing a few example references, a few recent review articles have summarized the progress made using this method~\cite{Amoroso:2022eow,Lin:2017snn,Ji:2020ect,Constantinou:2020hdm,Lin:2023kxn}.

\subsubsection{Connections among the Methods}

\begin{figure}[tb]
\centering
\includegraphics[width=0.65\textwidth]{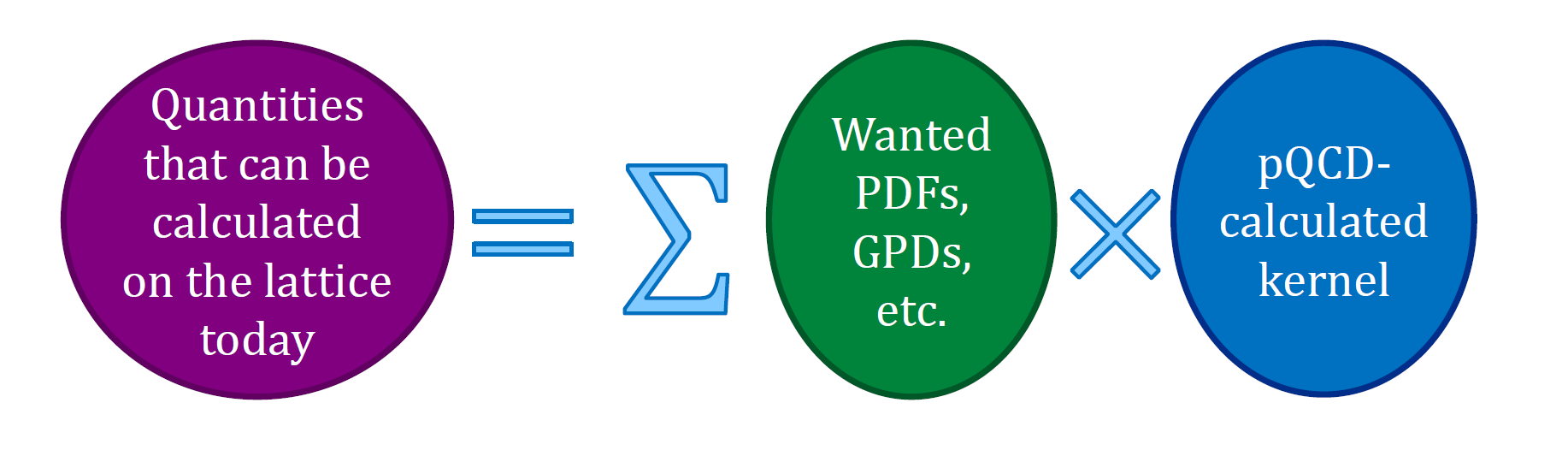}
\caption{
Various different $x$-dependent methods can be categorized into the quantities (usually three- or four-point matrix elements) that can be calculated at finite boosted hadron momentum initial and final states, and connect with the wanted quantities, such as PDFs, GPD, TMDs, as shown in the green ellipse on the right-hand side that often couple with a complicated perturbative calculated kernel.
\label{fig:x-dependent-method}}
\end{figure}

Before discussing further the recent developments in the most popular $x$-dependent LaMET methods, let us summarize the various $x$-dependent methods.
All the various methods can be summed up by the diagram as shown in Fig.~\ref{fig:x-dependent-method}.
There are proposals to calculate various matrix elements via three-point (one insertion of operators with quark-antiquark or gluon fields separated by some distance) or four-point functions (usually composed of two local currents separated by some distance) calculated on the lattice with some momentum dependence (the left purple object). 
These are related to the quantities we want to determine (the green ellipse), such as PDFs, GPDs, TMDs, but connected by some matching kernel that is usually calculable via perturbative QCD with the infrared differences of the physics canceled out. 
In some cases, one can directly derive the relationship directly.

For example, the LaMET/quasi-PDF and pseudo-PDF methods can be related to the lattice cross-sections method in the calculation of the unpolarized quark PDFs by choosing
\begin{equation}
{\cal O}_{q}(z)=Z_q(z^2)\overline{\psi}_q(z)\gamma\cdot {z}\,
{\cal P}e^{-ig\int_0^{1} z\cdot A(\lambda z)\,\dl\lambda}{\psi}_q(0)\,,
\end{equation}
with the renormalization constant $Z_q(z^2)$~\cite{Ma:2017pxb}. 
With $K^{q(0)}_{q}(x \nu,z^2,0,\mu)= 2 x \nu e^{i x \nu}$, one finds
\begin{equation}
\label{eq:lcsQuasi}
\int \frac{\dl \nu}{\nu}\, \frac{e^{-i x \nu}}{4\pi} \sigma_{q}(\nu,z^2,P^2)
\approx f_{q}(x,\mu)\, ,
\end{equation}
modulo $\mathcal{O}(\alpha_s)$ and higher twist corrections. 
By choosing $z_0=0$ and both $\vec{P}$ and $\vec{z}$ along the $z$-direction, one finds that $\nu=-z P_z$, and the left-hand side of Eq.~\ref{eq:lcsQuasi} is the quasi-quark distribution introduced in Ref.~\cite{Ji:2013dva} if the integral is performed by fixing $P_z$.
It is effectively the pseudo-quark distribution used in Ref.~\cite{Orginos:2017kos} if the integral is performed by fixing $z$.
That is, these two approaches for extracting PDFs are equivalent in the infinite-momentum $P_z \rightarrow \infty $ ($z \rightarrow 0$) limit if the matching coefficients are calculated at the lowest order in $\alpha_s$, neglecting all power corrections. 
However, at finite $P_z$ or $z$, the LaMET and pseudo-PDF (similar to all other short-distance operator product expansion methods) are two different approaches to extract parton distributions from coordinate-space correlation functions in large-momentum hadrons~\cite{Ji:2024oka,Ji:2022ezo}.

\subsection{Improved Renormalization Schemes}\label{sec.Renormalization}

\subsubsection{Hybrid-Renormalization Scheme}

Early renormalization methods for the LaMET matrix elements used in lattice parton calculations were the regularization-independent momentum-subtraction (RI/MOM) scheme~\cite{Martinelli:1994ty} in Refs.~\cite{Zhang:2018nsy,Zhang:2020gaj,Constantinou:2017sej,Stewart:2017tvs,Alexandrou:2017huk,Chen:2017mzz,Lin:2018pvv,Chen:2018xof,LatticeParton:2018gjr,Lin:2019ocg,Lin:2014zya,Chen:2016utp,Alexandrou:2015rja,Alexandrou:2016jqi,Chen:2017mzz,Alexandrou:2018eet,Fan:2018dxu,Liu:2018hxv,Alexandrou:2018pbm,Alexandrou:2019lfo,Fan:2020nzz} and the ratio scheme, similar to those used in the pseudo-PDF method~\cite{Radyushkin:2017cyf,Orginos:2017kos}. 
There also exist variants to the ratio renormalization scheme, such as using a vacuum matrix element~\cite{Braun:2018brg,Li:2020xml}, and the hadron matrix elements of a nonzero-momentum state~\cite{Fan:2020nzz}. 
Later, improved renormalized schemes, such as hybrid-ratio and hybrid-RI/MOM schemes~\cite{Ji:2020brr}, were proposed to normalize the matrix elements by zero-momentum boosted hadron matrix elements and RI/MOM nonperturbatively calculated renormalization constant, respectively, for short distance (say up to a distance $z_s\approx 0.3$~fm)~\cite{Ji:2020brr,Gao:2022uhg,Gao:2021dbh,LatticeParton:2022xsd};
for large distance, $z>z_s$, the bare matrix elements are instead multiplied by an exponential term designed to remove both the linear divergence and the renormalon ambiguity.
In such a scheme, the hybrid renormalized matrix elements $h^\text{R}_{H}(z,P_z)$ are given by
\begin{equation}\label{eq:hR}
 h^\text{R}_{H}(z,P_z) = \begin{cases}
 N \frac{h^\text{B}_{H}(z,P_z)}{Z(z)} &\text{if } z < z_s \\
 N e^{(\delta m+m_0)(z-z_s)} \frac{h^\text{B}_{H}(z,P_z)}{Z(z_s)} &\text{if } z \geq z_s
 \end{cases},
\end{equation}
where $Z(z)$ can be the bare matrix element at zero momentum $h^\text{B}_{H}(z,P_z=0)$ for the hybrid-ratio or RI/MOM factor or $Z^\text{RI/MOM}(z,\mu_\text{RI}, p_R^z=0)$ for the hybrid-RI/MOM scheme;
$\delta m$ and $m_0$ are the linear divergence and renormalon ambiguity, respectively~\cite{Zhang:2023bxs};
the normalization $N = Z(0)/h^\text{B}_{H}(z=0,P_z)$ sets the matrix element $h^\text{R}_{H}(0,P_z)=1$, satisfying conservation of charge for unpolarized PDFs or renormalized charges (for other structures).
The constants $Z(z_s)$ and $e^{-(\delta m+m_0)z_s}$ enforce continuity at $z=z_s$.
The $Z(z)$ is expected to contain the same factor $e^{-\delta mz}$; so at large $z$, one may fit $Z(z)$ to $Ae^{-\delta mz}$ in order to extract $\delta m$.
At short distances, one demands $z$, $h^R(z,0,\mu)= C_0(z, \mu)$, where $C_0(z,\mu)$ is the zeroth Wilson coefficient, so with the nonperturbative renormalization (NPR) factor and Wilson coefficients, we may fit $m_0$,
\begin{equation}\label{eq.m0formula2}
 \ln\left(\frac{e^{-\delta m\,z}C_0(z,\mu)}{Z(z)}\right)=m_0z+c
\end{equation} 
where $c$ is an unimportant shift, as shown in Ref.~\cite{Zhang:2023bxs}. 
One can also determine $\delta m(a)$ by using the combination of the static quark-antiquark potential and the free energy of a static quark at nonzero temperature~\cite{Gao:2021dbh,Ji:2020brr,Zhang:2017bzy,Petreczky:2021mef}.

\subsubsection{Self-Renormalization scheme}

There are some ambiguities in determining $m_0$ and $\delta m$ independently using matrix elements from a single lattice-spacing.
One can further improve the procedure by using multiple lattice spacings' $Z(z_s)$, a procedure called ``self-renormalization''~\cite{LatticePartonCollaborationLPC:2021xdx}.
The philosophy behind self-renormalization is that the divergences in the bare matrix elements at zero momentum can be determined by studying multiple lattice spacings fit to a functional form dictated by perturbative QCD.
The parameters determined in this process can then be used to divide out the divergences producing the renormalized matrix element at non-zero momentum. The functional form is
\begin{equation}\label{eq.functionalForm}
\ln\left(\frac{1}{Z(z)}\right) =
 \frac{kz}{a\ln(a\LambdaQCD)} + g(z) + f(z)a +
 \frac{3C_F}{4\pi\beta_0}
 \ln\left(\frac{\ln(a\LambdaQCD)}{\ln(\LambdaQCD/\mu)}\right) +
 \ln\left(1+\frac{d}{\ln(a\LambdaQCD)}\right),
\end{equation}
where $k$ is the linear divergence arising from the self-energy of the Wilson link that appears in the bare matrix elements, $a$ is the lattice spacing, $\LambdaQCD$ is the cutoff scale for QCD, $f(z)$ is a function describing the discretization effects of the lattice, $\mu$ is the final desired renormalization scale, $d$ is a parameter determined by demanding that the short-distance behavior ($z\lesssim 0.3$~fm) of the renormalized matrix element agrees with perturbation theory, $C_F$ is the quadratic Casimir for the fundamental representation of SU(3), and $\beta_0$ is the first coefficient of the QCD beta-function.
The remaining term, $g(z)$, encapsulates the renormalized matrix element.

The linear divergence and QCD cutoff ($k$ and $\LambdaQCD$) are independent of lattice spacing and are not treated as fit parameters but as global parameters.
One does not know the value of $d$, so the procedure starts by setting it to zero initially and later determines it by the required condition~\cite{LatticePartonCollaborationLPC:2021xdx,Holligan:2024wpv}.
The bare matrix elements are fitted to Eq.~\ref{eq.functionalForm} at fixed $z$ as a function of $a$ with $g$ and $f$ as the fitting parameters across a range of values for $k$ and $\LambdaQCD$.
For each pair of $k$ and $\LambdaQCD$ values, the average $\chi^2$ is computed across the range of $z$-values.
For a choice of $\LambdaQCD$, the optimum $k$ value (i.e. the one with the lowest average $\chi^2$) is then used for the subsequent steps in the procedure. 

Having determined the QCD cutoff, the linear divergence and the corresponding values of $g$ across the $z$-range, we now demand that the renormalized matrix element agrees with perturbation theory for distances $z\lesssim 0.3$~fm.
The short distance behavior is described by the Wilson coefficients, $C_0(z,\mu)$ at NLO or at NNLO~\cite{Yao:2022vtp,Li:2020xml,Izubuchi:2018srq}.
The term $g(z)$ encapsulates the renormalized matrix element, but we must take into account the renormalon ambiguity mentioned earlier.
The renormalized matrix element should obey at short distances
\begin{equation}
 h^R_\text{self}(z,P_z=0,a) =
 \exp\left[g(z)-m_0z\right] =
 C_0(z,\mu) + \mathcal{O}(z^2\LambdaQCD^2),
\end{equation}
where $m_0$ is the renormalon ambiguity.
The parameter $d$ is determined by fitting $g(z)-\ln(C_0(z,\mu))$ to a function linear in $z$: $m_0z+c$ by demanding $c=0$ to a desired accuracy in order to ensure that the short-distance behavior of the renormalized matrix element agrees with perturbation theory.
With each alteration of $d$, the $k$ parameter will also change but very weakly.

Having determined $d$, a full renormalization factor can be constructed as 
\begin{equation}\label{eq.ZRfactor}
\frac{1}{Z(z,a)} =
 \exp\left[\frac{kz}{a\ln(a\LambdaQCD)} +
 m_0z + f(z)a +
 \frac{3C_F}{4\pi\beta_0}
 \ln\left(\frac{\ln(a\LambdaQCD)}{\ln(\LambdaQCD/\mu)}\right) +
 \ln\left(1+\frac{d}{\ln(a\LambdaQCD)}\right)\right].
\end{equation}
Note that
\begin{equation}
 Z(z,a)h^B(z,a) = \exp\left[g(z)-m_0z\right],
\end{equation}
which, by construction, will agree with $C_0(z,\mu)$ at small $z$. 
One can then combine the self-renormalization factors with the hybrid-renormalization scheme, as shown in the earlier section.

ANL/BNL collaboration proposed a new procedure for the hybrid scheme in order to use larger $z$:
\begin{equation}\label{eq:mfit}
    \lim_{a\to0}e^{\delta m(a) (z-z_0)} { h_H^B(z,0,a)\over  h_H^B(z_0,0,a)} = e^{- {\bar{m}_0}(z-z_0)}  { C_0( \mu^2z^2) + \Lambda z^2\over C_0(\mu^2z^2_0) + \Lambda z^2_0}\,,
\end{equation}
where $z,z_0\gg a$, and the parameter $\Lambda\sim{\cal O}(\Lambda_\text{QCD}^2)$~\cite{Gao:2021dbh}.  
$\overline{m}_0$ was originally proposed in Refs.~\cite{LatticePartonCollaborationLPC:2021xdx,Green:2020xco} to be obtained by comparing the subtracted matrix elements of $O_\Gamma(z)$ or $W(z,0)$ with their $\overline{\text{MS}}$ operator product expansion (OPE);
however, it requires $z\lesssim 0.2$~fm~\cite{Ji:2020brr}, so there is a narrow window of $z$ that can be used.
They chose $z\ge z_0=0.24$~fm and found agreement between the $a=0.04$~fm and $a=0.06$~fm ratios at subpercent level up to $z\approx 1$~fm.
They further modified the hybrid scheme by correcting the $\Lambda z^2$ term in $\tilde h(z,0,\mu)$ at short $z$ as
\begin{multline}
\label{eq:hybren}
h_H^B(z, z_S,P^z,\mu,a) = \\
N{ h_H^B(z,P^z,a)\over  h_H^B(z,0,a)} {C_0(z^2\mu^2) + \Lambda z^2\over C_0( z^2\mu^2)} \theta(z_S-z)
    + Ne^{\delta m'(z - z_S)} { h_H^B(z, P^z, a)\over  h_H^B(z_S,0, a)} {C_0(z^2_S\mu^2) + \Lambda z^2_S \over C_0( z^2_S\mu^2) } \theta(z-z_S)\,,
\end{multline}
where $\delta m'=\delta m + \bar{m}_0$, and $N= h_H^B(0,0,a)/\tilde h(0,P^z,a)$ normalizes $ h_H^B(z, z_S,P^z,\mu,a)$ to one at $z=0$. 
See Ref.~\cite{Gao:2021dbh} for more details. 

\subsection{RGR and LRR Improvements}\label{sec:LeadingPower}

The renormalization scheme and the lightcone matching can each be augmented with the methods of renormalization group resummation (RGR)~\cite{Su:2022fiu} and leading renormalon resummation (LRR)~\cite{Zhang:2023bxs} to both the renormalization scheme and the lightcone matching.
RGR involves resumming the large logarithms that arise from the difference in renormalization scale and the intrinsic physical scale.
The method involves setting the renormalization scale such that the logarithms vanish and then evolving to the desired energy scale using the renormalization group equation.
The use of RGR enhances the presence of the renormalon divergence which arises from the asymptotic nature of perturbation series~\cite{Zichichi:1979gj};
this should be accounted for by including LRR in the calculations, which resums the series to account for the renormalon divergence.

One would start with the Wilson coefficients that was used to tune the $m_0$ parameter to improve the divergences in the renormalization.
The Wilson coefficients depend on the renormalization scale $\mu$, as well as the intrinsic physical scale, and the difference between the results in logarithmic terms that need to be resummed.
We perform the resummation using the renormalization-group equation (RGE)
\begin{equation}\label{eq.RGE}
 \frac{\dl{C_0(z,\mu)}}{\dl{\ln(\mu^2)}} = \gamma(\mu)C_0(z,\mu)
\end{equation}
where $\gamma(\mu)$ is the anomalous dimension, which has been calculated up to three loops~\cite{Braun:2020ymy}.
One can then set the energy scale such that the logarithms vanish and then evolve the Wilson coefficient to the desired energy scale by solving Eq.~\eqref{eq.RGE}:
\begin{equation}\label{eq.CNLORGR}
 C_0^\text{(N)NLO$\times$RGR}(z,\mu) =
 C_0^\text{(N)NLO}(z,\mathtt{z}^{-1}) \times
 \mathcal{I}(\mu,\mathtt{z}^{-1}),
\end{equation}
where $\mathtt{z}^{-1}\equiv 2e^{-\gamma_E}/z$, $\mathcal{I}(\mu,\mathtt{z}^{-1})$ is defined by
\begin{equation}
\mathcal{I}(\mu,\mathtt{z}^{-1}) =
 \exp\left(\int^{\alpha_s(\mu)}_{\alpha_s(\mathtt{z}^{-1})}\!\!\dl{\alpha'} \frac{\gamma(\alpha')}{\beta(\alpha')}\right),
\end{equation}
$\beta(\alpha_s)$ is the QCD $\beta$ function,
and $\gamma$ must be computed to the same order as $C_0(z,\mathtt{z}^{-1})$.
Note that RGR has not only been used in the LaMET calculations, but also in the short-distance factorization (SDF), such as pseudo-PDF method, analyses in the Mellin moment space~\cite{Gao:2021hxl}, where the resummation formulas were derived.
Note that RGR in SDF reveals that the perturbation theory breaks down beyond $z \approx 0.2$~fm, providing guidance on the maximum range of matrix elements that one can use with SDF method.
For the remainder of the section, we will focus on LaMET method, where more results on the further systematics are published.

The Wilson coefficients themselves are determined by a perturbation series in the strong coupling, which in general is not convergent to all orders, resulting in the renormalon ambiguity.
We can account for this by improving the Wilson coefficients with LRR.
Reference~\cite{Zhang:2023bxs} suggests modifying the Wilson coefficient according to
\begin{equation}\label{eq.CNLOLRR}
C^{\text{N}^{k}\text{LO+LRR}}_0(z,\mu) =
 C^{\text{N}^{k}\text{LO}}_0(z,\mu) +
 z\mu\left(C_\text{PV}(z,\mu) -
 \sum_{i=0}^{k-1}\alpha^{i+1}_s(\mu)r_i\right),
\end{equation}
where $k=1$ for NLO, $k=2$ for NNLO, and $C_\text{PV}$ the Cauchy principal value to regulate the poles in the integrand:
\begin{align}\label{eq.BorelInt}
 C_\text{PV}(z,\mu) =
 N_m\frac{4\pi}{\beta_0}\int^{\infty}_{0,\text{PV}}\dl u\exp(-\frac{4\pi u}{\alpha_s(\mu)\beta_0}) \times
 \frac{1}{(1-2u)^{b+1}}\left(1+c_1(1-2u)+c_2(1-2u)^2+\dots\right).
\end{align}
The improvements in the systematic errors in the renormalized matrix elements can be estimated by varying the energy scales, say 1.5 to 3~GeV with the central value calculated at $\mu=2$~GeV.
These errors then propagate into the Wilson coefficients and fitting of $m_0$ to the renormalized matrix elements and so on.
Example recent results on these systematic improvements can be found in Refs.~\cite{Holligan:2024umc,Holligan:2023jqh,LatticeParton:2022xsd,Gao:2023ktu}.

The matching process aligns the UV behavior of the quasi-PDF with that of the lightcone PDF $q^\text{v}_{\pi}(x,\mu)$.
The two quantities are related through the matching formula
\begin{equation}\label{eq.Matching}
 \tilde{f}(x,P_z) = \int^{1}_{-1}\frac{\dl {y}}{|y|}\mathcal{K}(x,y,\mu,P_z,z_s) f(y,\mu)
 +\mathcal{O}\left(\frac{\LambdaQCD^2}{P_z^2x^2(1-x)}\right),
\end{equation}
where $\mathcal{K}$ is the matching kernel.
The corrections to the matching process arise from the fact that the quasi-PDF is computed at finite momentum, whereas the lightcone PDF is defined at infinite momentum~\cite{Gao:2021dbh,Braun:2018brg}.
The full matching kernel is
\begin{equation}\label{eq.MatchingFull}
 \mathcal{K}(x,y,\mu,P_z,z_s) = \mathcal{K}^\text{h-ratio, h-RI/MOM, h-self} + \Delta\mathcal{K}^\text{LRR},
\end{equation}
depending on whether we renormalize in the hybrid-ratio, hybrid-RI/MOM or hybrid-self renormalization scheme.
The matching kernel has been computed up to NNLO for unpolarized quasi-PDFs renormalized in the hybrid-ratio scheme with LRR~\cite{Chen:2020ody,Li:2020xml,Su:2022fiu} as well as to NLO in the RI/MOM scheme in Refs.~\cite{Stewart:2017tvs,Chou:2022drv}. 
Note that when the RI/MOM matrix elements are computed at momentum $p_R=0$, the hybrid-ratio and hybrid-RI/MOM matching kernels coincide at NLO as was shown in Ref.~\cite{Chou:2022drv}.

The LRR modification~\cite{Zhang:2023bxs} in the matching kernel can be written as
\begin{equation}
\label{eq.MatchingLRR}
 \Delta\mathcal{K}^\text{LRR} = \int\frac{yP_z \dl{z}}{2\pi} e^{i(x-y)zP_z}
 \times z\mu\left[C_\text{PV}(z,\mu) - \sum_{i=0}^{k -1} \alpha^{i+1}_s(\mu)r_i\right].
\end{equation}
with $k=1$ for NLO and $k=2$ for NNLO.
Since the matching process can be numerically expensive, we convert the matching kernel $\mathcal{K}$ of Eq.~\eqref{eq.MatchingFull} into a matrix in $x$ and $y$, $\mathcal{K}_{xy}$, then multiply a vector of quasi-PDF values by the inverse $\mathcal{K}^{-1}_{xy}$.

The method of RGR can also be applied to the matching process; the algorithm has been derived in Ref.~\cite{Su:2022fiu}.
This time, the intrinsic physical scale is that of the parton ($\sim 2xP_z$). One can perform the lightcone matching at the scale $\mu=2xP_z$ so the logarithms vanish, and then evolve to the desired energy scale using the DGLAP (Dokshitzer-Gribov-Lipatov-Altarelli-Parisi)  equation:
\begin{equation}\label{eq.DGLAP}
 \diff{f(x,\mu)}{\ln(\mu^2)} =
 \int^1_x \frac{\dl{z}}{|z|} \mathcal{P}(z) f\left(\frac{x}{z},\mu\right),
\end{equation}
where $\mathcal{P}(z)$ is the DGLAP kernel, which has been calculated up to three loops~\cite{Moch:2004pa}.
It should be noted that the DGLAP evolution formula begins to break down at $x\approx 0.2$ for $P_z \approx 1.7$~GeV where $\alpha_s(\mu=2xP_z)$ becomes nonperturbative.

%% file: sec3-PDFs.tex
\section{Parton Distribution Functions}

In this section, we briefly discuss selected highlight results from the commonly used LaMET method for quark PDF and pseudo-PDF method for gluon PDFs, which used RI/MOM and ratio renormalization, respectively, with NLO matching kernel, unless otherwise specified.
In Sec.~3.4, we show selected results including the updated renormalization scheme and additional RGR and LRR in NLO and NNLO matching in LaMET method.

\subsection{Flavor Nonsinglet PDFs}

\subsubsection{Isovector Nucleon PDFs}
There has been rapid progress calculating the Bjorken-$x$ dependence of PDFs on the lattice since the proposal of Large-Momentum Effective Theory (LaMET, also called the ``quasi-PDF'' method)~\cite{Ji:2013dva,Ji:2014gla}.
LaMET relates equal-time spatial correlators, whose Fourier transforms (FTs) are called quasi-PDFs, to PDFs in the limit of infinite hadron momentum.
For large but finite momenta accessible on a realistic lattice, LaMET relates quasi-PDFs to physical ones through a factorization theorem, the proof of which was developed in Refs.~\cite{Ma:2017pxb,Izubuchi:2018srq,Liu:2019urm}.
Since the first lattice $x$-dependent PDF calculation~\cite{Lin:2014zya}, much progress has been made and calculations done.
For recent reviews, see Refs.~\cite{Lin:2017snn,Ji:2020ect,Ji:2020byp,Lin:2020rut,Lin:2023kxn}.
In this subsection, a few selected calculations at physical pion mass are shown.

\begin{figure}[htbp]
\includegraphics[width=.32\textwidth]{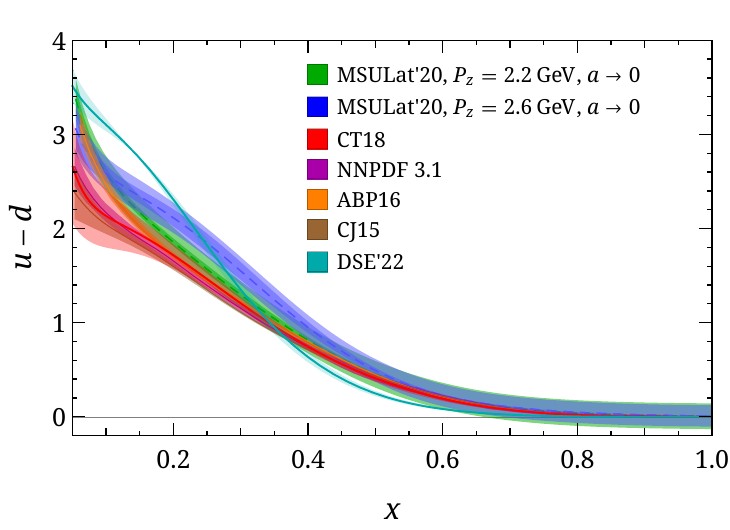}
\includegraphics[width=.32\textwidth]{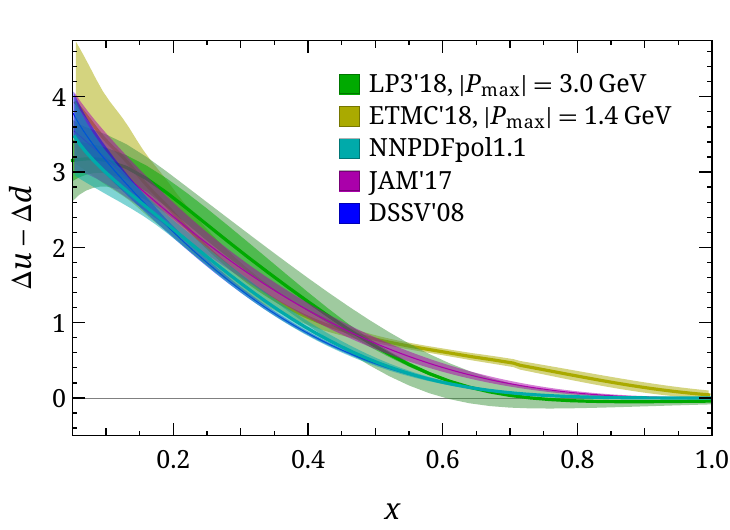}
\includegraphics[width=.32\textwidth]{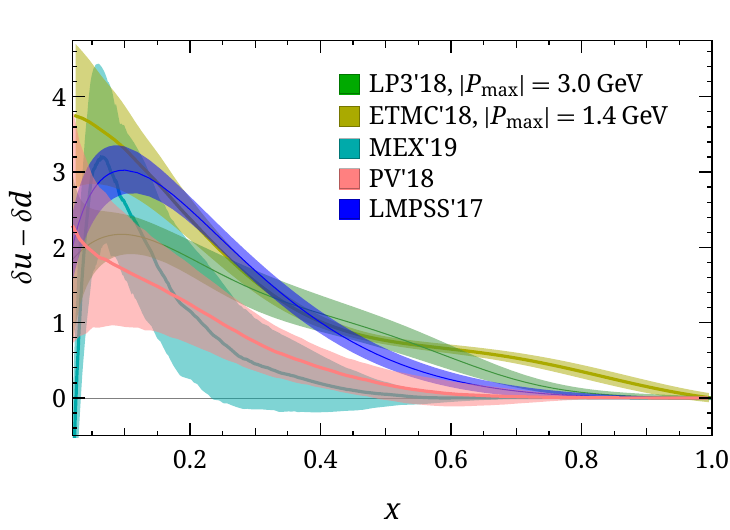}
\caption{
(left) The first lattice calculation of continuum-physical limit NLO nucleon isovector unpolarized PDFs~\cite{Lin:2020rut} is compared with global fits from Refs.~\cite{Hou:2019efy,Alekhin:2017kpj,Ball:2017nwa,Accardi:2016qay}.
Summary of the lattice calculation of isovector helicity (middle) and transversity (right)
with LP3 and ETMC physical--pion-mass NLO isovector quark taken from LP318~\cite{Lin:2018qky,Liu:2018hxv}, ETMC18~\cite{Alexandrou:2018eet,Alexandrou:2019lfo}, 
and global fits from NNPDFpol1.1~\cite{Nocera:2014gqa}, JAM17~\cite{Ethier:2017zbq}, DSSV08~\cite{deFlorian:2009vb} (helicity), MEX19~\cite{Benel:2019mcq}, PV18~\cite{Radici:2018iag}, LMPSS17\cite{Lin:2017stx} (transversity).
}
\label{fig:PDFresults}
\end{figure}

The most studied $x$-dependent structures are the nucleon unpolarized isovector parton distribution functions (PDFs).
The right-hand side of Fig.~\ref{fig:PDFresults} shows a state-of-the-art lattice calculation of the nucleon isovector unpolarized PDFs.
Most of the calculations are performed on a single ensemble (See Fig.~7 in Ref.~\cite{Lin:2020rut}) except the results from MSULat group~\cite{Lin:2020fsj}. 
The lattice results are calculated using ensembles with multiple sea pion masses with the lightest one around 135~MeV, three lattice spacings $a\in[0.06,0.12]$~fm, and multiple volumes with $M_\pi L$ ranging 3.3 to 5.5.
A simultaneous chiral-continuum extrapolation is performed on the RI/MOM-renormalized nucleon matrix elements with various Wilson-link displacements to obtain the physical-continuum matrix elements. 
Then one-loop perturbative matching is applied to the quasi-PDFs to obtain the lightcone PDFs and compare the LQCD results with a selection of global-fit PDFs.
The lattice nucleon isovector PDFs (with statistical and estimated systematic errors shown as inner/outter band) have nice agreement with those obtained from global fits,
CT18NNLO~\cite{Hou:2019efy},
NNPDF3.1NNLO~\cite{Ball:2017nwa},
ABP16~\cite{Alekhin:2017kpj}, and
CJ15~\cite{Accardi:2016qay}.
The errors increase toward the smaller-$x$ region for both lattice and global-fit PDFs, but overall, they agree within two standard deviations.

The middle and right of Fig.~\ref{fig:PDFresults} show a summary of lattice nucleon isovector helicity and PDFs at physical pion mass for the helicity and transversity PDFs, respectively, obtained in Refs.~\cite{Alexandrou:2018pbm,Lin:2018qky,Alexandrou:2018eet,Liu:2018hxv}.
The helicity lattice results are compared to two phenomenological fits, NNPDFpol1.1~\cite{Nocera:2014gqa} and JAM17~\cite{Ethier:2017zbq}, exhibiting nice agreement.
The lattice results for the transversity PDFs has better precision than the global analysis by PV18 and LMPSS17~\cite{Lin:2017stx}.
Note that none of these current polarized lattice calculations have taken the continuum limit ($a\rightarrow 0$) and have remaining lattice artifacts (such as finite-volume effects);
disagreement among the lattice results in the obtained distributions is not unexpected. 
Further studies of the systematic uncertainties, including multiple lattice spacings and volumes, from each collaboration will be needed.

\subsubsection{Pion and Kaon Valence-Quark PDFs}

Light pseudoscalar mesons, such as pion and kaon, play a fundamental role in QCD as they are the Nambu-Goldstone bosons associated with dynamical chiral symmetry breaking (DCSB).
While studies of pion and kaon structure both reveal physics of DCSB, a comparison between them also helps to reveal the relative impact of DCSB versus the explicit breaking of chiral symmetry by the quark masses. 
The pion and kaon PDFs can be measured by scattering a secondary pion ($\pi$) or kaon ($K$) beam over target nuclei ($A$), inducing the Drell-Yan process, $\pi (K)A \to X \mu^+ \mu^-$~\cite{Badier:1983mj,Betev:1985pf,Falciano:1986wk,Guanziroli:1987rp,Conway:1989fs}.
With a combined analysis of $\pi^{\pm}A$ Drell-Yan on the same nuclear target, the valence and sea distributions can be separated~\cite{Badier:1983mj}, provided that the nuclear PDF is known.
Currently, the nuclear PDFs are approximated by a combination of proton and neutron PDFs.
The valence-quark PDF of the pion for momentum fraction $x \gtrsim 0.2$ has been determined reasonably well~\cite{Badier:1983mj,Betev:1985pf,Conway:1989fs,Wijesooriya:2005ir,Aicher:2010cb}, subject to the systematic uncertainty in the PDF parametrization.
Combining $K^-A$ and $\pi^-A$ Drell-Yan data, the kaon valence PDF can be measured through the ratio~\cite{Badier:1980jq} $\bar{u}_v^{K^-}(x)/[\bar{u}_v^{\pi^-}(x)C(x)]$
where $\bar{u}_v^{K^-(\pi^-)}$ denotes the valence anti-up distribution in the $K^{-}$($\pi^{-}$).
The function $C(x)$ encodes the corrections needed due to the nuclear modification of the target PDFs, the omission of meson sea-quark distributions and the ignorance of the ratio $s_v^{K^-}(x)/\bar{u}_v^{K^-}(x)$.
In principle, the first two can be addressed by new experiments.
For example, the valence and sea PDFs for the pion and kaon at $x > 0.2$ can be separated in the $\pi^{\pm}$ and $K^{\pm}$ Drell-Yan experiments proposed by the COMPASS++/AMBER collaboration using the CERN M2 beamline~\cite{Denisov:2018unj}. 

The first lattice-QCD calculation of the pion and kaon valence-quark distribution functions was reported in Ref.~\cite{Zhang:2018nsy,Lin:2020ssv}.
The latest calculation was performed with multiple pion masses with the lightest one around 220~MeV, two lattice spacings $a=0.06$ and 0.12~fm, $(M_\pi)_\text{min} L \approx 5.5$, and high statistics ranging from 11,600 to 61,312 measurements.
This calculation uses a chiral-continuum extrapolation to obtain the renormalized matrix elements at physical pion mass, using a simple ansatz to combine the data from 220, 310 and 690~MeV: $h^R_{i}(P_z, z, M_\pi) = c_{0,i} + c_{1,i} M_\pi^2 + c_{a} a^2$ with $i=K, \pi$.
Mixed actions, with light and strange quark masses tuned to reproduce the lightest sea light and strange pseudoscalar meson masses, can suffer from additional systematics at $O(a^2)$;
such artifacts are accounted for by the $c_{a}$ coefficient.
All the $c_{a}$ were found to be to be consistent with zero.

\begin{figure}[htbp]
\includegraphics[width=.32\textwidth]{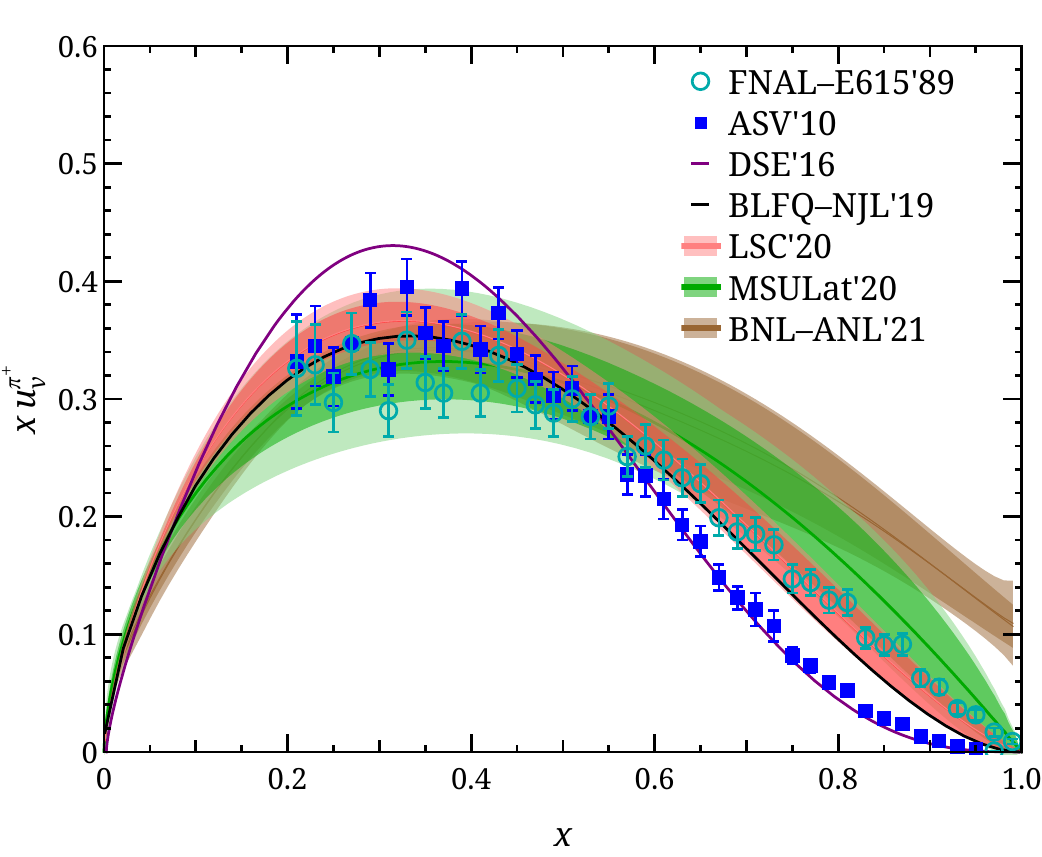}
\includegraphics[width=.33\textwidth]{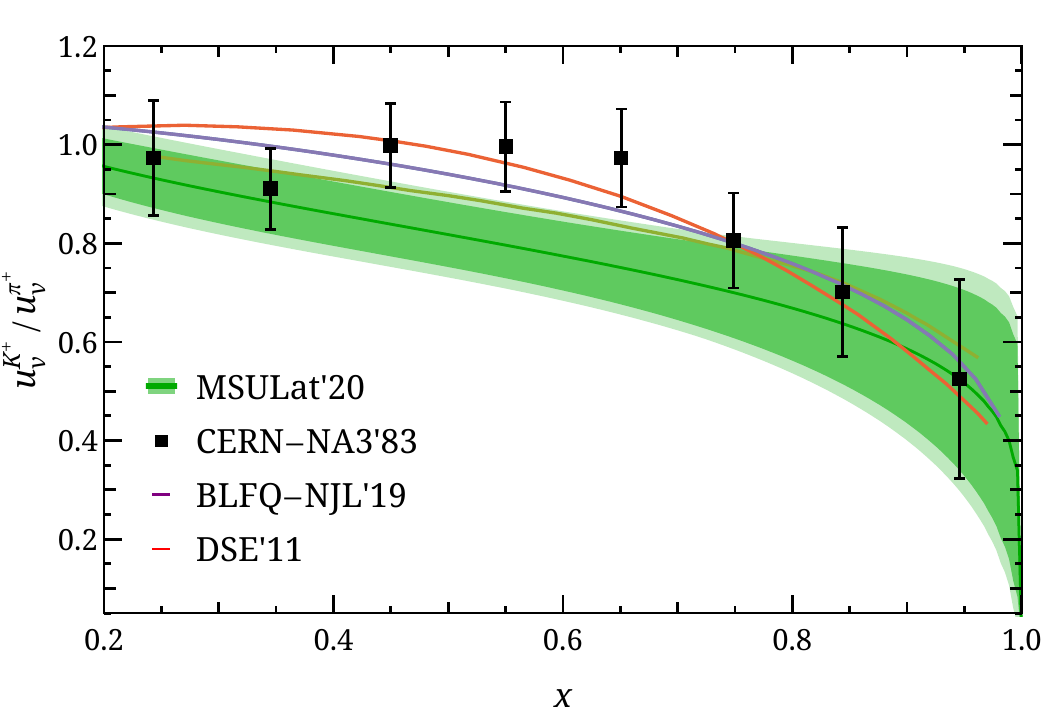}
\includegraphics[width=.32\textwidth]{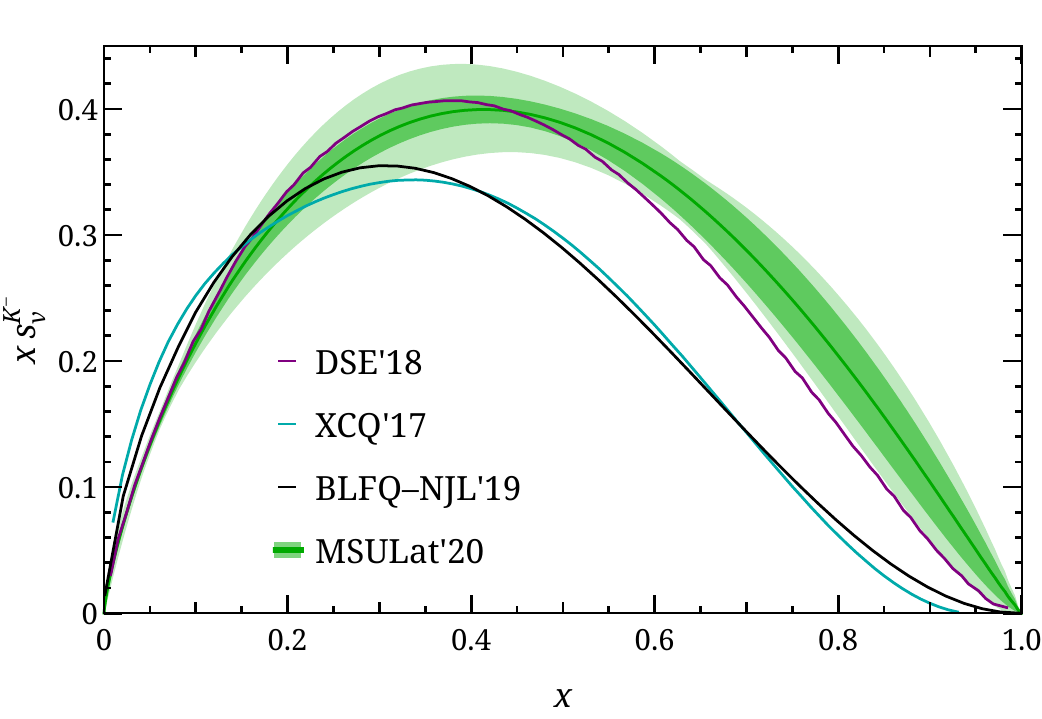}
\caption
{
(left) Lattice results on the valence-quark distribution of the pion,
$N_f=2+1$ BNL (NNLO with $a_\text{min}=0.04$~fm $P_z^\text{max}\approx 2.4$~GeV at $M_\pi=300$~MeV)~\cite{Gao:2021dbh} and $N_f=2+1+1$ MSULat (NLO with $a_\text{min}=0.06$~fm 
$P_z^\text{max}\approx 1.3$~GeV extrapolated to physical pion mass)~\cite{Lin:2020ssv} lattice groups using LaMET method, JLab and $N_f=2+1$ W\&M group (NLO with $a_\text{min}=0.09$~fm 
$P_z^\text{max}\approx 1.6$~GeV extrapolated to physical pion mass)~\cite{Sufian:2020vzb}, using the LCS method.
Results on the ratio of the light-quark valence distribution of kaon to that of pion (middle) and $x \overline{s}_v^K(x)$ as a function of $x$ (right) at a scale of $27\text{ GeV}^2$, both labeled ``MSULat'20'', along with results from relevant experiments and other calculations~\cite{Lin:2020ssv}.
The inner bands indicate statistical errors with the full range of $zP_z$ data while outter bands includes errors from using different data choices and fit forms.
}
\label{fig:mesonPDFs}
\end{figure}

Figure~\ref{fig:mesonPDFs} shows the final results for the pion valence distribution at physical pion mass ($u_v^{\pi^+}$) multiplied by Bjorken-$x$ as a function of $x$.
The inner bands indicate the statistical errors while the outer bands account for systematics errors from parametrization choices in the fits, the dependence on the maximum available Wilson-line displacement, etc.
The LQCD results were evolved to a scale of 27$\text{ GeV}^2$ using the NNLO DGLAP equations from the Higher-Order Perturbative Parton Evolution Toolkit (HOPPET)~\cite{Salam:2008qg} to compare with other results.
The LQCD result approaches $x=1$ as $(1-x)^{1.01}$ and is consistent with the original analysis of the FNAL-E615 experiment data.
On the other hand, there is tension with the $x>0.6$ distribution from the re-analysis of the FNAL-E615 experiment data using next-to-leading-logarithmic threshold resummation effects in the calculation of the Drell-Yan cross section (labeled as ``ASV'10''), which agrees better with the distribution from Dyson-Schwinger equations (DSE)~\cite{Chen:2016sno};
both prefer the form $(1-x)^2$ as $x \to 1$.
An independent lattice study of the pion valence-quark distribution~\cite{Sufian:2020vzb}, also extrapolated to physical pion mass, using the ``lattice cross sections'' (LCSs), reported similar results.
There has also been a first next-to-next-to-leading order (N$^2$LO) matching~\cite{Chen:2020ody,Li:2020xml} lattice calculation of the pion valence-quark PDF~\cite{Gao:2021dbh} with 300-MeV pion mass.
Both works take important steps toward precision PDFs from lattice QCD.
The left-hand side of Fig.~\ref{fig:mesonPDFs} illustrates selected lattice PDF predictions for the valence-quark PDF at $x\to 1$, where it can be compared against various nonperturbative approaches~\cite{Farrar:1975yb,Soper:1976jc,Bednar:2018mtf,Zhang:2018nsy,Ding:2019lwe,Novikov:2020snp,Courtoy:2020fex,Gao:2020ito,Alexandrou:2021mmi,Barry:2022itu}.

The middle of the Fig.~\ref{fig:mesonPDFs} shows the ratios of the light-quark distribution in the kaon to the one in the pion ($u_v^{K^+}/u_v^{\pi^+}$).
When comparing the LQCD result with the experimental determination of the valence quark distribution via the Drell-Yan process by NA3 Collaboration in 1982, good agreement is found between the LQCD results and the data~\cite{Lin:2020ssv}.
The LQCD result approaches $0.4$ as $x \to 1$ and agrees nicely with other analyses, such as constituent quark model,
the DSE approach (``DSE'11''),
and basis light-front quantization with color-singlet Nambu--Jona-Lasinio interactions (``BLFQ-NJL'19'').
The LQCD prediction for $x s_v^{K}$ is also shown in Fig.~\ref{fig:mesonPDFs} with the lowest three moments of $s_v^{K}$ being $0.261(8)_\text{stat}(8)_\text{syst}$, $0.120(7)_\text{stat}(9)_\text{syst}$, $0.069(6)_\text{stat}(8)_\text{syst}$, respectively;
the moment results are within the ranges of the QCD-model estimates
from chiral constituent-quark model
(0.24, 0.096, 0.049)
and DSE~\cite{Chen:2016sno} (0.36, 0.17, 0.092).

\subsection{Flavor-Singlet PDFs}

\subsubsection{Gluon PDFs}
The nucleon gluon PDF $g(x)$ is an important input to calculate the cross section for processes in $pp$ collisions, such as the cross section for Higgs boson production and jet production at the Large Hadron Collider (LHC)~\cite{CMS:2012nga,Kogler:2018hem}, and the direct $J/\psi$ photoproduction at Jefferson Lab~\cite{mammeiproposal}.
In contrast with the quark PDFs, gluon-PDF calculations are less calculated due to their notoriously noisier matrix elements on the lattice.
To date, there have only been a few exploratory gluon-PDF calculations for unpolarized nucleon~\cite{Fan:2018dxu,Fan:2020cpa,HadStruc:2021wmh,Fan:2022kcb}, pion~\cite{Fan:2021bcr,Good:2023ecp} and kaon~\cite{Salas-Chavira:2021wui}, and polarized nucleon~\cite{HadStruc:2022yaw} using the pseudo-PDF~\cite{Balitsky:2019krf} and quasi-PDF~\cite{Zhang:2018diq,Wang:2019tgg} methods.
Most of these calculations, like many exploratory lattice calculations, are done only using one lattice spacing at heavy pion mass (exceptions in Refs.~\cite{Fan:2022kcb,Good:2023ecp}).

The very first LQCD gluon-PDF exploratory study was only done in 2018~\cite{Fan:2018dxu}, since to gluon quantities are much noisier than quark disconnected loops, and calculations with very high statistics are necessary.
The calculations were done using overlap valence fermions on gauge ensembles with 2+1 flavors of domain-wall fermions with $a\approx 0.1$~fm, $P_\text{max}=0.93$~GeV and at $M_\pi^\text{sea}=330$~MeV.
The gluon operators were calculated for all spacetime lattice sites and a high-statistics of 207,872 two-point functions measurements were taken with valence quarks at the light sea and strange masses.
The coordinate-space gluon quasi-PDF matrix element ratios were then compared to the corresponding ones of the gluon PDF based on two global fits at NLO: the \textsc{PDF4LHC15} combination~\cite{Butterworth:2015oua} and the \textsc{CT14}~\cite{Dulat:2015mca}.  
The lattice matrix elements at 330-MeV pion mass are compatible with the ones from global fits within the statistical uncertainty up to $z \approx 0.66$~fm. 
The pion gluon quasi-PDFs were also studied for the first time in Ref.~\cite{Fan:2018dxu}, revealing features similar to those observed for the proton.

\begin{figure}[!t]
\includegraphics[width=.32\textwidth]{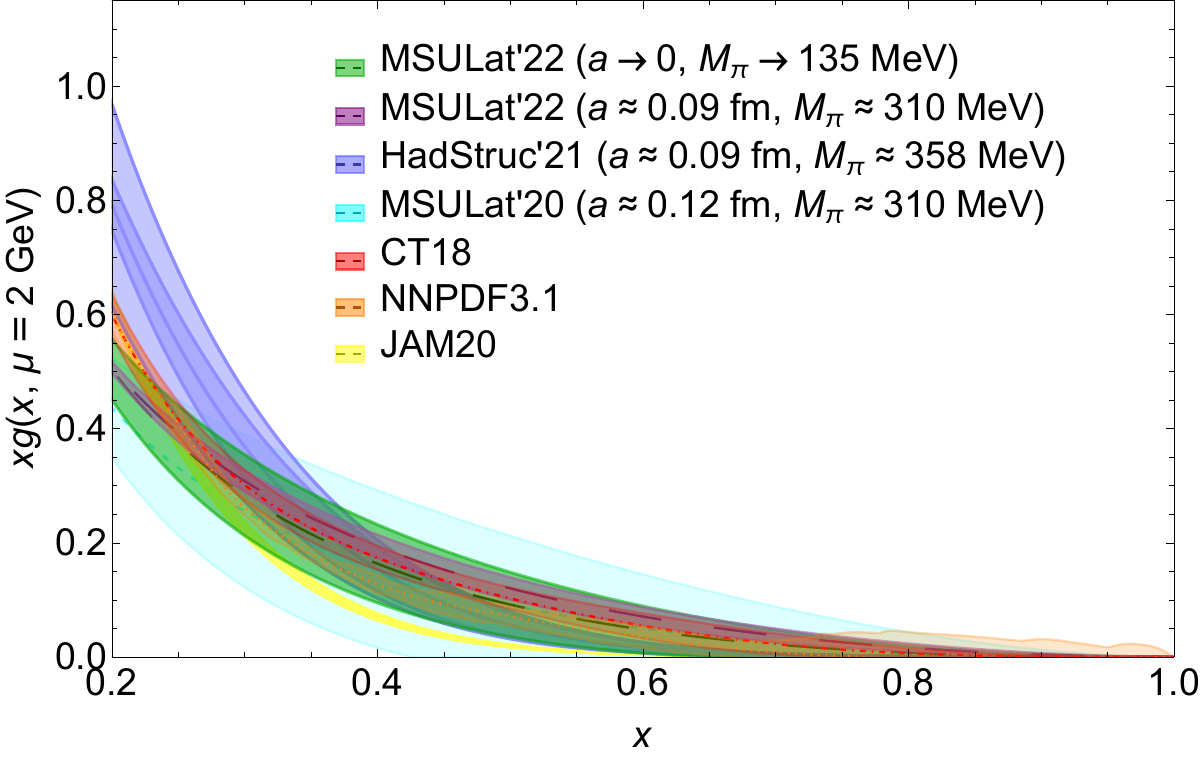}
\includegraphics[width=.32\textwidth]{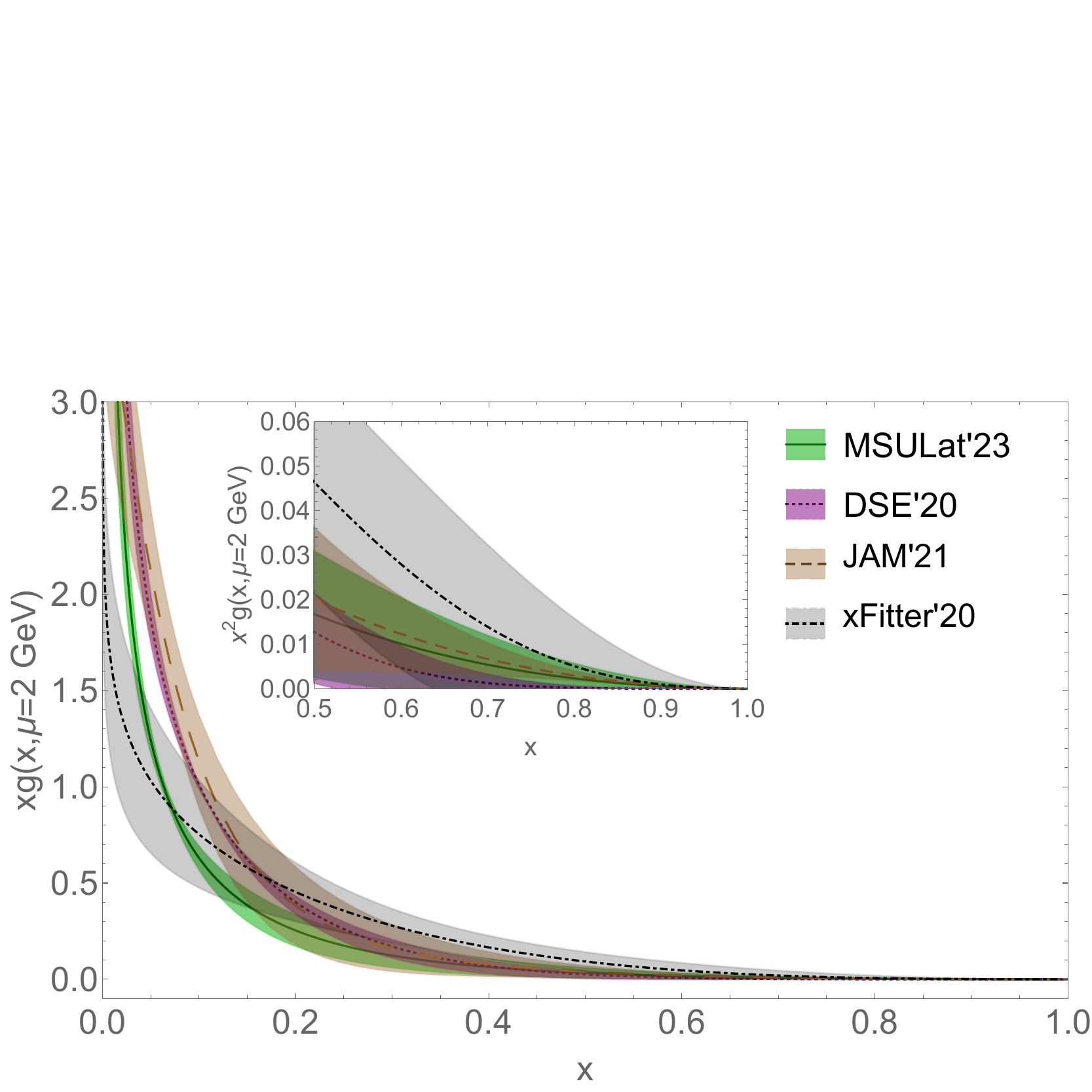}
\includegraphics[width=0.32\textwidth]{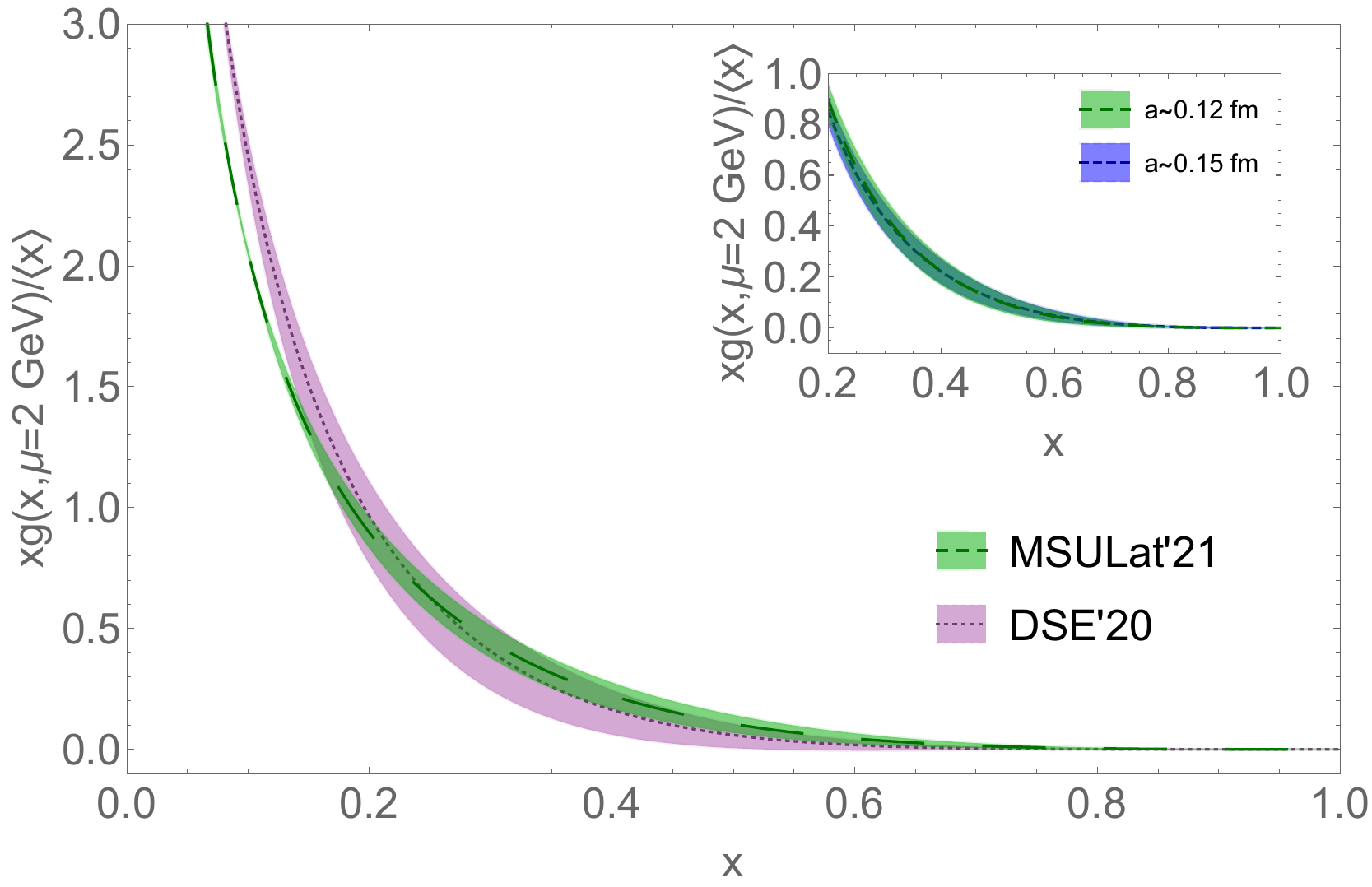}
\caption{
(right)
The $N_f=2+1+1 $unpolarized gluon PDF,
$xg(x,\mu)$, obtained from the NLO fit to the lattice data at pion masses $M_\pi=135$ (extrapolated)~\cite{Fan:2022kcb},
compared with the NNLO CT18 and NNPDF3.1 gluon PDFs.
The $x > 0.3$ PDF results are consistent with the NNLO CT18 and NNPDF3.1 unpolarized gluon PDFs in $\overline{\text{MS}}$ scheme at $\mu=2$~GeV~\cite{Fan:2022kcb}.
(middle) A comparison of the $a \approx 0.12$~fm $M_\pi \approx 220$~MeV $xg(x, \mu)$ result compared with the NLO pion gluon PDFs from xFitter'20~\cite{Novikov:2020snp} and JAM'21~\cite{Barry:2018ort,Cao:2021aci}, along with the DSE'20~\cite{Cui:2020tdf} at $\mu = 2$~GeV in the $\overline{\text{MS}}$ scheme~\cite{Good:2023ecp}.
The inset shows $x^2g(x, \mu)$ to highlight the agreements between the PDFs.
All the results are fairly consistent in the regions where $x > 0.2$.
(right)
The kaon gluon PDF $xg(x, \mu)/\langle x \rangle_g$ as a function of $x$ obtained from the fit to the lattice data on ensembles with lattice spacing $a\approx\{0.12,0.15\}$~fm (inset plot), pion masses $M_\pi\approx310$~MeV at $a\approx 0.12$~fm, compared with the kaon gluon PDF from DSE'20 at $\mu=2$~GeV in the $\overline{\text{MS}}$ scheme~\cite{Salas-Chavira:2021wui}.
\label{fig:gluonPDF}
}
\end{figure}

A first continuum--physical-limit unpolarized $N_f=2+1+1$ gluon nucleon PDF study using three lattice spacings: 0.09, 0.12 and 0.15~fm with pion mass ranging from 220 to 700~MeV with $P_z^\text{max}\approx 3$~GeV using the pseudo-PDF method was reported in Ref.~\cite{Fan:2022kcb}.
They calculated the reduced pseudo-ITD using the fitted lattice matrix elements and studied their pion-mass and lattice-spacing dependence, finding it to be mild compare with the statistical errors.
The reduced pseudo-ITDs were extrapolated to physical-continuum limit before extracting the NLO gluon parton distribution $xg(x)/\langle x \rangle_g$ in the $\overline{\text{MS}}$ scheme at 2~GeV.
Using the nonperturbatively renormalized nucleon momentum fraction calculated on same lattice ensembles, $xg(x)$ was obtained for the first time on the lattice. 
It was found both the lattice 0.09-fm and continuum-physical limit $xg(x)$ to be in good agreement with CT18 and NNPDF3.1 NNLO global-fit results for $x \in [0.2,1]$ in $\overline{\text{MS}}$ scheme at 2~GeV, as shown on the left-hand side of Fig.~\ref{fig:gluonPDF}.
There is some tension with the gluon PDF from JAM20~\cite{Moffat:2021dji} for $x < 0.6$, but its gluon PDF also behaves quite differently from the CT18 and NNPDF results, even with smaller errors. 
We look forward to updates by the global-fit community to resolve this discrepancy.
Comparison with prior lattice calculations of $xg(x)$ from $N_f=2+1$ HadStruc~\cite{HadStruc:2021wmh} ($a\approx0.09$~fm $M_\pi\approx 360$~MeV and $P_z^\text{max}\approx 2.5$~GeV) and $N_f=2+1+1$ MSULat~\cite{Fan:2020cpa} ($a\approx 0.1$~fm $M_\pi\approx 310$~MeV and $P_z^\text{max}\approx 2.2$~GeV) are also shown in the plot.

Using nonperturbatively renormalized pion gluon moments, Ref.~\cite{Good:2023ecp} provides the latest state-of-the-art pion gluon PDFs using the pseudo-PDF method. 
The pion gluon PDF was studied using lattice spacings 0.12 and 0.15~fm with three pion masses, 220, 310 and 690-MeV;
the pion-mass and lattice-spacing effects in this study were found to be negligible within the large statistical errors of the gluon matrix elements.
The $xg(x, \mu)$ pion PDFs are shown in the middle of Fig.~\ref{fig:gluonPDF}, focusing on the gluon PDF result at the lightest pion mass, 220~MeV, and comparing with the global fits~\cite{Barry:2018ort,Cao:2021aci,Novikov:2020snp} and phenomenological results from the Dyson-Schwinger equation (DSE)~\cite{Cui:2020tdf}.
The inset plot weights an additional factor of $x$ and provides a further zoomed-in view of large $x$.
The lattice pion gluon PDF results are consistent with the results from JAM and the DSE for $x>0.2$, while the xFitter results are consistent with lattice ones for $x > 0.15$ with slight tension for $0.3 < x < 0.375$.
The discrepancies in the small-$x$ region are likely due to lack of precision data in the small-$x$ region on the global-fit side and the lack of larger-momentum lattice matrix elements, which would provide better constraint of the results.
Overall, with the current accuracy in lattice calculation and global fits, there is reasonable agreement among them in the mid-to large-$x$ region.

The right-hand side of Fig.~\ref{fig:gluonPDF} shows the only lattice kaon gluon PDF results~\cite{Salas-Chavira:2021wui}.
Only the kaon gluon PDF from smallest lattice spacing, $a\approx 0.12$~fm, is compared with the DSE results~\cite{Cui:2020tdf} and found consistent within one-sigma error for $x>0.15$. 
Reference~\cite{Salas-Chavira:2021wui} also compares the kaon result at smaller lattice spacing with the pion results obtained from the same ensembles;
it is noted that the kaon gluon PDF is slightly smaller than the one obtained for the pion, similar to its corresponding quark valence PDF and DSE~\cite{Cui:2020tdf} results. 
The kaon gluon PDF moments, $\langle x^2 \rangle_g/\langle x \rangle_g$ and $\langle x^3 \rangle_g/\langle x \rangle_g$, are predicted to be 0.0779(94) and 0.0187(42), which is in good agreement with corresponding results from DSE~\cite{Cui:2020tdf}: 0.075 and 0.015.
The pion $\langle x^2 \rangle_g/\langle x \rangle_g$ and $\langle x^3 \rangle_g/\langle x \rangle_g$ 310-MeV results give 0.092(15) and 0.0250(75), which have some tension with those from DSE~\cite{Cui:2020tdf}, JAM~\cite{Barry:2018ort,Cao:2021aci} and xFitter~\cite{Novikov:2020snp}, which are 0.076, 0.103, 0.158 and 0.015, 0.024, 0.048, respectively. 
Future study including finer lattice spacing and lighter pion mass will be important to refine this calculation and provide better predictions on this poorly known meson quantity.

\subsubsection{Strange and Charm PDFs}

In order to distinguish the flavor content (strange or charm) of the PDFs, experiments use nuclear data, such as neutrino scattering off heavy nuclei, and the current understanding of medium corrections in these cases is limited.
Thus, the uncertainty in the strange and charm PDFs remains large.
In many global-analysis cases, the assumptions $\overline{s}(x)=s(x)$ and $\overline{c}(x)=c(x)$ and other assumptions to reduce the number of fitting parameters are often made;
many assumptions, regardless of physical basis, can fit the data well due to their large uncertainty.
At the LHC, strangeness can be extracted through the $W+c$ associated-production channel, but their results are rather puzzling.
For example, ATLAS got the ratios of averaged strange and antistrange to the antidown distribution, $(s+\overline{s})/(2\overline{d})$, to be $0.96^{+0.26}_{-0.30}$ at $Q^2=1.9\text{ GeV}^2$ and $x=0.023$~\cite{Aad:2014xca}.
CMS performed a global analysis with deep-inelastic scattering (DIS) data and the muon-charge asymmetry in $W$ production at the LHC to extract the ratios of the total integral of strange and antistrange to the sum of the antiup and antidown, at $Q^2=20\text{ GeV}^2$, finding it to be $0.52^{+0.18}_{-0.15}$~\cite{Chatrchyan:2013mza}. 
Future high-luminosity studies may help to improve our knowledge of the strangeness.
In the case of the charm PDFs, there has been a long debate concerning the size of the ``intrinsic'' charm contribution, as first raised in 1980~\cite{Brodsky:1980pb} and developed in subsequent papers~\cite{Brodsky:1984nx,Harris:1995jx,Franz:2000ee}\footnote{We refer interested readers to Ref.~\cite{Brodsky:2015fna} and references within~\cite{Hobbs:2013bia,Jimenez-Delgado:2014zga} for a review of intrinsic charm.}.
It is important to distinguish this contribution from radiative contributions in future NNLO and N3LO global PDF analyses.
$c(x)-\overline{c}(x)$ provides an important check of the intrinsic-charm contribution to the proton.
Again, the current experimental data are too inconclusive to discriminate between various proposed QCD models, and future experiments at LHC or EIC could provide useful information in settling this mystery.

\begin{figure}[tb!]
\centering	
\includegraphics[width=.45\textwidth]{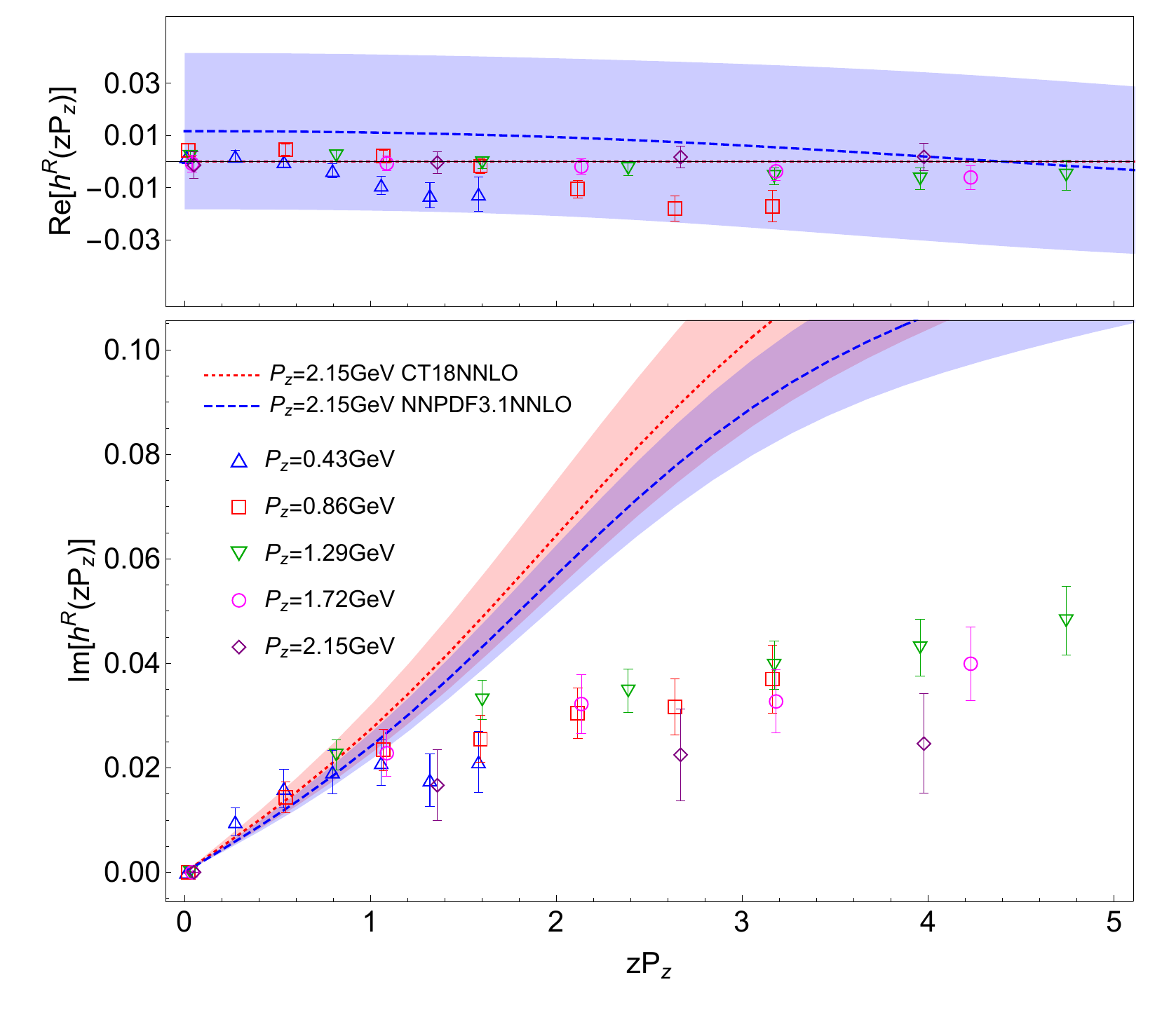}
\includegraphics[width=.45\textwidth]{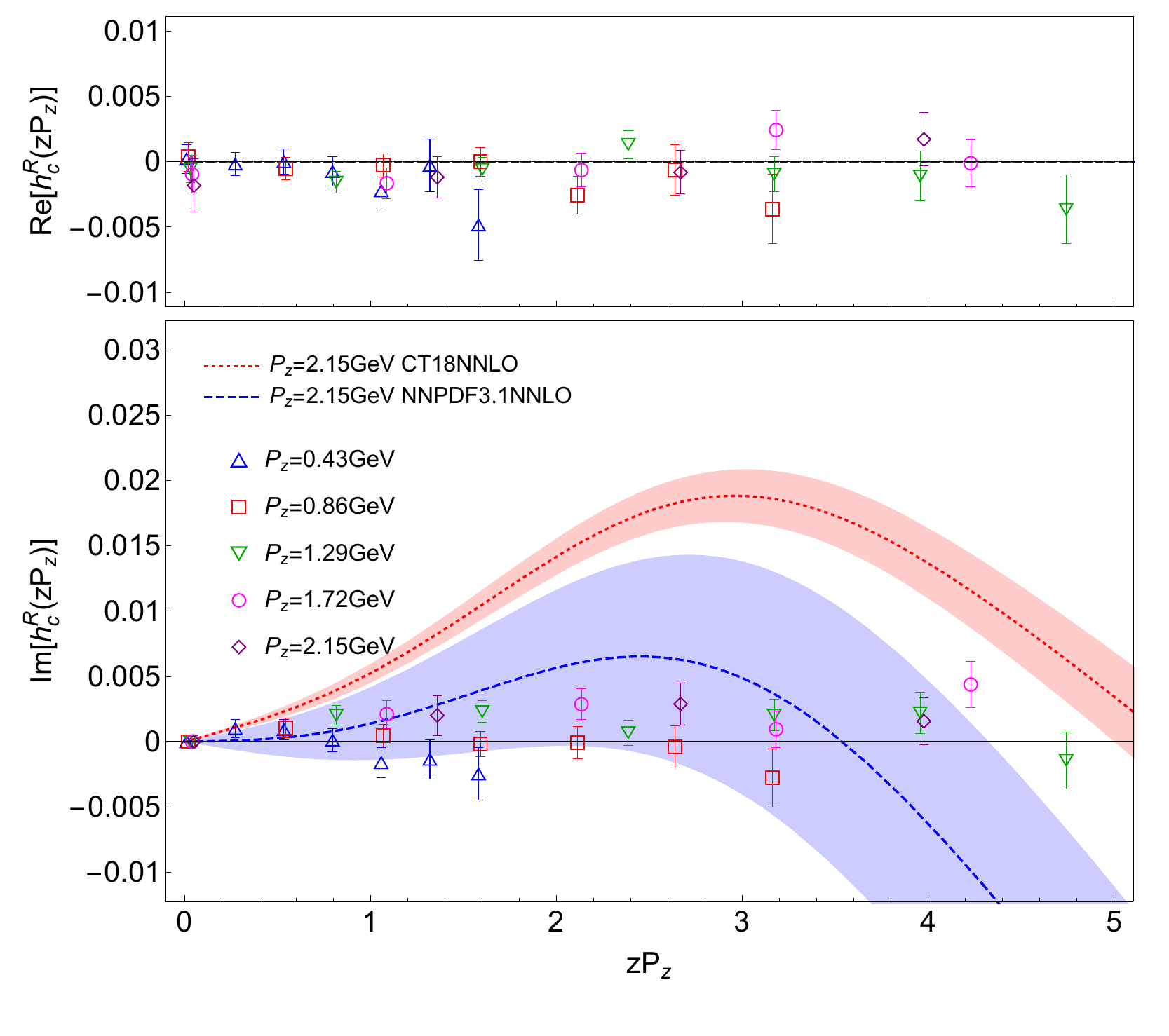}
\caption{
\label{fig:lamet_global_fit_msu}
The real (top) and imaginary (bottom) parts of the strange (left) and charm (right) quasi-PDF matrix elements in coordinate space from $N_F=2+1+1$ MSULat calculations at $a \approx 0.1$~fm physical pion mass with $P_z \in [0.43, 2.15]$~GeV, along with those from CT18 and NNPDF NNLO in RI/MOM renormalized scale of 2.3~GeV~\cite{Zhang:2020dkn}.
The CT18 analysis assumes $s(x)=\bar{s}(x)$, so their results are exactly zero after matching and Fourier transformation.
MSULat real matrix elements at $P_z>1$~GeV are consistent with zero, supporting strange-antistrange symmetry, while the imaginary ones are smaller than global-fit results.
The real charm quasi-PDF matrix elements are also consistent with zero, while the charm imaginary ones within the bounds of NNPDF3.1, but smaller than CT18 results.}
\end{figure}

The first LQCD calculations of the strange and charm parton distributions using LaMET approach were also reported in Ref.~\cite{Zhang:2020dkn} using non-physical pion mass calculations to extrapolate its results to physical pion mass value at a single-lattice ($a \approx 0.12$) ensemble.
The left-hand side of Fig.~\ref{fig:lamet_global_fit_msu} shows the the real strange-quark matrix elements that are proportional to the integral of the difference between strange and antistrange ($\int dx \left(s(x)-\bar{s}(x)\right) \cos(xzP_z)$) and they are consistent with zero at 95\% confidence level for most $zP_z$ points at $P_z>1$~GeV, indicating that the strange quark-antiquark asymmetry is likely very small.
The CT18NNLO PDFs assumes a symmetric $s-\bar{s}$ distribution, so are exactly zero under the transformation with the renormalization scale used, consistent with the lattice findings.
The imaginary matrix elements are proportional to $\int \dl x \left(s(x)+\bar{s}(x)\right) \sin(xzP_z)$ calculated at different boost momenta and there are some small discrepancies, but overall are consistent with each other at large momentum and seem to be approaching a universal curve.
The quasi-PDF matrix elements from both CT18 and NNPDF3.1 are consistent with the lattice results within 2 standard deviations up to $zP_z \approx 2$, and deviate from the results at large $zP_z$, suggesting deviations at moderate to small-$x$ in the PDFs. 
The same work also studied charm matrix elements and compared the charm results with the global-fit PDFs, as shown on right-hand side of Fig.~\ref{fig:lamet_global_fit_msu}.
Note that CT18 and NNPDF3.1 both assume $c(x)=\bar{c}(x)$;
therefore, both of them have vanishing real matrix elements, which is consistent with those for the charm quasi-PDF.
The lattice imaginary charm matrix elements have much smaller magnitudes than the strange, similar to the strange-charm relation observed by global PDF fitting, such as CT18 and NNPDF3.1.
The charm-PDF errors from global fits are significantly different, because the CT18 charm PDF is generated by perturbatively evolving from light-quark and gluon distributions at $Q_0 = 1.3$~GeV, while NNPDF3.1 numerically fit the charm distribution.
The lattice imaginary matrix elements are close to zero at small $zP_z$; at large $zP_z$ they are about a factor of 10 smaller than the strange ones and are within the bounds of the NNPDF3.1 results.

\begin{figure*}[th!]
\centering	
\includegraphics[width=.45\textwidth]{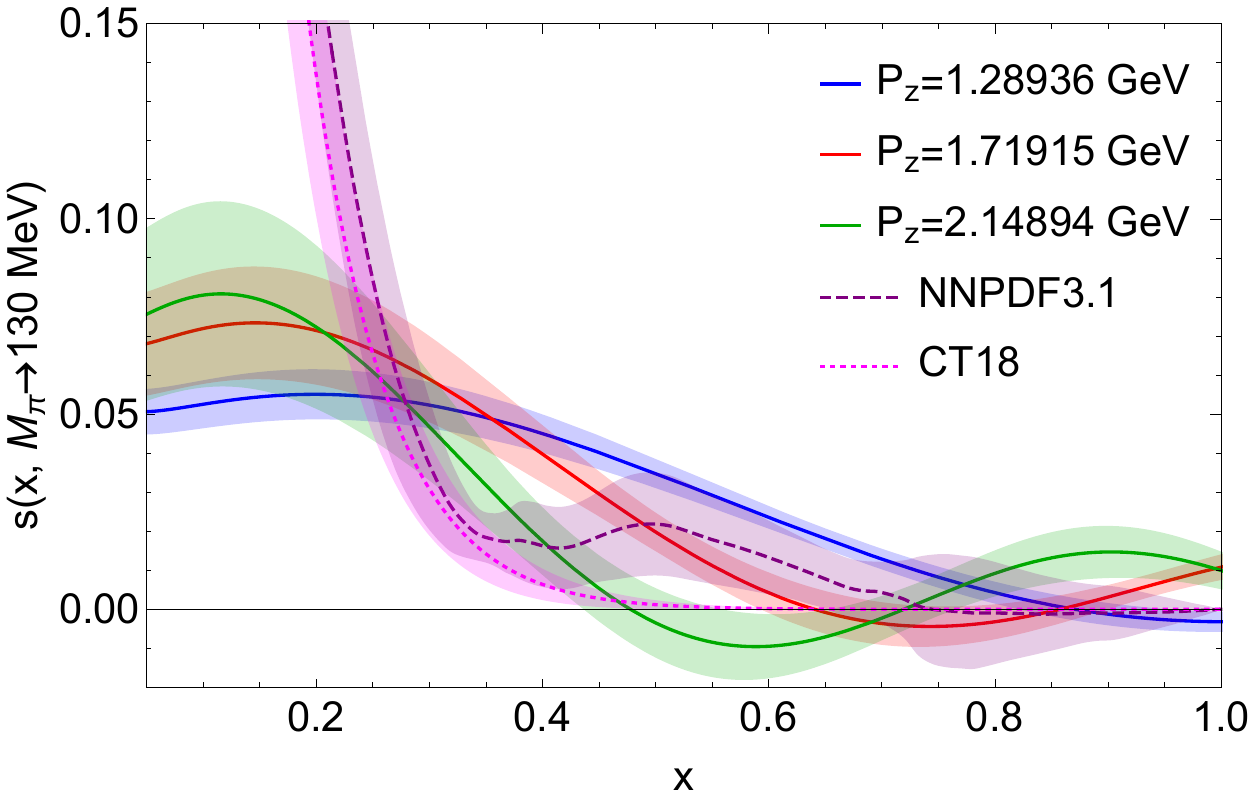}
\includegraphics[width=.45\textwidth]{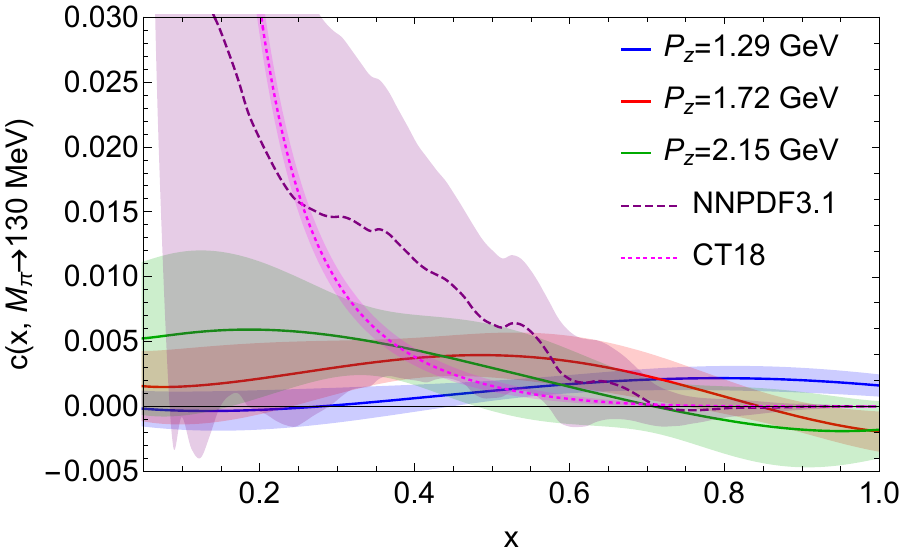}
\caption{\label{fig:lamet_fx_a12_msu_phy} 
The $x$-dependent strange (left) and charm (right) PDF from a naive truncated inverse Fourier transformation and inverse matching for $P_z=\{1.29, 1.72, 2.15\}\text{ GeV}$ extrapolated to physical pion mass~\cite{Zhang:2020dkn} .}
\end{figure*}

To extract the correct $x$-dependence for the strange and charm PDFs, one needs to compute both disconnected quark matrix elements and gluon matrix elements on the lattice, renormalize them with the mixing renormalization matrix, then apply the full matching in momentum space~\cite{Wang:2019tgg}. 
A direct inverse Fourier transformation from the global-fit PDF shows that in coordinate space, the strange PDF matrix element does not converge until very large $zP_z$. 
This indicates that to obtain a reliable $x$-dependence for the strange PDF, one needs go to even larger $zP_z$ in coordinate space beyond the capability of a lattice calculation.
A direct inverse Fourier transformation of the lattice data truncated at $z=8$, then applied the inverse matching kernel to obtain the lightcone distribution as shown in Fig.~\ref{fig:lamet_fx_a12_msu_phy}.
Because the lack of larger-$z$ contributions, which one expects to be large, the distribution in the smaller-$x$ region would suffer larger systematic undertainties in the lattice calculations. 
The momentum dependence is also significant, although their matrix elements are consistent in coordinate space;
this is likely due to different momenta correspond to different truncation ranges of $zP_z$.
Note that the lattice strange and charm imaginary matrix elements mix with the gluon operator;
therefore, one will need to apply the full nonperturbative renormalization to the strange PDFs alongside the gluon matrix elements, as well as the NPR factors for gluon and gluon-quark mixing~\cite{Wang:2019tgg}.
Future studies will be necessary to discern the full strange and charm PDF structure from lattice calculations.

\subsection{Strange and Antistrange Asymmetric PDF Input to Global Fits}
\label{subsec:StrangeAsymFit}

The experimental uncertainty of the strange quark and antiquark PDFs remains large.
In the global fits, they contribute in subleading channels and in neutrino-nucleus DIS experiments with substantial uncertainties.
DIS and LHC experiments, while not in a certain disagreement on the amount of strangeness in the proton, exert contradicting pulls on it, such that the extracted results depend on the type of global analysis used~\cite{CMS:2013pzl,ATLAS:2014jkm,Alekhin:2017olj,Hou:2019efy,Bailey:2020ooq,Faura:2020oom}.
Some of these tensions are relieved by allowing $s(x) \neq \bar{s}(x)$ at the $Q_0$ scale, as is done in some~\cite{Bailey:2020ooq,NNPDF:2021njg} but not all analyses.
The strangeness contributes a large part of the PDF uncertainty to precision $W$ and $Z$ measurements at the LHC~\cite{Nadolsky:2008zw}, so understanding its behavior is important.
Lattice QCD can already provide competitive constraints on the strangeness asymmetry $s_-(x)\equiv s(x)-\bar s(x)$ and reduce some of the uncertainties that are not constrained by experiment~\cite{Hou:2022sdf}.
The left-hand side of Figure~\ref{fig:LRR-RGR-PDFs} shows that the uncertainty on $s_-(x)$ in the CT18As NNLO fit is notably reduced upon adding the first LQCD constraints on $s_-(x)$ at $x\in[0.3,0.8]$.
In the figure, the uncertainty in the lattice data points at $x>0.3$ is quite small compared to the error band of CT18As determined from the global fit, so that including the lattice data in the CT18As\_Lat fit greatly reduces the $s_-$-PDF error-band size in the large-$x$ region.
The reduction of the CT18As\_Lat error band at $x <0.3$ depends on the chosen parametrization form of $s_-(x)$ at $Q_0=1.3$~GeV
Hence, it is important to have more precise lattice data, extended to smaller $x$.
The figure also illustrates the projected reduction in the CT18As\_Lat uncertainty on $s_-(x)$ if the current uncertainties on lattice data are reduced by half.

\subsection{Systematic Improvements}

In the $x$-dependent PDFs, improvements have been made in the development of the hybrid renormalization scheme~\cite{Ji:2020brr} (compared to the pure ratio or pure regularization-independent momentum-subtraction (pure RI/MOM)-schemes), lightcone matching up to two-loop order and the inclusion of Wilson-line extrapolation in the renormalized matrix elements, which reduces the presence of unphysical oscillations in the PDF.
However, the presence of large logarithms and the renormalon ambiguity both in the renormalization and matching processes were not addressed in many of the published works yet.
The large logarithms and renormalon ambiguity can be handled with renormalization group resummation (RGR)~\cite{Su:2022fiu} and leading renormalon resummation (LRR)~\cite{Zhang:2023bxs}.
The RGR method accounts for the fact that lightcone matching depends on the longitudinal momentum of the parton as well as the final renormalization scale, and the difference between them results in large logarithms, which require resummation.
The RGR technique, as applied to the matching process, chooses the renormalization scale such that the logarithms vanish, and the result is evolved to the final desired scale with the DGLAP equations.
The LRR method is applied to the renormalization of the bare matrix elements by demanding that the short-distance behavior agrees with the theoretical predictions of the operator-product expansion (OPE).
The Wilson coefficients are computed as a perturbation series in the strong coupling;
however, the series is not convergent due to the presence of an infrared renormalon (IRR).
The LRR method resums the terms due to the IRR and improves the convergence of the perturbative calculation.
The two methods of RGR and LRR in both the renormalization and matching processes performed up to next-to-next-to-leading-order (NNLO) result in greatly reduced systematic uncertainties in the computed PDF.
The method of RGR in the matching process has been applied to the pion PDF~\cite{Su:2022fiu} as well as the nucleon transversity PDF~\cite{LatticeParton:2022xsd}.
The RGR and LRR methods were applied simultaneously to the pion~\cite{Zhang:2023bxs,Holligan:2024umc} and helicity~\cite{Holligan:2024wpv} and transversity PDFs~\cite{Gao:2023ktu} for the renormalization and lightcone matching processes.
Two examples are described below to provide context for the impact of these systematics.

\begin{figure}
\centering
\includegraphics[width=0.32\textwidth]{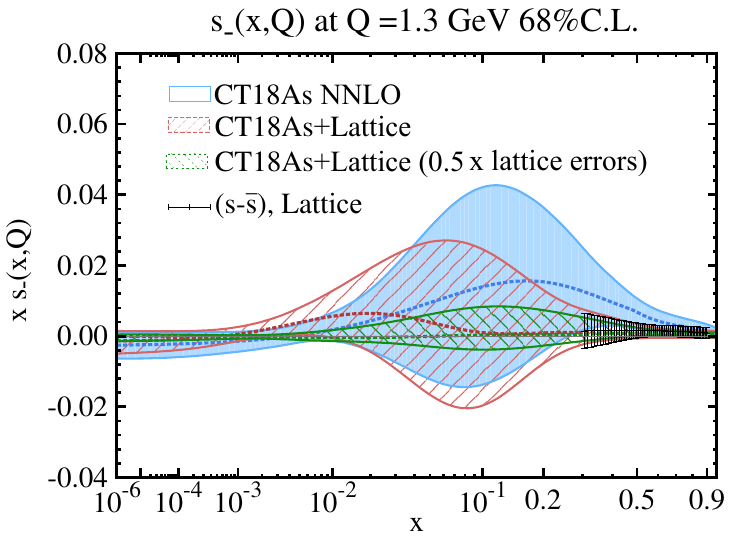}
\includegraphics[width=0.32\textwidth]{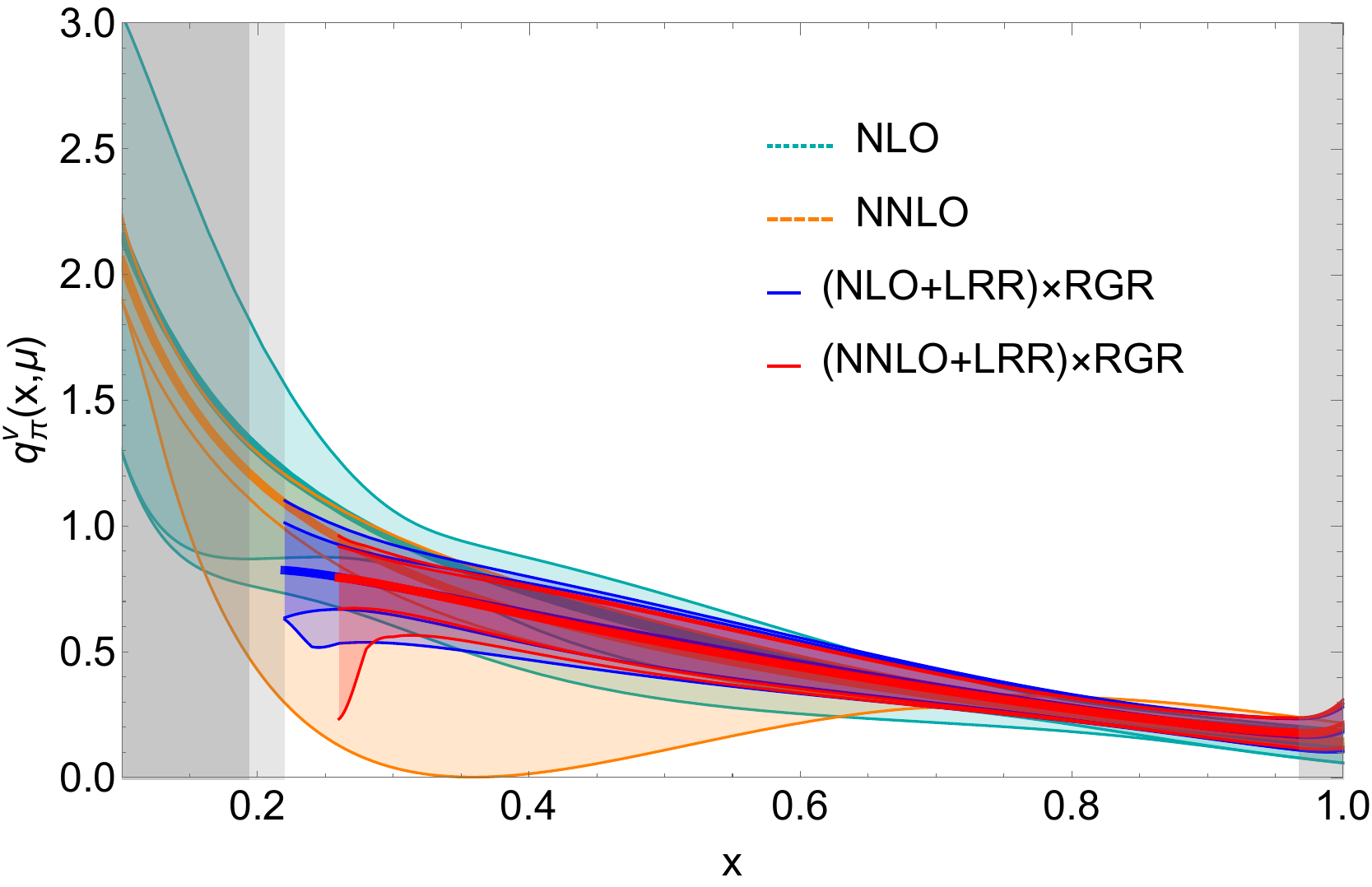}
\includegraphics[width=0.32\textwidth]{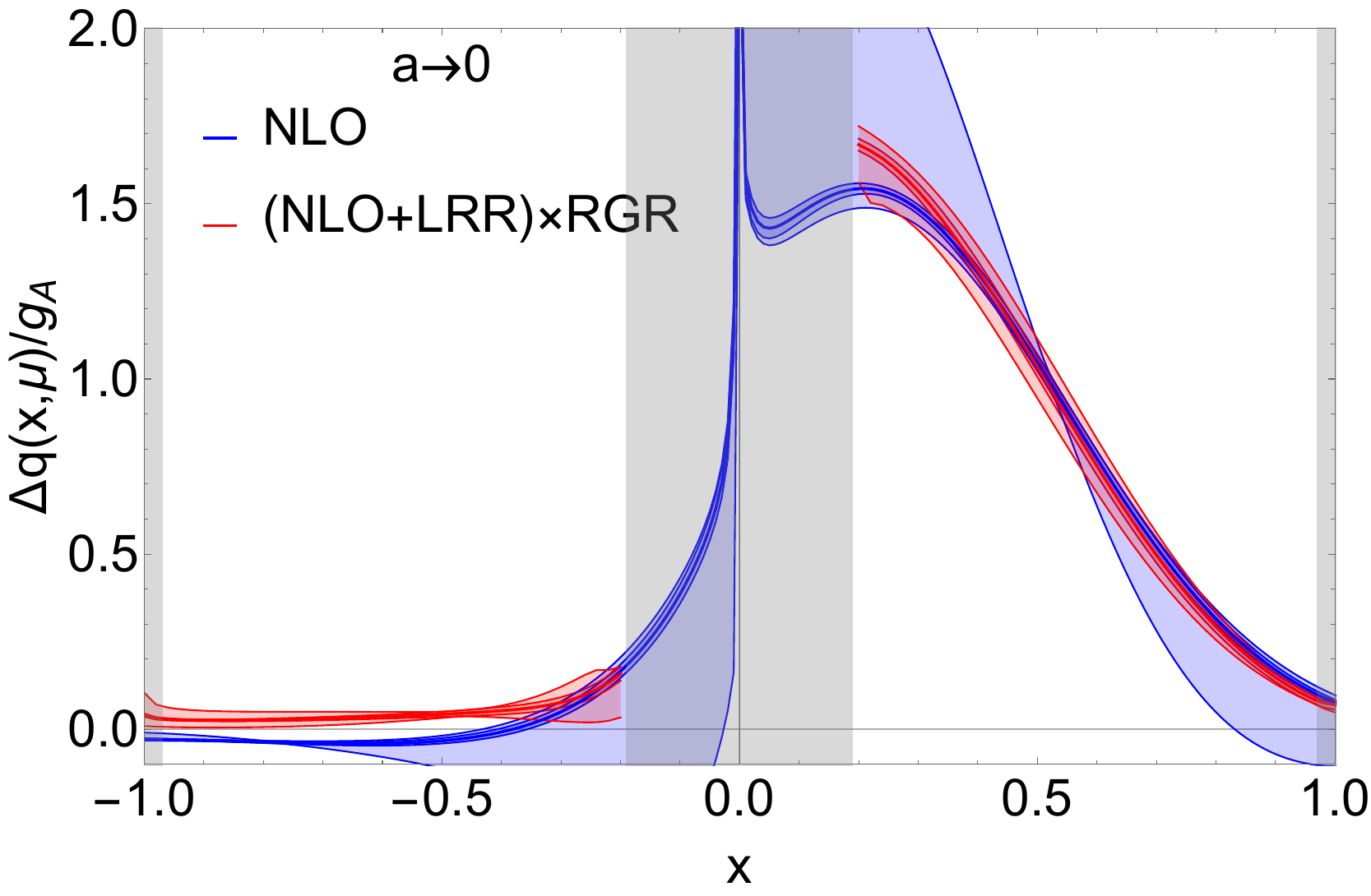}
\caption{
(left)
Impact of constraints from lattice QCD (black dashed area) on the difference between strange quark and antiquark PDFs in a recent CT18As NNLO fit~\cite{Hou:2022sdf}.
The red (green) error bands are obtained with the current (reduced by 50\%) LQCD errors.
(middle)
The lightcone $N_f=2+1+1$ pion valence-quark PDFs at $a\approx 0.09$~fm physical pion mass as a function of $x$ renormalized in hybrid-ratio scheme with \N\ (cyan), \NLR\ (blue), \NN\ (orange) and \NNLR\ (red) improvement with $P_z\approx 1.72$~GeV. 
In each PDF, the inner darker band shows statistical error, while the outer lighter band shows combined statistical and systematic errors (due to scale variation).
The dark-gray region is where the LaMET calculation begins to break down, and the light-gray region is where the RGR matching begins to break down.
(right)
Lightcone NLO $N_f=2+1+1$ nucleon helicity PDFs at \N\ (blue) and \NLR\ (red) with pion mass $M_{\pi}\approx 310$~MeV with $P_z\approx 1.75$~GeV in the continuum limit~\cite{Holligan:2024wpv}.
The inner error bars are statistical, and the outer error bars are combined statistical and systematic errors.
}
\label{fig:LRR-RGR-PDFs}
\end{figure}

Reference~\cite{Holligan:2024umc} applied both RGR and LRR methods to different renormalization schemes to examine their effects on the pion valence-quark PDFs and their systematic errors at physical pion mass.
The pion matrix elements were calculated with a boost momentum $P_z=8\times\frac{2\pi}{L}\approx 1.72$~GeV with the number of measurements for each source-sink separation up to $O(10^6)$.
The study showed that the renormalized matrix elements in both the hybrid-RI/MOM and hybrid-ratio scheme were consistent with each other within the statistical errors, but the former have slightly higher central value across all the Wilson-line displacements studied.
The systematic errors from scale variation in the renormalized matrix elements in the hybrid-RI/MOM and hybrid-ratio schemes were greatly reduced by the simultaneous application of RGR and LRR at one- and two-loop level respectively.
However, the application of RGR on its own at either level increased the systematic errors, due to its enhancement of the renormalon ambiguity.
It was also found that pion valence-quark PDF in hybrid-RI/MOM and hybrid-ratio scheme were consistent with each other within one sigma. 
In studying the $x$-dependent PDFs, the middle plot of Fig.~\ref{fig:LRR-RGR-PDFs}
shows pion valence-quark PDFs for one- and two-loop treatments are compared with
\N\ (dotted cyan), \NLR\ (solid blue), \NN\ (dashed orange) and \NNLR\ (solid red) improvements.
One would normally expect a calculation to yield more precise results as one goes to higher order (e.g. \N\ to \NN);
however, this is not the case, since the systematic errors increase from \N\ to \NN.
It is necessary to account for the effects of large logarithms and the renormalon divergence with the methods of RGR and LRR, respectively.
The systematic errors decrease from \NLR\ to \NNLR, showing that, in this case, the calculation becomes more precise.
The left-hand side of Fig.~\ref{fig:LRR-RGR-PDFs} is a demonstration that it is necessary to control the sources of systematic errors if higher-order terms are to be included in the lightcone matching.
Since the systematic errors decrease from \NLR\ to \NNLR\ by 10\% to 15\%, one can see the benefits of including higher-order terms in the matching and renormalization.
This demonstrates that it is necessary to include both RGR and LRR if we wish to renormalize and match to two-loop level.

The first lattice calculation of the nucleon isovector helicity PDF in the LaMET framework that uses the hybrid scheme with self-renormalization at lattice spacings $a\in\{0.1207,0.0888,0.0582\}$~fm with pion mass of $M_\pi \approx 315$~MeV~\cite{Holligan:2024wpv}.
The right-hand side of Fig.~\ref{fig:LRR-RGR-PDFs} shows the helicity PDFs at the continuum limit with \N\ (\NLR) improvement plotted as a blue (red) band.
The inner error bars are statistical and the outer error bars are combined statistical and systematic errors, the latter of which is computed with scale variation. 
The regions $x>0$ and $x<0$ correspond to $\Delta u(x,\mu)-\Delta d(x,\mu)$ (``quark region'') and $\Delta \overline{u}(x,\mu)-\Delta \overline{d}(x,\mu)$ (``antiquark region''), respectively.
One can see that the antiquark region becomes much flatter when going from \N\ to \NLR.
The two cases have opposite signs in the antiquark region for all lattice spacings, whereas the \NLR\ case is more reliable due to its improved control of systematic errors.
In addition, it predicts $\Delta\overline{u}(x)-\Delta\overline{d}(x)>0$ which is consistent with results from global fits in Refs.~\cite{STAR:2014afm,PHENIX:2015ade,deFlorian:2009vb}.

%% file: sec4-GPDs.tex
\section{Generalized Parton Distributions}\label{sec:GPDs}

Generalized parton distributions (GPDs) provide hybrid momentum and coordinate-space distributions of partons and bridge the standard nucleon structure observables: form factors and collinear PDFs.
Thus, GPDs not only depend on fraction of the hadron momentum carried by the parton, $x$ but also the square of the invariant momentum transfer $t$ and the longitudinal momentum transfer $\xi$.
GPDs bring the energy-momentum tensor matrix elements within experimental grasp through electromagnetic scattering and can be viewed as a hybrid of PDFs, form factors, and distribution amplitudes.
For example, the forward limit of the unpolarized and helicity GPDs leads to the $f_1(x)$ and $g_1(x)$ PDFs, respectively.
Taking the integral over $x$ at finite values of the momentum transfer results in the form factors and generalized form factors.
In the case of the unpolarized GPDs, for example, one obtains the Dirac ($F_1$) and Pauli ($F_2$) form factors.
More importantly, GPDs provide information on the spin and mass structure of the nucleon.
Several limits of the GPDs have physical interpretations, for instance, the spin decomposition of the proton using Ji's sum rule~\cite{Ji:1996ek}.
GPDs also provide information about gravitational form factors, which
have been interpreted as a measure of the pressure and shear forces inside hadrons~\cite{Polyakov:2002wz, Polyakov:2002yz,Polyakov:2018zvc,Lorce:2018egm};
a detailed recent review on this topic can be found in Ref.~\cite{Burkert:2023wzr}. 

Experimentally, one can access GPDs via, for example, deeply virtual Compton scattering (DVCS)~\cite{Muller:1994ses, Ji:1996ek, Radyushkin:1996nd, Ji:1996nm, Collins:1998be} and hard exclusive meson production (DVMP)~\cite{Radyushkin:1996ru, Collins:1996fb, Mankiewicz:1997uy} processes, but with limited information.
Some information is available in the intermediate to high-$x$ range from fixed-target DVCS and in the low-$x$ region from HERA measurements.
The upcoming JLab 12-GeV program will offer more data on GPDs, as will the future EIC.
However, successfully disentangling the GPDs from the experimental data is very challenging due to the strict requirements in luminosity, center-of-mass energy, and hadron beam parameters~\cite{AbdulKhalek:2021gbh}. 
Lattice QCD can play an important and complementary role to provide crucial information on GPDs in different kinematic regions that is not available (say $\xi=0$) or those regions that are difficult to reach experimentally. 
Furthermore, the recent $x$-dependent lattice results provides even more important input to GPDs than PDFs since the inverse problem in GPDs is way more severe than those in PDFs. 
We started this section with the selected results of nucleon isovector (Sec.~4.1) and pion valence (Sec.~4.2) GPD directly calculated at physical pion mass results, and showed example results with the improved renormalization and RGR+LRR improvement in Sec.~4.3.  

\subsection{Nucleon Isovector GPDs}

The unpolarized and longitudinally polarized GPDs can be parametrized in terms of the quark and gluon correlation functions involving matrix elements of operators at a lightlike separation between the parton fields~\cite{Ji:1996nm},
\begin{align}
\label{eq:def_vec}
W_{\Lambda \Lambda'}^{[\gamma^+]} & =\frac{1}{2\sqrt{1-\xi^2}} \int \frac{\dl y}{2 \pi}^- e^{i k^+ y^-} \langle p', \Lambda'\mid \bar{\psi}\left(-\frac{y}{2} \right) \gamma^+ \, {\cal U}\left(-\frac{y}{2}, \frac{y}{2}\right) \psi\left(\frac{y}{2} \right) \mid p, \Lambda \rangle_{y^+ =y_T=0} \nonumber \\
& = \frac{1}{2 P^+\sqrt{1-\xi^2}} \bar u(p', \Lambda') \left[ \gamma^+ H(x,\xi,t) + \frac{i \sigma^{+j} \Delta_j}{2M} E(x,\xi,t)\right] u(p, \Lambda) ,
\\
\label{eq:def_axi}
\widetilde{W}_{\Lambda \Lambda'}^{[\gamma^+\gamma_5]} & = \frac{1}{2\sqrt{1-\xi^2}} \int \frac{\dl y}{2 \pi}^- e^{i k^+ y^-} \langle p', \Lambda'\mid \bar{\psi}\left(-\frac{y}{2} \right) \gamma^+ \gamma_5 \, {\cal U}\left(-\frac{y}{2}, \frac{y}{2}\right) \psi\left(\frac{y}{2} \right) \mid p, \Lambda \rangle_{y^+=y_T=0}
\nonumber \\
& = \frac{1}{2 P^+\sqrt{1-\xi^2}} \bar u(p', \Lambda') \left[ \gamma^+ \gamma_5 \widetilde{H}(x,\xi,t) + \frac{ \gamma_5 \Delta^+}{2M} \widetilde{E}(x,\xi,t) \right] u(p, \Lambda) , \nonumber \\
\end{align}
where the kinematic variables are:
the lightcone longitudinal momentum fraction $x=k^+/P^+$ (with $P=(p+p')/2$);
the lightcone component of the longitudinal momentum transfer between the initial and final proton $\xi=-\Delta^+/(2P^+)$ 
(with $\Delta = p'-p$); 
and the transverse component, ${\Delta}_T={ p}_T'-{ p}_T$.
the latter is taken into account through a invariant variable, $t= \Delta^2= M^2 \xi^2/(1-\xi^2) - \Delta_T^2/(1-\xi^2), t<0$).
At leading twist-two level, four quark-chirality--conserving (chiral-even) GPDs, $H$, $E$, $\widetilde{H}$ and $\widetilde{E}$, defined in Eqs.~\eqref{eq:def_vec} and \eqref{eq:def_axi}, parametrize the quark-proton correlation functions.

Information on GPDs from lattice QCD has been available via their form factors and generalized form factors, using the operator product expansion (OPE), for many decades;
recent efforts have been put toward direct calculation at the physical pion mass (more details can be found in Ref.~\cite{Constantinou:2020hdm}).
As in PDFs, such information is limited due to the suppression of the signal as the order of the Mellin moments increases and the momentum transfer between the initial and final state increases.
Significant progress has been made towards new methods to access the $x$- and $t$-dependence of GPDs ($t=-Q^2$), which is driven by the advances in PDF calculations.
In lattice QCD, there are several challenges in calculating GPD using these new methods.
The extraction of $x$-dependent GPDs is more challenging than collinear PDFs, because GPDs require momentum transfer, $Q^2$, between the initial (source) and final (sink) states.
Another complication is that GPDs are defined in the Breit frame, in which the momentum transfer is equally distributed to the initial and final states;
such a setup increases the computational cost, as separate calculations are necessary for each value of the momentum transfer.
Nevertheless, there has been progress made extracting $x$-dependent GPDs using lattice QCD.

\begin{figure}[tb]
\includegraphics[width=0.32\textwidth]{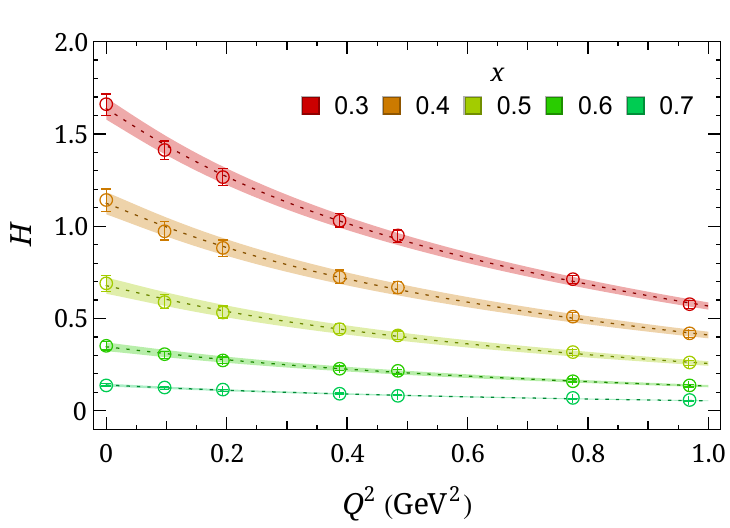}
\includegraphics[width=0.32\textwidth]{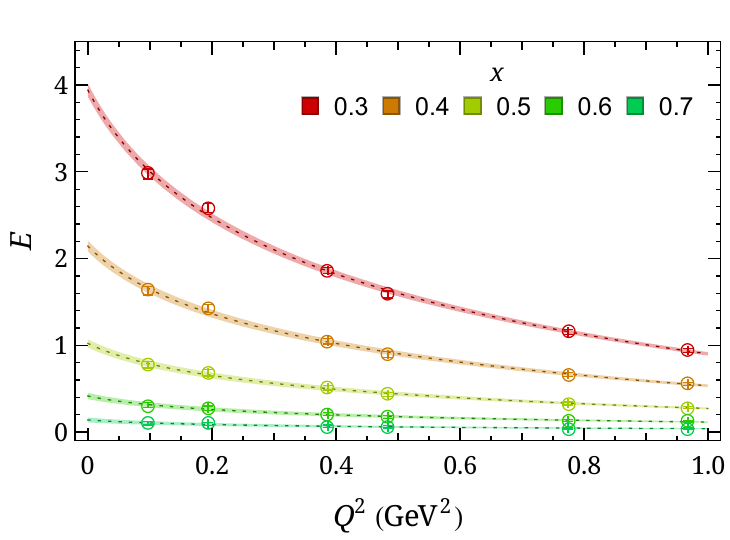}
\includegraphics[width=0.32\textwidth]{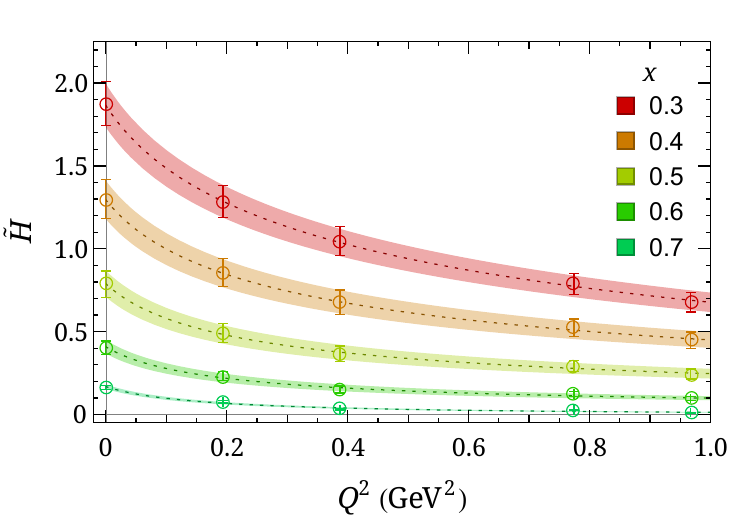}
\caption{
Nucleon isovector $H$ (left), $E$ (middle) and $\tilde{H}$ (right) GPDs at $\xi=0$ with $z$-expansion to $Q^2$ at selected $x$ values.
\label{fig:3DGPD}}
\end{figure}

The first lattice $x$-dependent GPD calculations were carried out in 2019 using LaMET method, to study the pion valence-quark GPD at zero skewness with multiple transfer momenta ($P_z^\text{max} 1.7$~GeV) with $N_f=2+1+1$ $a\approx 0.1$~fm pion mass $M_\pi \approx 310$~MeV~\cite{Chen:2019lcm}.
There is a reasonable agreement with traditional local-current form-factor calculations at similar pion mass, but the current uncertainties remain too large to show a clear preference among different model assumptions about the kinematic dependence of the GPD.
In 2020, another effort is made to study the Bjorken-$x$ dependence of the $N_f=2+1+1$ isovector nucleon GPDs, $H$, $E$ and $\tilde{H}$~\cite{Alexandrou:2020zbe} also using LaMET with single momentum transfer at pion mass $a\approx 0.09$~fm $M_\pi \approx 260$~MeV with $P_z^\text{max} \approx 1.7$~GeV. 
This work also presented results at one nonzero skewness, which has an additional divergence near $x=\xi$ due to the matching.
About the same time and following year, Refs.~\cite{Lin:2020rxa,Lin:2021brq} reported the first lattice-QCD calculations of the $N_f=2+1+1$ NLO RI/MOM-renormalized unpolarized and helicity nucleon GPDs with boost momentum around 2.0~GeV at $a\approx 0.09$~fm physical pion mass with multiple transfer momenta, allowing study of the three-dimensional structure and impact-parameter--space distribution.
The $z$-expansion is used interpolate the $Q^2$ dependence of the lightcone GPD functions, and as shown in Fig.~\ref{fig:3DGPD}, the fit describes the lattice data well.
The authors used the interpolated $Q^2$ dependence to study the tomography.

To ensure that the lattice inputs can generate reliable tomography results, Refs.~\cite{Lin:2020rxa,Lin:2021brq} take the integral over $x$ of the $H$, $E$ and $\tilde{H}$ GPDs extracted from the lattice to obtain the moments. These are compared with previous lattice calculations using traditional form factors and generalized form factors at or near the physical pion mass.
The zero-skewness limit of GPD function is related to the Mellin moments by taking the $x$-moments~\cite{Ji:1998pc,Hagler:2009ni}:
\begin{align}
\label{eq:GFFs}
\int_{-1}^{+1}\!\!\dl x \, x^{n-1} \, H(x, \xi, Q^2)&=
\sum\limits_{i=0,\text{ even}}^{n-1} (-2\xi)^i A_{ni}(Q^2) + (-2\xi)^{n} \, C_{n0}(Q^2)|_{n\text{ even}},\nonumber \\
\int_{-1}^{+1}\!\!\dl x \, x^{n-1} \, E(x, \xi, Q^2)&=
\sum \limits_{i=0,\text{ even}}^{n-1}	(-2\xi)^i B_{ni}(Q^2) - (-2\xi)^{n} \, C_{n0}(Q^2)|_{n\text{ even}}\,, \nonumber \\
\int_{-1}^{+1}\!\!\dl x \, x^{n-1} \, \tilde{H}(x, \xi, Q^2)&=
\sum\limits_{i=0,\text{ even}}^{n-1} (-2\xi)^i \tilde{A}_{ni}(Q^2) + (-2\xi)^{n} \, \tilde{C}_{n0}(Q^2)|_{n\text{ even}}, 
\end{align}
where the unpolarized and polarized generalized form factors (GFFs) are $A_{ni}(Q^2)$, $B_{ni}(Q^2)$ and $C_{ni}(Q^2)$, and $\tilde{A}_{ni}(Q^2)$, 
and $\tilde{C}_{ni}(Q^2)$. 
When $n=1$, we get the Dirac and Pauli electromagnetic, and axial form factors $F_1(Q^2) = A_{10}(Q^2)$, $F_2(Q^2) = B_{10}(Q^2)$,
$G_A(Q^2) = \tilde{A}_{10}(Q^2)$;
$n=2$, GFFs $A_{20}(Q^2)$, $B_{20}(Q^2)$ and $\tilde{A}_{20}(Q^2)$.

\begin{figure}[tb]
\centering
\includegraphics[width=0.32\textwidth]{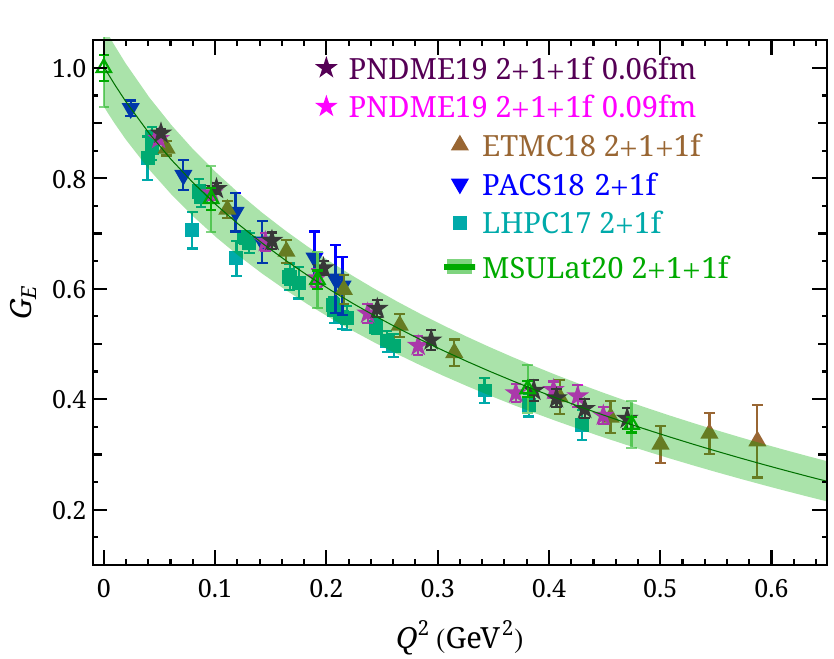} 
\includegraphics[width=0.32\textwidth]{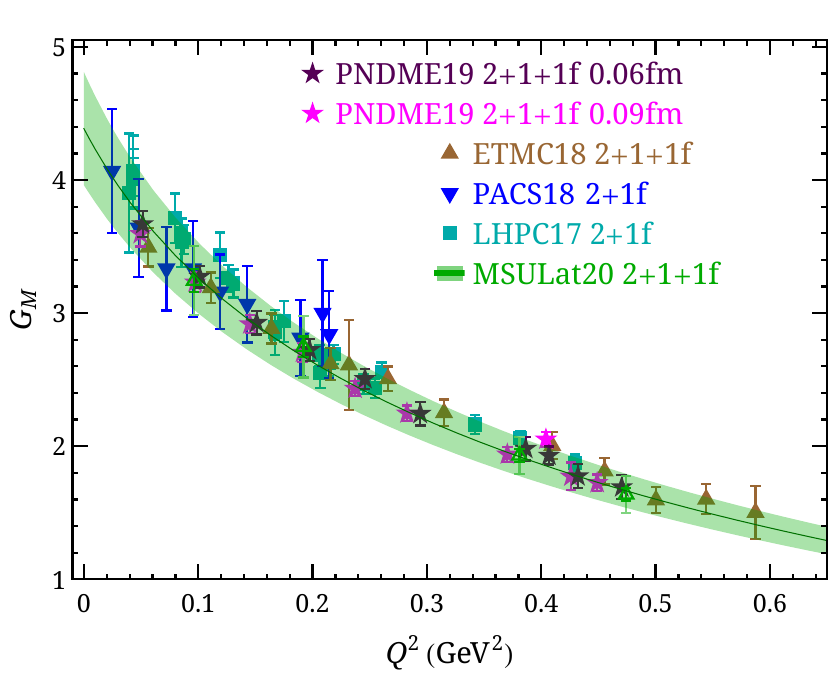}
\includegraphics[width=0.32\textwidth]{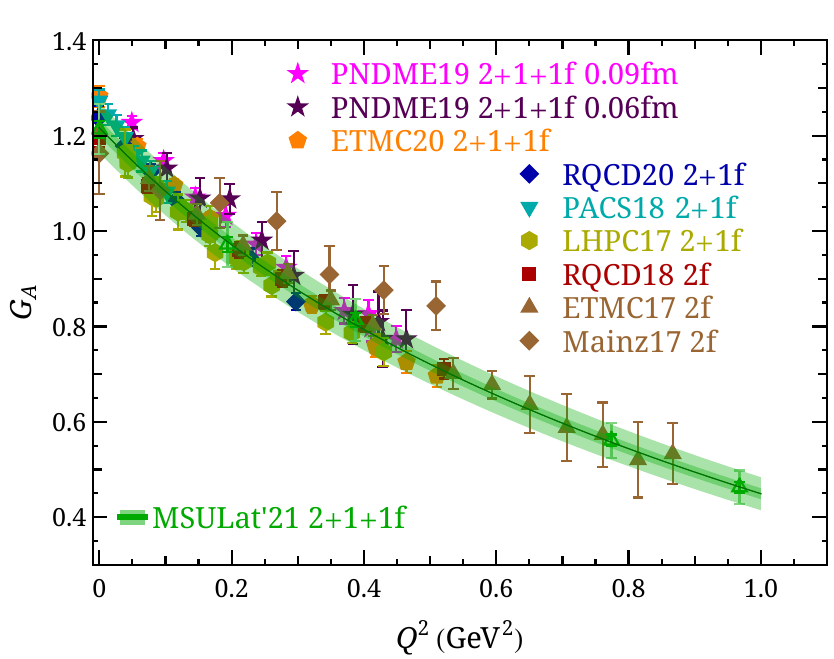}
\includegraphics[width=0.32\textwidth]{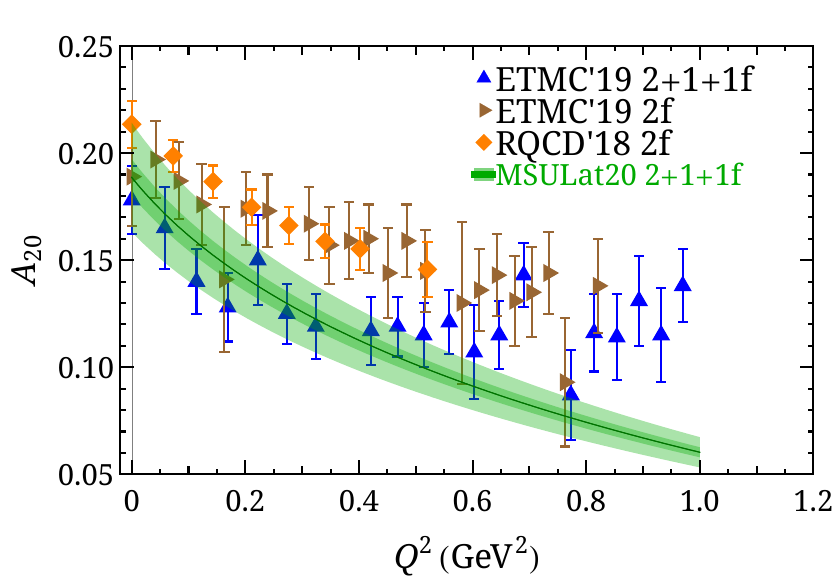} 
\includegraphics[width=0.32\textwidth]{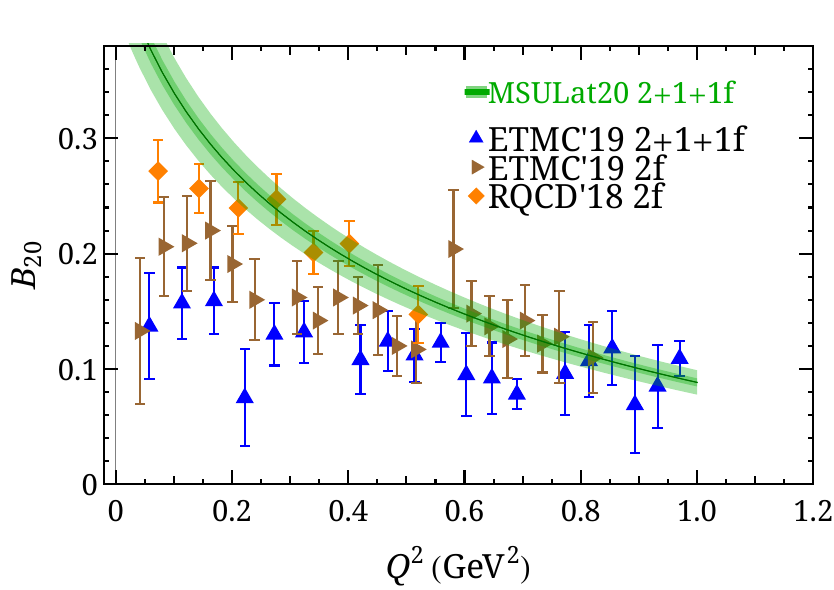}
\includegraphics[width=0.32\textwidth]{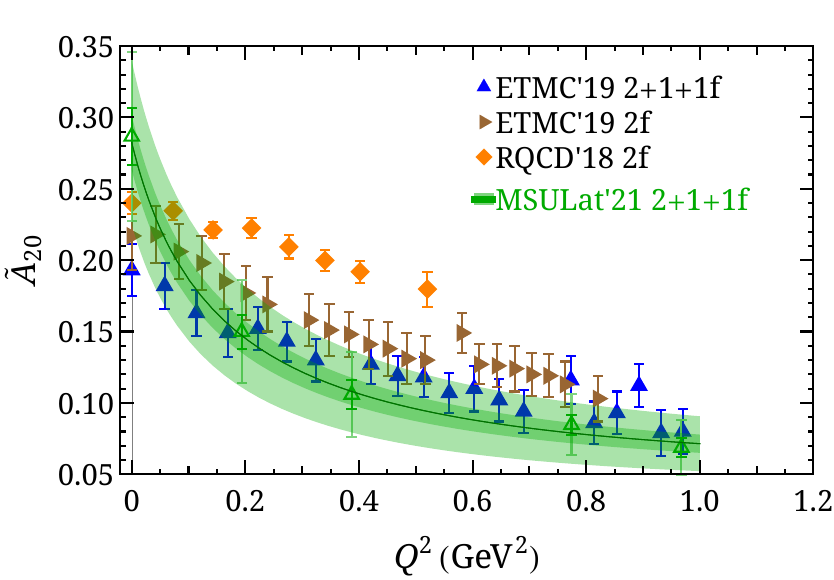}
\caption{
\label{fig:LatGFF}
(top) The nucleon isovector electric (left), magnetic (middle) and axial (right) form factor results obtained from $x$-dependent GPDs~\cite{Lin:2020rxa,Lin:2021brq} (labeled as green bands in the plots) as functions of transfer momentum $Q^2$, and comparison with other lattice works calculated near physical pion mass~\cite{Green:2014xba,Rajan:2017lxk,Hasan:2017wwt,Capitani:2017qpc,Alexandrou:2017hac,Bali:2018qus,Alexandrou:2018sjm,Jang:2018djx,Shintani:2018ozy,RQCD:2019jai,Alexandrou:2020okk}. 
(bottom) The unpolarized nucleon isovector GFFs $\{A,B\}_{20}(Q^2)$ and $\tilde{A}_{20}(Q^2)$ obtained from $x$-dependent GPDs~\cite{Lin:2020rxa,Lin:2021brq} (labeled as green bands in the plots by taking $n=2$ in Eq.~\ref{eq:GFFs}, compared with other lattice results calculated near physical pion mass as functions of transfer momentum $Q^2$~\cite{Bali:2018zgl,Alexandrou:2019ali}.
}
\end{figure}

The upper row of Fig.~\ref{fig:LatGFF} shows the $n=1$ moments of MSULat's $x$-dependent GPDs calculated at physical pion mass~\cite{Lin:2020rxa,Lin:2021brq}, alongside prior LQCD OPE calculations obtained by other lattice collaborations.
To compare with more lattice results, the Sachs electric ($G_E$) and magnetic ($G_M$) form factors are plotted by using the $F_{1,2}$ obtained from the $x$-integral in $G_E(Q^2)=F_1(Q^2)+q^2F_2(Q^2)/(2 M_N)^2$ and $G_M(Q^2)=F_1(Q^2)+F_2(Q^2)$.
In each plot, the green point and green band indicate MSULat's results from $x$-dependent GPDs, while the OPE calculations are shown as points in various colors.
We can see nice agreement among the methods and between different LQCD calculations for the single lattice-spacing results.
The lower row of Fig.~\ref{fig:LatGFF} shows $n=2$ moment of MSULat's $x$-dependent GPD results with those obtained from GFFs at or near the physical pion mass using OPE methods~\cite{Alexandrou:2019ali,Bali:2018zgl}.
We note that even with the same OPE approach by the same collaboration, the two data sets for $A_{20}$ in the ETMC calculation exhibit some tension.
This is an indication that the larger lattice systematic uncertainties are more complicated for these GFFs.
Given that the blue points correspond to finer lattice spacing, larger volume and larger $M_\pi L$, we expect that the blue points have suppressed systematic uncertainties.
MSULat's moment result $A_{20}(Q^2)$ is in better agreement with those obtained using the OPE approach at small momentum transfer $Q^2$, while $B_{20}(Q^2)$ is in better agreement with OPE approaches at large $Q^2$.
The comparison between the $N_f=2$ ETMC data and $N_f=2$ RQCD data reveals agreement for $A_{20}$ and $B_{20}$. 
However, the RQCD data have a different slope than the ETMC data, which is attributed to the different analysis methods and systematic uncertainties.
Both MSULat's results and ETMC's are done using a single ensemble;
future studies to include other lattice artifacts, such as lattice-spacing dependence are important to account for the difference in the results.

\begin{figure}[tb]
\begin{center}
\includegraphics[width=0.85\textwidth]{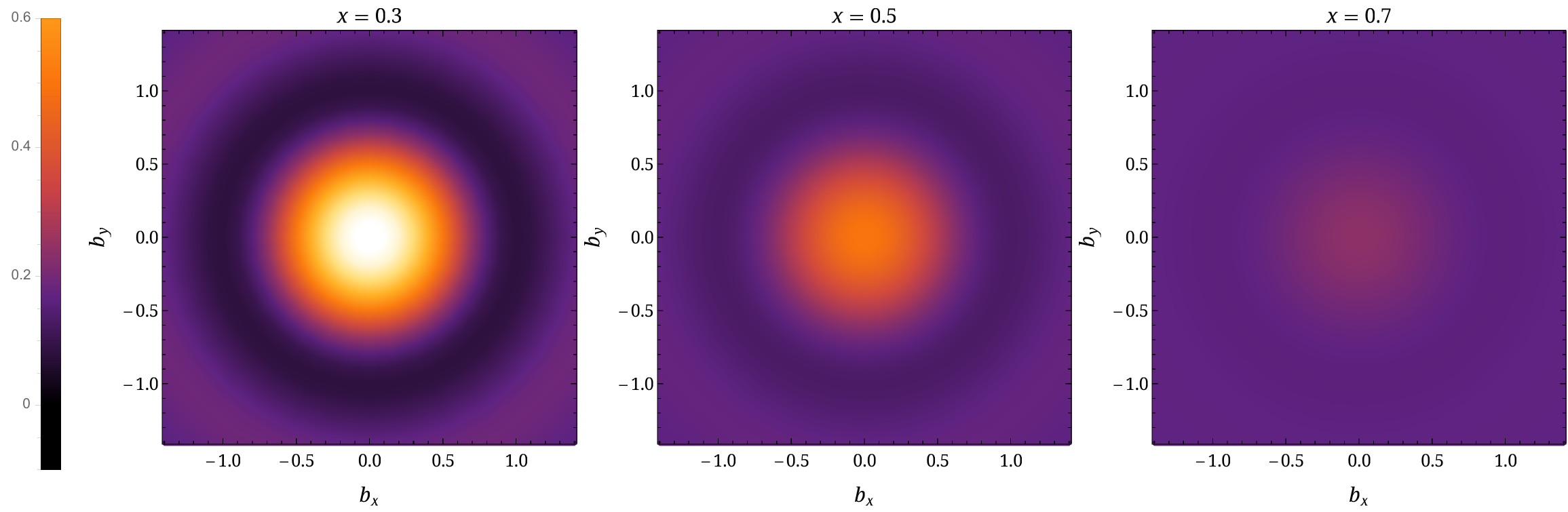}\\
\includegraphics[width=0.85\textwidth]{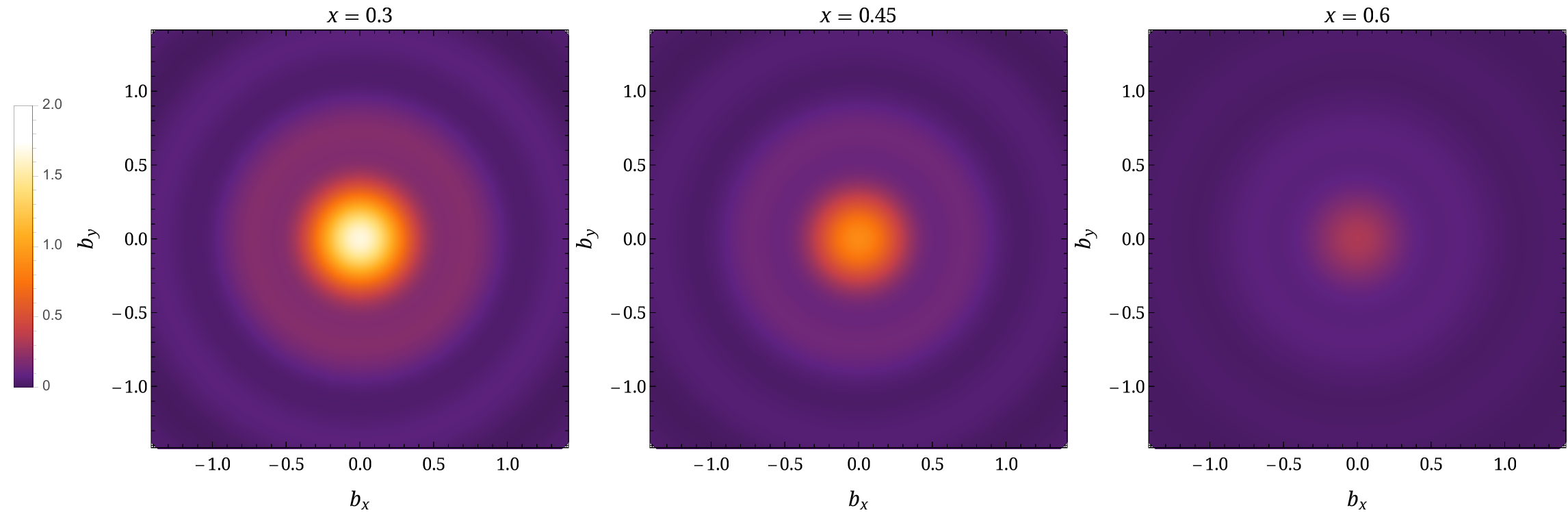}\\
\includegraphics[width=0.85\textwidth]{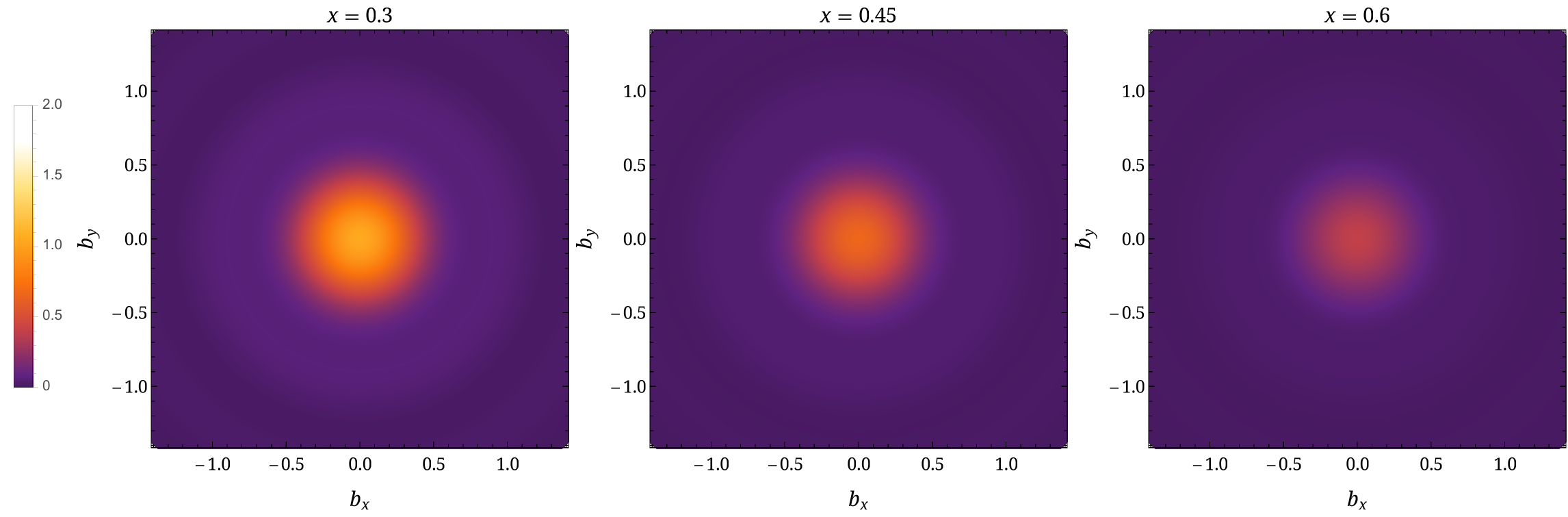}
\end{center}
\caption{
Nucleon (left and middle) and pion (right) tomography: three-dimensional impact-parameter--dependent parton distribution as a function of $x$ and $b$ using lattice $H$, $\tilde{H}$ and $H^\pi$ GPD functions at physical pion mass~\cite{Lin:2020rxa,Lin:2021brq,Lin:2023gxz}.
}
\label{fig:b-density}
\end{figure}

MSULat group further took the Fourier transform of the non--spin-flip $Q^2$-dependent GPD $H(x,\xi=0,Q^2)$ to calculate the impact-parameter--dependent distribution~\cite{Burkardt:2002hr}
\begin{equation}\label{eq:impact-dist}
\mathsf{q}(x,b) =
\int \frac{ \dl{\mathbf{q}}}{(2\pi)^2} H(x,\xi=0,t=-\mathbf{q}^2) e^{i\mathbf{q}\,\cdot \, \mathbf{b} },
\end{equation}
where $b$ is the transverse distance from the center of momentum~\cite{Lin:2020rxa}.
Figure~\ref{fig:b-density} shows the first LQCD results for the impact-parameter--dependent 2D distributions at $x=0.3$, 0.5 and 0.7.
The impact-parameter--dependent distribution describes the probability density for a parton with momentum fraction $x$ at distance $b$ in the transverse plane, providing $x$-dependent nucleon tomography using LQCD for the first time.
Similar tomography results for the helicity GPD, $\tilde{H}(x,\xi=0,Q^2)$ can be found in Ref.~\cite{Lin:2021brq}.

\subsection{Pion Valence-Quark GPDs}

\begin{figure*}[t]
\includegraphics[width=0.32\textwidth]{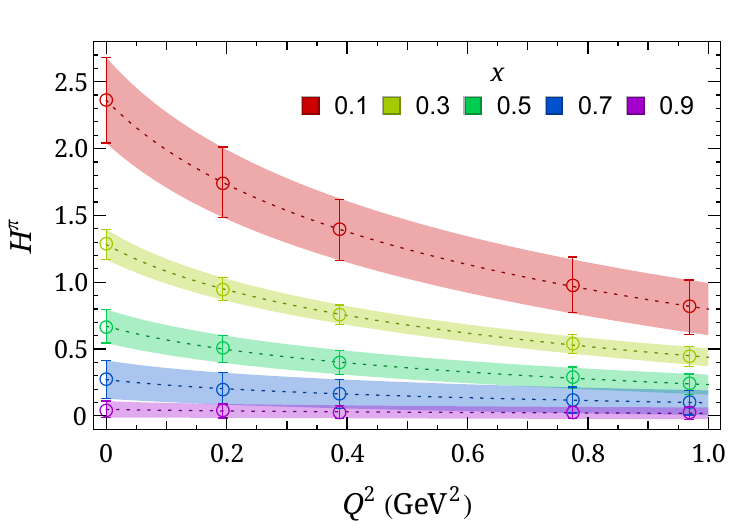}
\includegraphics[width=0.32\textwidth]{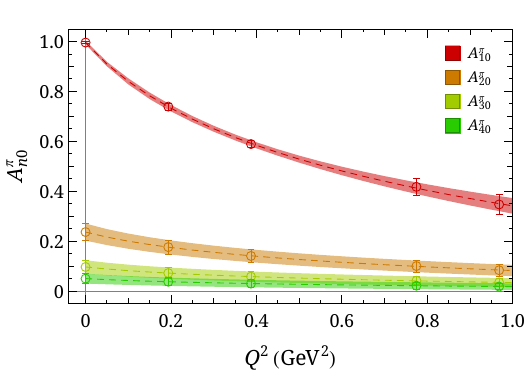}
 \includegraphics[width=0.32\textwidth]{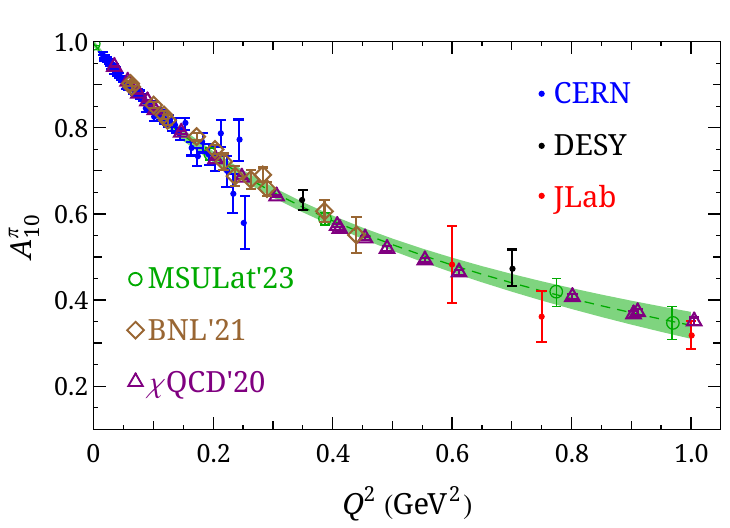}
\caption{\label{Fig:HGPD-moments}
(left) Pion valence-quark GPD at physical pion mass as a function of transfer momentum at selected Bjorken-$x$ indicated in the bands.
The $z$-expansion is used to interpolate between the five transfer momenta.
(middle) Lowest four unpolarized pion GFFs $A_{n0}$ with $n\in[1,4]$ obtained from taking the moment integral using the pion GPD function obtained from this work.
(right) Selected pion form factor $F_\pi(Q^2)$ at the physical pion mass flavours of light quarks from different lattice groups (labeled ``$\chi$QCD'20''~\cite{Wang:2020nbf}, ``BNL'21''~\cite{Gao:2021xsm}), together with the result obtained in this work (labeled ``MSULat'23'') and experimental data \cite{NA7:1986vav}.
The leading moments of the pion GPD are in agreement with prior lattice works and existing experimental data~\cite{JeffersonLab:2008jve,JeffersonLab:2008gyl,Horn:2007ug,JeffersonLabFpi-2:2006ysh,JeffersonLabFpi:2000nlc}.
}
\end{figure*}

Similar efforts have also been made to make progress on pion tomography on the lattice;
the pion is much harder to study experimentally due to its decay.
MSULat also reported the first LQCD $x$-dependent pion valence-quark generalized parton distribution (GPD) calculated directly at $N_f=2+1+1$ $a\approx 0.09$-fm physical pion mass using LaMET with $P_z^\text{max}\approx 1.7$~GeV next-to-next-to-leading order perturbative matching correction~\cite{Lin:2023gxz}.
The pion valence distribution in the zero-skewness limit is renormalized in hybrid scheme with Wilson-line mass subtraction at large distances in coordinate space, followed by a procedure to match it to the $\overline{\text{MS}}$ scheme;
$H^\pi$ as a function of $Q^2$ at selected values of $x$ can be found on the left-hand side of Fig.~\ref{Fig:HGPD-moments}.
The integral of the $H^\pi$ GPD functions is taken to generate leading moment to make comparisons with past LQCD and experimental determinations of the pion form factors.
This found consistent agreement among them, as shown on the right-hand side of Fig.~\ref{Fig:HGPD-moments}.
MSULat predicts the higher GPD moments and reveals the $x$-dependent tomography of the pion for the first time using lattice QCD (see middle of Fig.~\ref{Fig:HGPD-moments}).
This work also shows GPD has a probability-density interpretation in the longitudinal Bjorken $x$ and the transverse impact-parameter distributions (as shown in Fig.~\ref{fig:b-density}).

\subsection{Recent Developments in Systematic Improvements}

There have been new developments in reducing the computational cost in the $x$-dependent GPDs calculations.
ANL-BNL-ETMC collaboration introduced the ``asymmetric'' frame (rather than directly using Breit frame, as shown in the above examples) and were able to calculating more transfer momenta without requiring too much computational resources~\cite{Bhattacharya:2022aob,Bhattacharya:2023jsc}.
This is achieved by using the Lorentz covariant parametrization of the matrix elements in terms of Lorentz-invariant amplitudes, allowing them to connect matrix elements in different frames.
Nucleon isovector polarized and polarized GPDs functions of many momentum transfer at 260-MeV pion mass are reported in Refs.~\cite{Bhattacharya:2022aob,Bhattacharya:2023jsc}, and the pion valence-quark GPD distribution at 300-MeV pion mass in Ref.~\cite{Ding:2024hkz}.
There should be more results at physical pion mass on more GPD functions and tomography in the near future.

Since the aforementioned numerical studies of GPDs, there have been developments in the framework of LaMET including renormalization-group resummation (RGR)~\cite{Su:2022fiu} and leading-renormalon resummation (LRR)~\cite{Zhang:2023bxs}.
RGR is designed to resum the logarithms that arise from the differing intrinsic physical scale and the final renormalization scale of the parton.
The method is to set the energy scale such that the logarithmic terms vanish and then evolve to the desired scale with the renormalization group.
This process can be applied both to the renormalization of the bare matrix elements as well as the perturbative matching.
LRR is designed to resum the divergence arising from the infrared renormalon (IRR) which plagues perturbation series~\cite{Zichichi:1979gj} and whose effect is more pronounced with the application of RGR alone.
The first application of LRR was to the pion PDF in Ref.~\cite{Zhang:2023bxs} and showed that LRR in combination with RGR results in greatly reduced systematics.
Both of these improvements have been applied to distribution amplitudes (DA)~\cite{Holligan:2023rex}, PDFs~\cite{Su:2022fiu,Zhang:2023bxs,Holligan:2024umc,Gao:2023ktu} 
but only MSULat group has applied these improvement to nucleon isovector $H$ and $E$ GPDs~\cite{Holligan:2023jqh} at physical pion mass ensemble with momentum transfers $Q^2=[0, 0.97]$~GeV$^2$ at skewness $\xi=0$ as well as $Q^2\in 0.23$~GeV$^2$ at $\xi=0.1$, renormalized in the $\MSbar$ scheme at scale $\mu=2.0$~GeV, with two- and one-loop matching, respectively. \footnote{Soon after, the authors of Ref.~\cite{Ding:2024hkz} also applies both improvements in their 300-MeV lattice pion GPD $Q^2=[0, 1.69]$~GeV$^2$.}

\begin{figure*}[htp]
 \centering
 \includegraphics[width=0.4\linewidth]{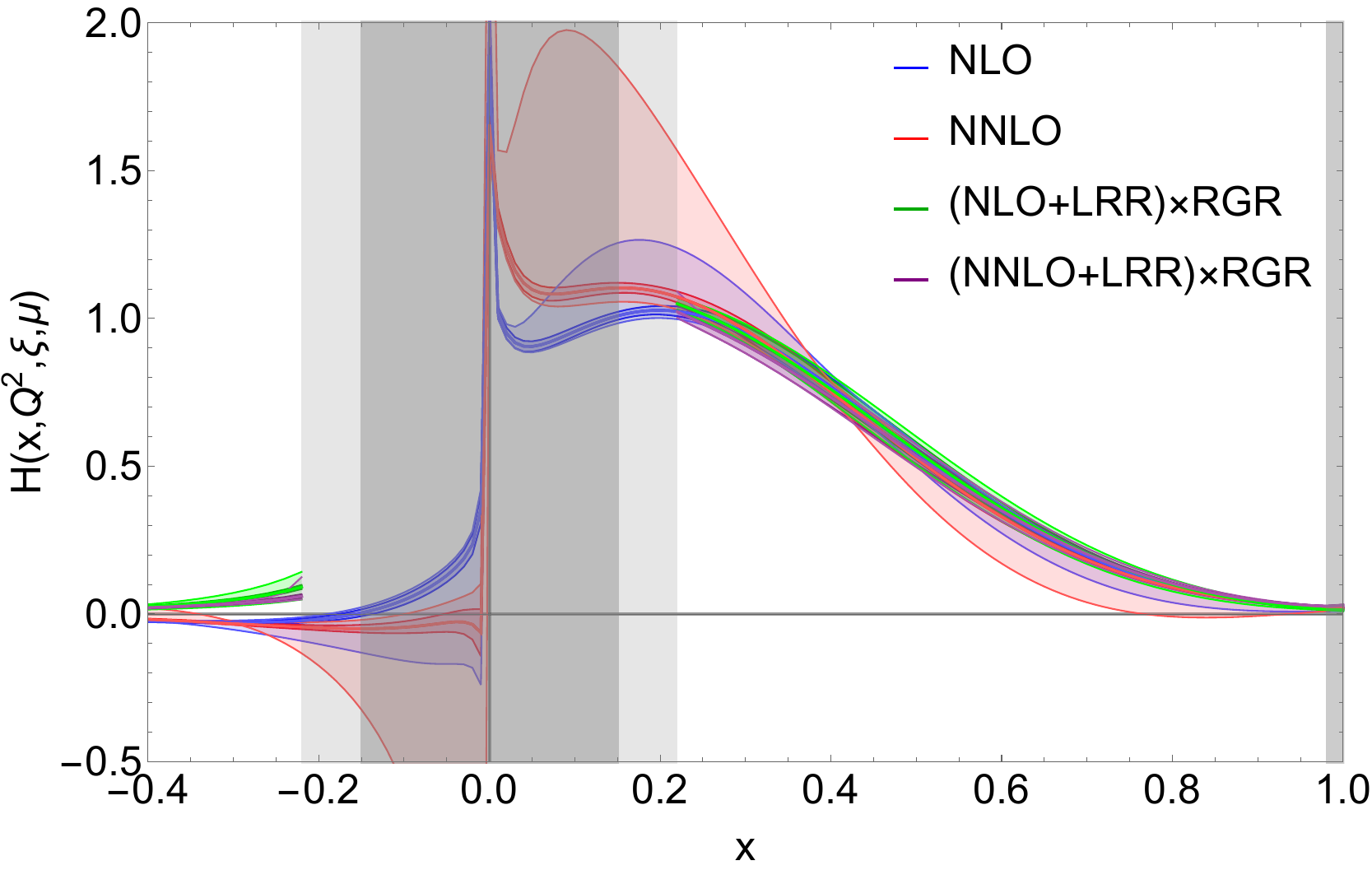}
 \includegraphics[width=0.4\linewidth]{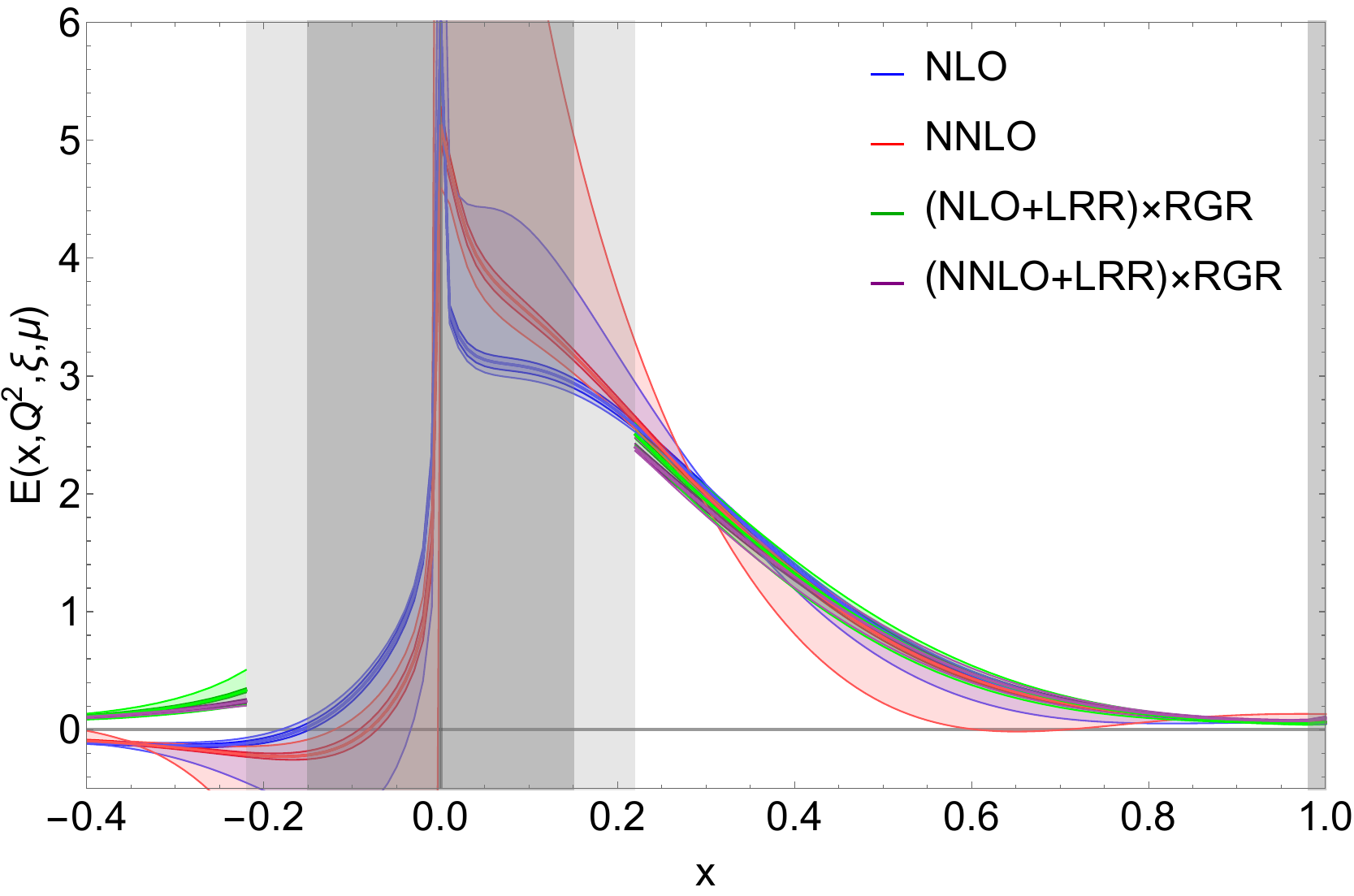}
 \caption{
 Lightcone $H$ and $E$ GPDs (left and right, respectively) with \N\ (blue), \NN\ (red), \NLR\ (green) and \NNLR\ (purple) evaluated at $Q^2=0.39$~GeV$^2$ and $\xi=0$~\cite{Holligan:2023jqh}.
 The inner bands are statistical errors;
 the outer bands are combined statistical and systematic errors, derived from the scale variation. 
 The dark-gray regions are the $x$-values at which the LaMET calculation breaks down.
 In addition, when RGR is applied, the matching formula breaks down for $|x|\lesssim 0.2$, which is shaded in light gray.
 }\label{fig:xi0GPD-Q2-0p39}
\end{figure*}

Figure~\ref{fig:xi0GPD-Q2-0p39} shows example results from the $N_f=2+1+1$ unpolarized $H$ and $E$ GPDs at $a \approx 0.09$-fm physical pion mass with $P_z^\text{max}\approx 2$~GeV in the \N, \NN, \NLR\ and \NNLR\ cases for $Q^2=0.39$~GeV$^2$.
The inner error bars are statistical and the outer error bars are combined statistical and systematic errors the latter computed in the same way as in the $Q^2=0$ case. 
As indicated in the figure, the systematics are at a minimum in the \NNLR\ scheme both for $H$ and $E$ GPDs.
The upper and lower systematic errors increase from \N\ to \NN\ for almost the whole interval $x\in[0.2, 0.8]$ which shows that the need to account for both the large logarithms and the renormalon divergence persists across different $Q^2$ values.
Also, the systematic errors decrease by up to 40\% from \NLR\ to \NNLR\ in the interval $x\in[0.3,0.9]$ both for $H$ and $E$ GPDs.
This shows the benefits of going up to two loops in the matching process.
The central values for all four schemes are in general agreement, showing that the main improvement afforded by RGR and LRR is a reduction in systematic errors.
It is also evidence for convergence in the matching procedure, since the central values for \NLR\ and \NNLR\ are close.
It is to be expected that the improved systematic errors persist across $Q^2$ values since the RGR and LRR improvements are universal and should be applicable in all LaMET calculations.
The fact that the systematics increase from \N\ to \NN\ and decrease from \NLR\ to \NNLR, shows that the handling of systematic uncertainties must keep pace with higher orders in the matching and renormalization processes.
It was found that the simultaneous application of RGR and LRR significantly reduces the systematic errors in renormalized matrix elements and distributions for both the zero and nonzero skewness GPDs, and that it is necessary to include both RGR and LRR at higher orders in the matching and renormalization processes.
More detailed impacts on other momentum transfers for the GPD function and non-zero skewness results can be found in Ref.~\cite{Holligan:2023jqh}.
The application of these methods to GPDs would offer a much more precise calculation of such quantities which would correspond to a more precise calculation of tomography in the future.

%% file: sec5-summary-future.tex
\section{Summary and Future Outlook}

There has been much exciting progress made on the lattice in extracting the parton distributions of hadrons.
Many of the widely calculated hadronic quantities, such as charges, moments and form factors, have made significant progress in the study of their systematics and are entering an era of precision structures for lattice QCD.
The field also overcame longstanding obstacles;
one of the greatest being the ability to study the Bjorken-$x$ dependence of parton distributions, which has now been widely studied with LaMET method and its variants. 
However, there remain challenges to be overcome in the lattice calculations, such as reducing the noise-to-signal ratio in flavor-singlet matrix elements, extrapolating lattice-calculated quantities to the physical-continuum limits, and most importantly, increasing hadronic boosts to suppress systematic uncertainties and reach the large- and smaller-$x$ regions.
Computational resources place significant limitations on the achievable precision, as sufficiently large and fine lattices are necessary to suppress finite-size and higher-twist contaminating contributions.
Novel ideas can bypass these limitations faster with more support from a diverse workforce in the future.
With sufficient support in both computational and human resources, lattice QCD can fill in the gaps where experiments are difficult or not yet available, improve the precision of global fits, and provide better SM inputs to aid new-physics searches across several QCD frontiers.
Snowmass~\cite{Constantinou:2022yye} and EIC Theory Alliance~\cite{Abir:2023fpo} whitepapers and references within provide more details on the rapid advances in LQCD calculations of PDFs, GPDs and other QCD quantities (such meson distribution amplitudes and transverse-momentum--dependent distributions) and have more complete references to relevant work.

\begin{figure}[htbp]
\centering
\includegraphics[width=.9\textwidth]{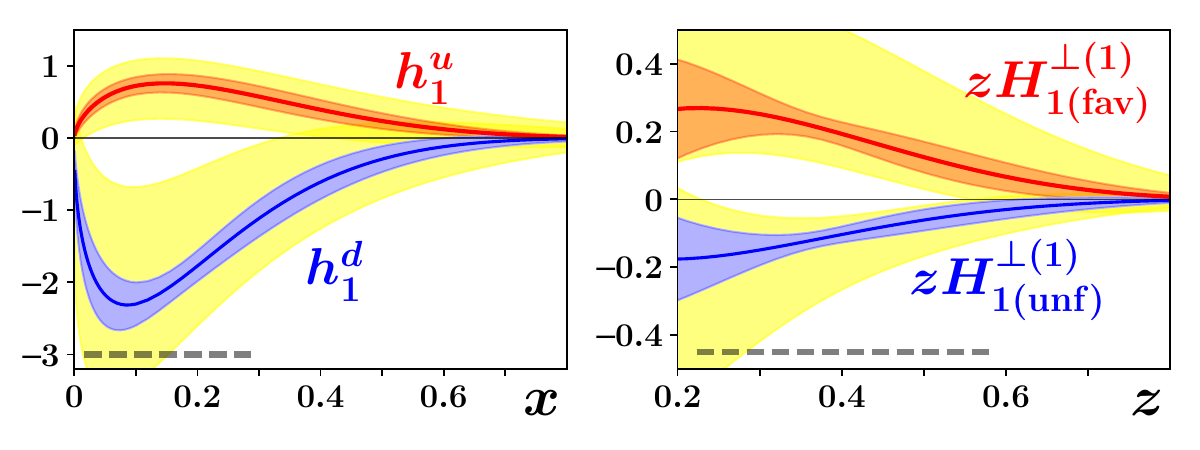}
\caption{
Example of impacts with LQCD input on the global QCD analysis.
Transversity PDFs $h_1^{u,d}$ (left) and favored $zH_{1\text{ (fav)}}^{\perp (1)}$ and unfavored $zH_{1\text{ (unf)}}^{\perp (1)}$ Collins FFs (right) for the SIDIS$+$Lattice-QCD $g_T$ constraint (red and blue bands) at $Q^2=2$~GeV$^2$, compared with the SIDIS-only fit uncertainties (yellow bands).
The range of direct experimental constraints is indicated by the horizontal dashed lines~\cite{Lin:2017stx}.
}
\label{fig:gT-constrained-PDF}
\end{figure}

There have also been multiple new developments in incorporating LQCD inputs into the global QCD analysis framework to compensate supplement kinematic regions where data is either sparse, not as precise or suffers greater systematic uncertainties.
Using precision moments from lattice QCD as inputs can effectively constrain the PDFs.
Consider the transversely polarized PDF, the least known leading twist-2 PDF, for example. 
Reference~\cite{Lin:2017stx} used the lattice-averaged isovector $g_T$ to constrain the global-analysis fits of SIDIS charged-pion production data from proton and deuteron targets, including their $x$, $z$ and {$P_\perp$} dependence, with a total of 176 data points collected from measurements at HERMES and COMPASS.
This gives in principle eight linear combinations of transversity TMD PDFs and Collins TMD FFs for different quark flavors, from which we attempt to extract the $u$ and $d$ transversity PDFs and the unflavored Collins FFs, together with their respective transverse-momentum widths, as shown on the left-hand side of Fig.~\ref{fig:gT-constrained-PDF}.
With lattice calculations alone, one only has limited information about the area under the integral of the up and down-quark distribution, and knows nothing about the shapes of the $x$-dependence.
Without the lattice constraints, the distribution is consistent with zero within 2 sigma;
with both the constraints from the lattice tensor charge and experimental analysis from global-QCD analysis, one is able to make world-best predictions for the large-$x$ transversity for both up and down quarks. 
Similarly, one can also perform a global analysis of GPDs by combining LQCD calculations and experimental measurements including PDFs, form factors and DVCS measurements by parametrizing the GPDs in terms of their moments for zero and small-skewness GPDs~\cite{Guo:2023ahv,Guo:2022upw}.

\begin{figure}[tb]
\centering
\includegraphics[width=0.85\textwidth]{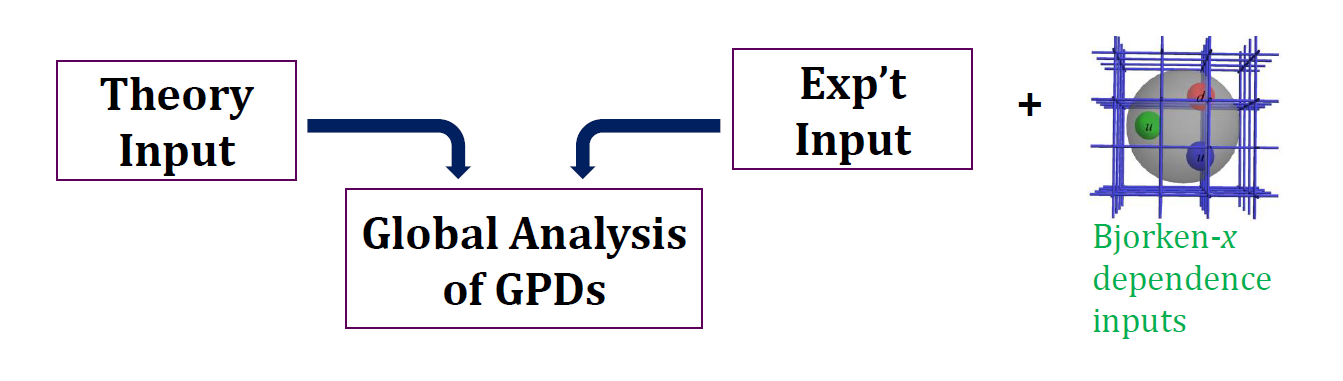}
\caption{
Future global QCD analysis can take in not only experimental data but also lattice $x$-dependent inputs to better constrain the PDFs, GPDs, or other structure, especially in kinematic regions where experimental data is not as precise and/or difficult to obtain.
\label{fig:future-global-fit.png}}
\end{figure}

Now with the past decade of the new $x$-dependent quantities, LQCD calculations can provide a wider range of inputs in the global-QCD analysis. Sec.~\ref{subsec:StrangeAsymFit} provides examples of lattice calculation of the strange-antistrange asymmetry for $x \in [0.3,0.8]$ (a conservative choice due to the boost momentum used being smaller than 2~GeV). 
Nevertheless, one can predict that by reducing the LQCD uncertainty in this region by another factor of 2, there will be even greater improvement in the strange-quark distribution, even down to as low as $x=10^{-6}$, where the LQCD cannot reliably calculate directly.
There are many quantities where lattice $x$-dependent parton distributions can make impacts, for example, charm-anticharm asymmetry and gluon PDFs at Bjorken-$x$ above 0.4.
On the GPDs, LQCD can provide zero-skewness results, as discussed in Sec.~\ref{sec:GPDs}, and explore a few nonzero-skewness GPD results.
Given that GPDs are more complicated to extract from experimental data than PDFs, it is not hard to imagine that LQCD inputs to GPDs can have greater impact on the ongoing efforts in GPD determination~\cite{Almaeen:2022imx,Abir:2023fpo,Burkert:2022hjz} with a pipeline as shown in Fig.~\ref{fig:future-global-fit.png}.

With more complex tasks ahead of the hadronic community, machine learning and AI tools have also been widely adopted by lattice community as well.
For the parton-distribution--related quantities, early work by MSULat~\cite{Zhang:2019qiq} attempted to battle the noise-to-signal problems in gluonic observables, and for large momentum and long Wilson-line displacements by using machine learning to improve the signal.
Machine learning has also been used to extract various parton distributions from lattice-calculated LaMET and pseudo-PDF matrix elements; see Refs.~\cite{Zhang:2020gaj,Gao:2023ktu,Khan:2022vot} 
for a few selected examples.
Efforts in global extraction of Compton form factors and GPDs from deeply virtual exclusive scattering and other experimental processes using machine learning has been in progress~\cite{Almaeen:2022imx,Almaeen:2024guo}.
A proposed pipeline of QCD phenomenology and LQCD inputs into global GPD extraction with the aid of AI was proposed by EXCLAIM (EXCLusives with AI and Machine learning)  collaboration~\cite{Liuti:2024zkc}.
Similar ideas and efforts to explore the unknowns of hadron distributions will be carried out and will lead to greater understanding of the three-dimensional structure of hadrons and tomography in the near future.

\begin{figure}[tb]
\centering
\includegraphics[width=0.85\textwidth]{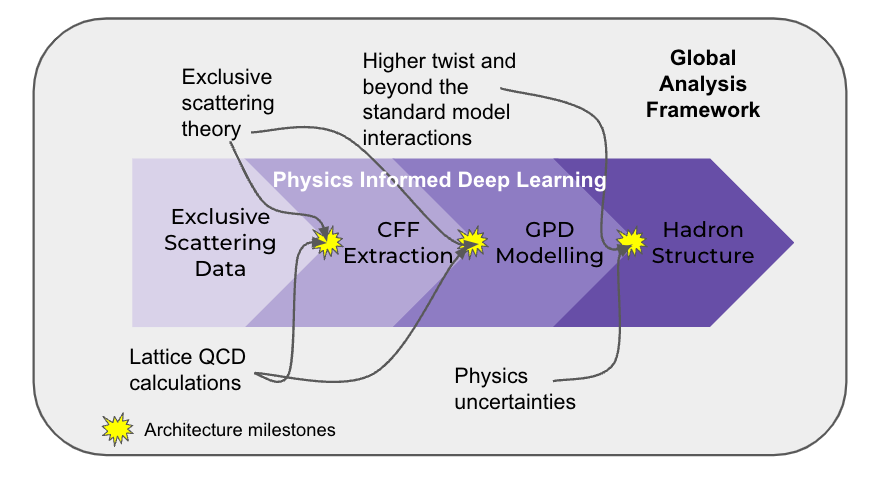}
\caption{
Proposed pipeline of a physics-informed deep-learning framework that goes from exclusive scattering data to information on hadron structure by EXCLAIM collaboration~\cite{Liuti:2024zkc}
\label{fig:GPD-AI-pipeline}}
\end{figure}